\documentclass[12pt]{iopart}

\usepackage{graphicx}% Include figure files
\usepackage{iopams}
\usepackage{lscape}
\usepackage{color, colortbl}
\definecolor{Gray}{gray}{0.9}
\newcolumntype{g}{>{\columncolor{Gray}}r}

\renewcommand{\vec}[1]{\ensuremath{\mathbf{#1}}}
\newcommand{\Mod}[1]{\ \ensuremath{ (\mathrm{mod}\ #1) }}

\newcommand{\lamx}{\lambda_x}
\newcommand{\lamy}{\lambda_y}
\newcommand{\lamz}{\lambda_\parallel}
\newcommand{\ellrt}{\ell_{r|L}}
\newcommand{\ellzt}{\ell_{Z|L}}
\newcommand{\ellra}{\ell_{r|B}}
\newcommand{\ellza}{\ell_{Z|B}}
\newcommand{\ellrza}{\ell_{rZ|B}}
\newcommand{\ellxp}{\ell_{x|P}}
\newcommand{\ellyp}{\ell_{y|P}}
\newcommand{\ellzp}{\ell_{z|P}}
\newcommand{\taucp}{\tau_{c|P}}

\newcommand{\taucpl}{\tau_{\mathrm{life}|P}}
\newcommand{\tauca}{\tau_{c|B}}
\newcommand{\kxp}{k_{x|P}}
\newcommand{\kyp}{k_{y|P}}
\newcommand{\dr}{\Delta r}
\newcommand{\dz}{\Delta Z}
\newcommand{\drij}{\Delta r_{ij}}

\newcommand{\krt}{k_{r|L}}
\newcommand{\kzt}{k_{Z|L}}
\newcommand{\tauct}{\tau_{c|L}}

\newcommand{\tautp}{\tau_{\mathrm{eddy}|P}}
\newcommand{\tautt}{\tau_{\mathrm{eddy}|L}}
\newcommand{\tauta}{\tau_{\mathrm{eddy}|B}}

\newcommand{\kra}{k_{r|B}}
\newcommand{\kza}{k_{Z|B}}
\newcommand{\apsf}{\alpha_\mathrm{PSF}}
\newcommand{\vz}{v_\zeta}
\newcommand{\vZ}{v_Z}
\newcommand{\vr}{v_r}

\newcommand{\dvr}{\delta v_r}

\newcommand{\vpol}{v_\mathrm{pol}}
\newcommand{\dt}{\Delta t}
\newcommand{\ddt}{\Delta t_1}
\newcommand{\dto}{\Delta t_0}
\newcommand{\dtpeak}{\Delta t_{\mathrm{peak}|B}}
\newcommand{\dtpeakt}{\Delta t_{\mathrm{peak}|L}}
\newcommand{\dtpeaka}{\Delta t_{\mathrm{peak}|B}}
\newcommand{\ApeakB}{A_{\mathrm{peak}|B}}
\newcommand{\sigmap}{\sigma_{\mathrm{amp}|P}}
\newcommand{\vth}{v_{\mathrm{th}i}}
\newcommand{\dII}{\overline{\delta I/I}}
\newcommand{\dnn}{\overline{\delta n/n}}

\newcommand{\apsfi}{\alpha_{\mathrm{PSF}i}}
\newcommand{\meann}{\langle n(t)\rangle}
\newcommand{\tauerr}{\hat{\tau}_\mathrm{err}}

\begin{document}

\title[]{Experimental determination of the correlation properties of plasma turbulence using 2D BES systems}

\author{M F J Fox$^{1,2,3}$, A R Field$^{3}$,  F van Wyk$^{1,3,4}$, Y-c Ghim$^{5}$, A A Schekochihin$^{1,2}$, and the MAST Team}\ead{Michael.Fox@physics.ox.ac.uk}
\address{$^{1}$Rudolf Peierls Centre for Theoretical Physics, University of Oxford, Oxford, OX1 3UP, UK.}
\address{$^{2}$Merton College, Oxford, OX1 4JD, UK.}
\address{$^{3}$CCFE, Culham Science Centre, Abingdon, OX14 3DB, UK.}
\address{$^{4}$STFC Daresbury Laboratory, Daresbury, Cheshire, WA4 4AD, UK.} 
\address{$^{5}$Department of Nuclear and Quantum Engineering, KAIST, Daejeon 305-701, Republic of Korea.}

\date{\today}% It is always \today, today,

\begin{abstract}
A procedure is presented to map from the spatial correlation parameters of a turbulent density field (the radial and binormal correlation lengths and wavenumbers, and the fluctuation amplitude) to correlation parameters that would be measured by a Beam Emission Spectroscopy (BES) diagnostic. The inverse mapping is also derived, which results in resolution criteria for recovering correct correlation parameters, depending on the spatial response of the instrument quantified in terms of Point-Spread Functions (PSFs). Thus, a procedure is presented that allows for a systematic comparison between theoretical predictions and experimental observations. This procedure is illustrated using the MAST BES system and the validity of the underlying assumptions is tested on fluctuating density fields generated by direct numerical simulations using the gyrokinetic code GS2. The measurement of the correlation time, by means of the cross-correlation time-delay (CCTD) method, is also investigated and is shown to be sensitive to the fluctuating radial component of velocity, as well as to small variations in the spatial properties of the PSFs.
\end{abstract}

% ----------------------------------------------------------------------------------------------------------
% ----------------------------------------------------------------------------------------------------------
% ----------------------------------------------------------------------------------------------------------
\maketitle

\section{Introduction}\label{sec:introduction}
\begin{figure}
	\centering
	\includegraphics[width=\textwidth]{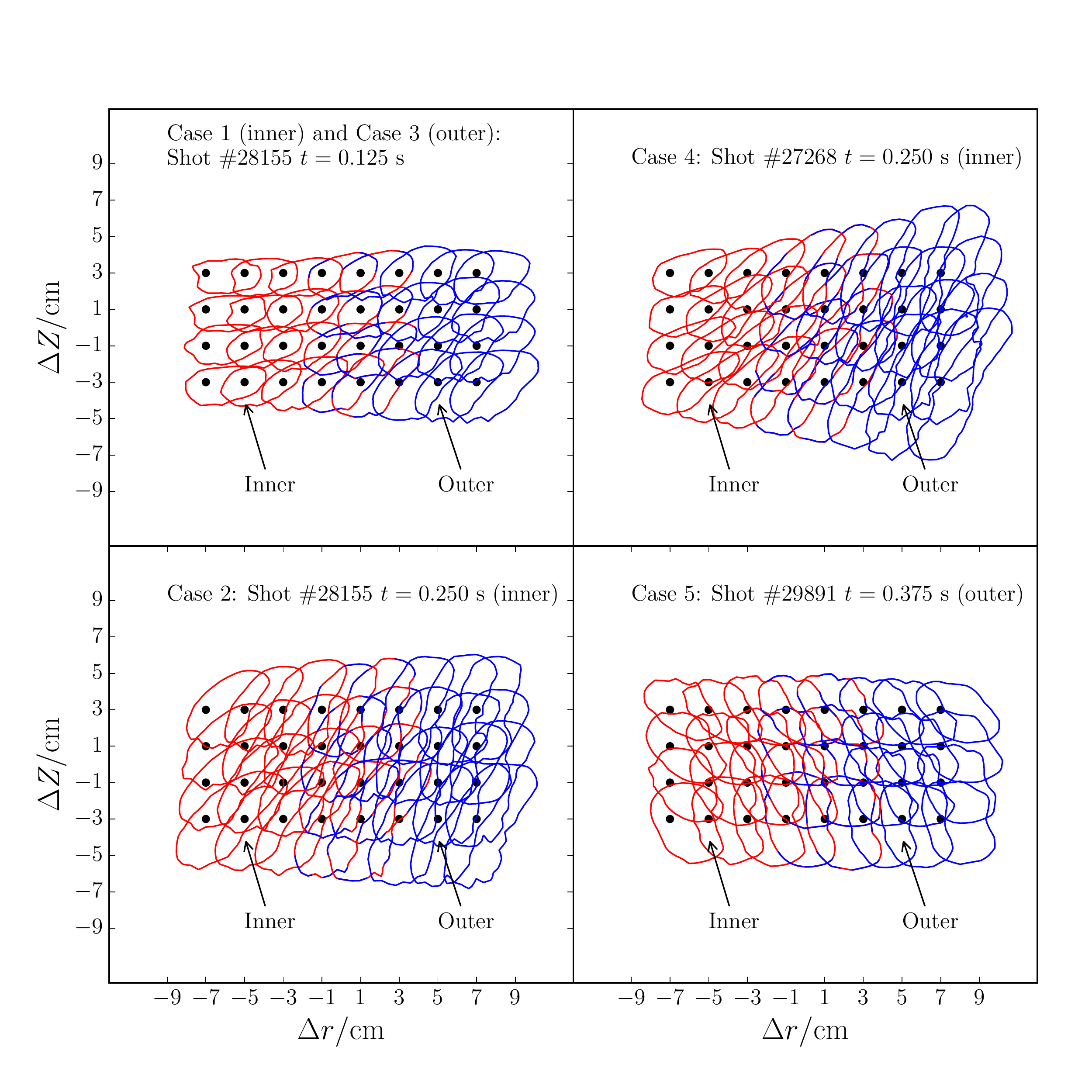}
	\caption[Example PSFs]{Example $e^{-1}$ amplitude contours of PSFs~\cite{Ghim2010} for five specific cases taken from MAST shots and described in Section~\ref{sec:PSFsInMAST}. The dots mark the locations of the focal points of the detector channels. The characterisation of the PSFs is described in Section~\ref{sec:MeasuringrealPSFs}. \label{fig:psf_alpha}}
\end{figure}
Turbulence plays an important role in the transport of particles, momentum and heat in tokamak plasmas~\cite{Liewer1985, Tynan2009}, therefore measurements of the turbulent fluctuating quantities (density, temperature, electric potential, magnetic field and velocity) are essential to validate our understanding of the physical processes responsible for this transport. For example, of particular interest is the effect of velocity shear on the spatial structure of turbulence~\cite{Fedorczak2012a, Shafer2012, Ghim2014} and the relationship between this and improved confinement regimes~\cite{Burrell1997, Highcock2010}. There are also interesting questions of a fundamental nature, e.g., whether the turbulence is in critical balance~\cite{Goldreich1995,Barnes2011,Ghim2013}, or how the transition to turbulence occurs~\cite{vanWyk2016}.

The link between theory and experiment is made through the spatial and temporal correlation functions of the turbulent fields, which are often characterised using correlation lengths and times. However, correlation functions of experimentally measured fluctuating quantities are not directly comparable to the correlation functions of the physical fields that we are interested in. Thus, Beam Emission Spectroscopy (BES) systems~\cite{Fonck1990,Paul1990,Mckee1999,Smith2010,Field2012,Nam2012} measure the fluctuating intensity, $\delta I_i$, of the Doppler-shifted $D_\alpha$ emission from excited neutral beam atoms, which is nontrivially related to the density field in the plasma via the Point-Spread Functions (PSFs) of the diagnostic~\cite{Ghim2012PPCF}, see Figure~\ref{fig:psf_alpha},
\begin{equation}\label{PSF_def}
\delta I_i = \int P_i(r-r_i,Z-Z_i) \beta \delta n(r,Z) \mathrm{d}r \mathrm{d}Z,
\end{equation}
where $\delta n(r,Z)$ is the fluctuating (laboratory-frame) density field inside the plasma at the focal plane of the BES optics, $r$ and $Z$ are the radial and poloidal coordinates, respectively, $P_i(r-r_i,Z-Z_i)$ is the PSF for channel $i$ with focal point at $(r_i,Z_i)$ and $\beta$ is a coefficient weakly dependent on the atomic physics of the line emission~\cite{Ghim2010,Hutchinson2002} ($\beta=1$ is assumed throughout this paper~\cite{Ghim2012PPCF}). The physical field we are interested in is the plasma-frame density field, which can be reached from the laboratory-frame density field by a transformation into field-aligned and rotating (with the mean plasma flow) coordinates. It is clear from (\ref{PSF_def}) that if the PSFs are taken to be delta functions then the measured intensity will be directly proportional to the density field in the laboratory frame. However, the PSFs of BES systems have typical widths greater than the ion gyro-radius, $\rho_i$, and, therefore, are of a similar size to the ion-scale turbulence that is being measured. This raises the following two questions, which will be answered in this paper:
\begin{enumerate}
\item What is the difference between the correlation parameters of the density field in the plasma frame ($P$), namely, the radial, $\ellxp$, and binormal, $\ellyp$, correlation lengths, the radial, $\kxp$, and binormal, $\kyp$, wavenumbers, the correlation time, $\taucp$, and the root-mean square (RMS) fluctuation amplitude, $\overline{\delta n/n}$, and the respective correlation parameters of the intensity field measured by the BES system ($B$): the radial, $\ellra$, and poloidal, $\ellza$, correlation lengths, the radial, $\kra$, and poloidal, $\kza$, wavenumbers, the correlation time, $\tauca$, and the RMS fluctuation amplitude, $\overline{\delta I/I}$? I.e., what is the effect of the PSFs?
\item Given BES measurements of the intensity field's correlation parameters, is it possible to reconstruct the plasma-frame correlation parameters of the density field?
\end{enumerate}

The rest of this paper, focused on answering these questions, is organised as follows. We start, in Section~\ref{sec:measurement}, by describing how the correlation parameters of the intensity field are measured from the BES signal. Then, in Section~\ref{sec:physical_fields}, we derive the relation between the correlation parameters in the plasma frame and in the laboratory frame, which, in the absence of PSF effects, would fully describe how to reconstruct the plasma-frame correlation parameters. In Section~\ref{sec:psfs}, we discuss how the PSFs are calculated, taking our examples from the BES system on the Mega-Ampere Spherical Tokamak (MAST), characterising the PSFs using principal-component analysis, and introducing a simplified Gaussian model of the PSFs. These Gaussian-model PSFs are then used in Section~\ref{sec:PSF_analytic} to calculate analytically the effect of the PSFs on the measured laboratory-frame correlation parameters. Then, in Section~\ref{sec:numeric}, we test the validity of these calculations by comparing the predictions of Section~\ref{sec:PSF_analytic} to the correlation parameters measured from synthetic-BES data generated by evaluating (\ref{PSF_def}) numerically using the real PSFs and a model of a fluctuating density field. Having established that this comparison is reasonably successful, in Section~\ref{sec:inversion} we present equations that allow one to reconstruct the plasma-frame spatial correlation parameters and the fluctuation amplitude from BES measurements. In Section~\ref{sec:testing}, we test this reconstruction procedure by applying real PSFs to density-fluctuation data generated by a non-linear, local, gyrokinetic simulation of MAST turbulence using the GS2 code~\cite{Kotschenreuther1995}, and successfully map from the spatial correlation parameters of this synthetic-BES data to the spatial correlation parameters of the fluctuating density field. In Section~\ref{sec:gs2temporal}, we find that the PSFs have an effect also on the measurement of the correlation time and establish that this can be due to the presence of a fluctuating radial component of velocity; however, the Gaussian model of PSFs is shown to be unable to account correctly for this radial velocity effect. Finally, in Section~\ref{sec:conclusion}, we summarise, and discuss the implications of, our results.

Some technical details of our models and procedures are given in the Appendices. In \ref{sec:poloidal_corr_lengths}, we test our improved method for measuring the poloidal correlation length.  In \ref{sec:toymodel_appendix}, we present the model of fluctuating fields that we use for the tests in Section~\ref{sec:numeric}. In \ref{sec:appendix_CCTD}, we calculate the correlation time and apparent poloidal velocity inferred from our assumed form of the correlation function, with and without including PSF effects. In \ref{sec:psf_testing}, we quantify the differences between the real and Gaussian-model PSFs.

% ----------------------------------------------------------------------------------------------------------
% ----------------------------------------------------------------------------------------------------------
% ----------------------------------------------------------------------------------------------------------
\section{Correlation parameters of the measured BES signal}\label{sec:measurement}
A BES system consists of an array of detector channels each of which receives photons from a spatially localised region within the plasma. This region is assumed to lie in a plane described by the radial $r$ and poloidal $Z$ coordinates (the plane of detection). BES systems are designed to resolve fluctuations that vary on the turbulent-fluctuation timescale of a few microseconds. Therefore, both spatial and temporal properties of the turbulence can be investigated. In this section, we describe the operational definitions of the parameters that characterise the correlation function of the signals detected by a BES diagnostic. 

Each BES detector channel $i$ measures the intensity of photons as a function of time,
\begin{equation}\label{intensity_decomp}
I_i(t) = \langle I_i(t) \rangle + \delta I_i(t),
\end{equation}
where we have split the signal into its temporal mean $\langle I_i(t) \rangle$ and fluctuating part $\delta I_i(t) \equiv I_i(t) - \langle I_i(t) \rangle$. The detector channel $i$ can be associated with a viewing location in the radial-poloidal plane given by the coordinates $(r_i, Z_i)$, which are determined by the focal point of the optical system for that channel.

All the parameters of the turbulence that we will be considering are determined from measurements of the covariance function of the fluctuating part of the time series. The covariance between two detector channels $(i,j)$ is defined as
\begin{equation}\label{covfun}
C_{ij}^\mathrm{cov}(\dt) =\frac{\langle \delta I_i(t) \delta I_j(t+\dt) \rangle}{ \langle I_i(t) \rangle \langle I_j(t+\dt) \rangle},
\end{equation}
where $\dt$ is the time delay between the signals. Thus, the covariance function is a function of the time delay $\dt$ and of the spatial separation of the two channels $(\Delta r_{ij}, \Delta Z_{ij})= (r_i-r_j, Z_i-Z_j)$. When the auto-covariance ($i=j$) is calculated, (\ref{covfun}) needs to be corrected for auto-correlated photon noise, as described in~\cite{Ghim2012PPCF}. The correlation function is defined as the normalised covariance function
\begin{equation}\label{correlation_def}
C_{ij}(\dt) =\frac{\langle \delta I_i(t) \delta I_j(t+\dt) \rangle}{\sqrt{\langle \delta I_i^2(t)\rangle \langle \delta I_j^2(t+\dt) \rangle}}.
\end{equation}

% ----------------------------------------------------------------------------------------------------------
% ----------------------------------------------------------------------------------------------------------
% ----------------------------------------------------------------------------------------------------------

\subsection{Fluctuation amplitude}\label{sec:meas_fluctuationamp}
The fluctuation amplitude of the signal is the mean, over a set of $N$ channels, of the square root of the auto-covariance function (\ref{covfun}) at $\dt = 0$,
\begin{equation}
\overline{\delta I/I} \equiv \sqrt{ \frac{1}{N} \sum_{i=0}^{N-1}\frac{\langle\delta I_i^2(t)\rangle } {\langle I_i(t) \rangle^2}} = \sqrt{\frac{1}{N} \sum_{i=0}^{N-1} C_{ii}^\mathrm{cov}(0)}. \label{flucAmpDef}
\end{equation}

% ----------------------------------------------------------------------------------------------------------
% ----------------------------------------------------------------------------------------------------------
% ----------------------------------------------------------------------------------------------------------

\subsection{Correlation time}\label{sec:meas_corr_time}
\begin{figure}
\centering
\includegraphics[width=0.8\textwidth]{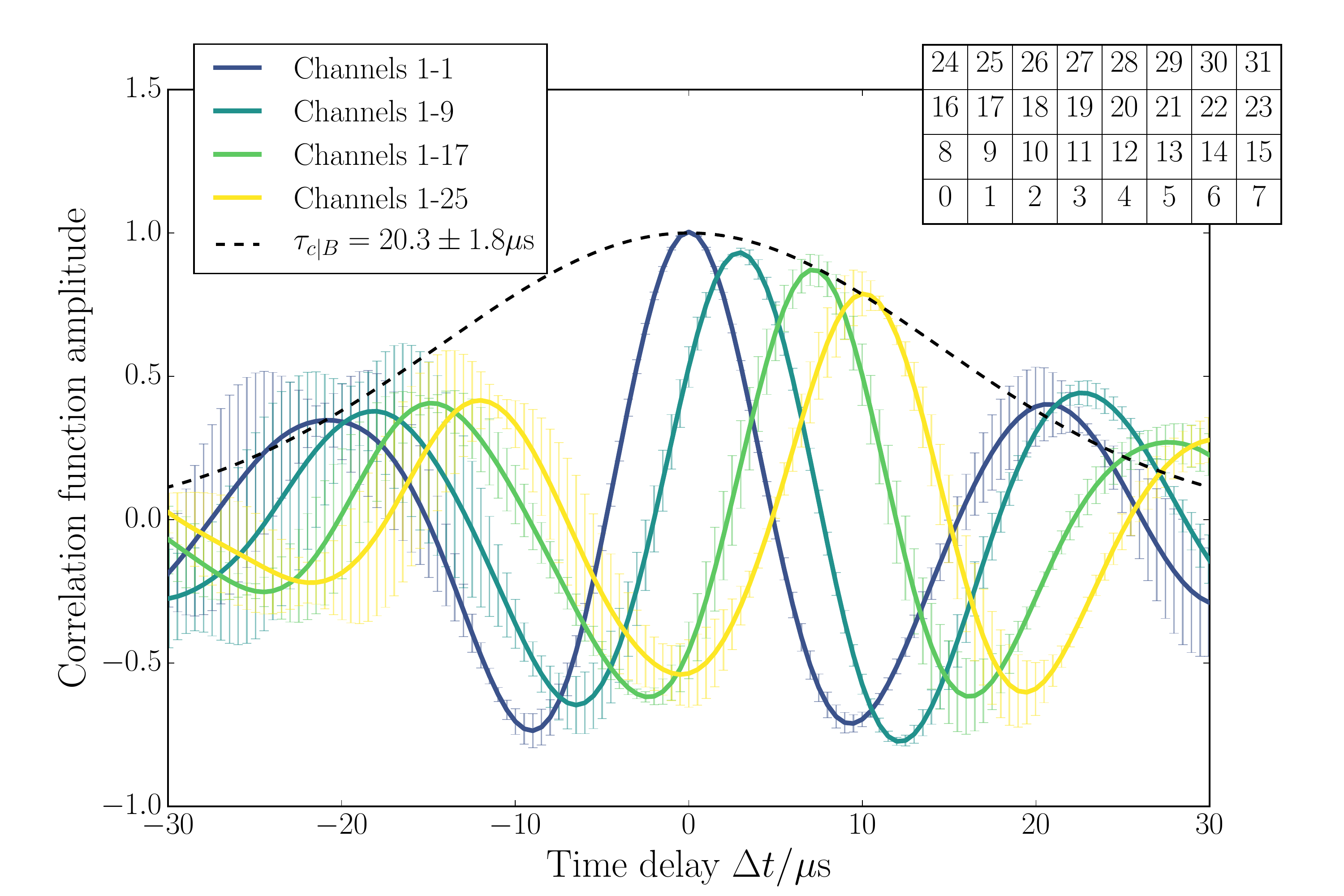}
\caption[Example time delay correlation functions]{Time delay auto- and cross-correlation functions taken from shot $\# 28155$ at $t=126\ \mathrm{ms}$ using channels 1, 9, 17 and 25, with channel 1 as the reference channel (see inset for their positions in the BES array). The black dashed line is the result of the CCTD technique where (\ref{correlation_timedelay_fit}) is fit to the global maxima (peaks) of the cross-correlation functions.\label{fig:example_time_delay}}
\end{figure}

The standard technique~\cite{Durst1992, Ghim2012PPCF} for extracting a correlation time from BES measurements relies on the toroidal rotation of the plasma with velocity $\vz$ to advect turbulent structures, which are field-aligned and anisotropic~\cite{Liewer1985,Ritz1988,Koch2007} (the correlation length parallel to the magnetic field is much greater than the perpendicular lengths), past the plane of detection. The structures, therefore, appear to move in the poloidal direction in the plane of detection with velocity $\vz \tan\alpha$, where $\alpha = \arctan(B_Z/B_\zeta)$ is the pitch angle of the magnetic field, with $B_Z$ the poloidal and $B_\zeta$ the toroidal components of the magnetic field. Provided that the minimum passing time of a turbulent structure, $\tau_\mathrm{min}=\dz_\mathrm{min}/(\vz\tan\alpha)$, where $\dz_\mathrm{min}$ is the minimum poloidal distance between detector channels, is less than the correlation time of the turbulence, the turbulent structure will be seen to decay as it passes multiple detector channels.

By considering the correlation function between poloidally, but not radially, separated detector channels (such that $\dr_{ij} = 0$), the decay in the amplitude can be observed, an example of which is shown in Figure~\ref{fig:example_time_delay}. To measure this decay, we first identify the amplitude, $\ApeakB(\dz_{ij})$, and the time delay, $\dtpeak(\dz_{ij})$, of the peak of the time-delay correlation function, $C(\dz_{ij}, \dt)$, for each value of poloidal separation, $\dz_{ij}$, by fitting the function
\begin{equation}\label{peak_finder}
f_\mathrm{peak}(\dt) \equiv \ApeakB(\dz_{ij}) \exp\left( - \frac{[ \dt - \dtpeak (\dz_{ij})]^2}{\tau_\mathrm{eff}^2} \right)
\end{equation}
to the data points selected in the vicinity of the peak, where the fitting parameter $\tau_\mathrm{eff}$ is an effective decay time of the peak. We have introduced the subscript $B$ to identify parameters that describe the covariance/correlation properties of the measured intensity field (i.e., the BES signal). Then, using the pairs of amplitudes and time delays $(\ApeakB(\dz_{ij}), \dtpeak(\dz_{ij}))$ for each $\dz_{ij}$, we find the correlation time of the BES-measure intensity signal, $\tauca$, by minimising
\begin{equation}\label{correlation_timedelay_fit}
\chi^2 = \sum_{\dz_{ij}} \left[ \ApeakB(\dz_{ij}) - \exp\left( - \frac{\dtpeak^2(\dz_{ij})}{\tauca^2} \right) \right]^2.
\end{equation}
This technique is known as the cross-correlation time-delay (CCTD) method~\cite{Durst1992} and the parameter $\tauca$ can be considered, by definition, to be the correlation time of the BES signal, although, as we will see, the interpretation of this in terms of physical turbulent fields is complicated (see Section~\ref{sec:gs2temporal} and \ref{sec:toy_CCTD}).

% ----------------------------------------------------------------------------------------------------------
% ----------------------------------------------------------------------------------------------------------
% ----------------------------------------------------------------------------------------------------------

\subsection{Apparent poloidal velocity}\label{sec:meas_app_pol_vel}

Using the CCTD technique, it is also possible to measure the apparent poloidal velocity, given that we know the distances between the viewing locations of the detector channels $\dz_{ij}$, and the time delays of the peaks $\dtpeak (\dz_{ij})$ of the cross-correlation functions. Then,
\begin{equation}
v_{\mathrm{pol}|B}\equiv \frac{ \dz_{ij} }{\dtpeak},
\end{equation}
assuming that neither the toroidal velocity nor pitch angle change during the measurement time\footnote{We use $v_{\mathrm{pol}|B}$ here to distinguish this apparent velocity, mainly due to $\vz\tan\alpha$, from the `true' poloidal velocity $\vZ$, which we introduce later in Section~\ref{sec:plasma_frame}.}. Practically, as multiple values of $\dz_{ij}$ are available, a linear fit is used to find $v_{\mathrm{pol}|B}$. The efficacy of this measurement method is supported by successful cross-diagnostic comparisons with charge-exchange recombination spectroscopy~\cite{Ghim2012PPCF} and, more recently, with Doppler backscattering~\cite{Hillesheim2015}.

% ----------------------------------------------------------------------------------------------------------
% ----------------------------------------------------------------------------------------------------------
% ----------------------------------------------------------------------------------------------------------

\subsection{Two-dimensional spatial correlation parameters}\label{sec:ztd_measurement}
We now consider the spatial properties of the correlation function (\ref{correlation_def}) by setting the time delay to zero, $\dt=0$. We also make the assumption that the turbulence is homogeneous, drop the indices $i,j$, and treat the correlation function as a function of $\dr$ and $\dz$ only. In order to extract the spatial correlation parameters, the following function, which is similar in form to that used in~\cite{Shafer2012}, is  fitted to the correlation function (\ref{correlation_def}):
\begin{equation}\label{fit_fun}
C_{\mathrm{fit}}(\dr, \dz) = p + (1-p)\exp\left(-\frac{\dr^2}{\ellra^2}-\frac{\dz^2}{\ellza^2}\right) 
\cos(\kra \dr + \kza \dz),
\end{equation}
where $\ellra$ is the radial correlation length, $\ellza$ is the poloidal correlation length, $\kra$ is the radial wavenumber, and $\kza$ the poloidal wavenumber. The parameter $p$ is used with experimental data to account for offsets caused by global MHD modes~\cite{Ghim2012PPCF} or beam-fluctuation effects~\cite{Moulton2015}. We also define, for later convenience, the tilt angle of the correlation function as
\begin{equation}\label{theta_B_def}
\Theta_B = -\arctan\left(\frac{\kra}{\kza}\right).
\end{equation}

As the spatial distribution of viewing locations is often sparse for BES systems (the MAST BES has only $8\times4$ radial-poloidal channels), the fit of (\ref{fit_fun}) to the measured correlation function (\ref{correlation_def}) can be insufficiently constrained. Therefore, we impose a further constraint by fixing the product $\kza\ellza$ in the above fitting procedure. The value of this product is determined from the shape of the time-delayed auto-correlation function, using the fact that a turbulent perturbation is advected past a single detector channel by the bulk velocity of the plasma (Section~\ref{sec:meas_corr_time}), therefore encoding the spatial structure of the perturbation in the temporal domain of the detected signal~\cite{Taylor1938, Smith2012}. 

The procedure for determining $\kza\ellza$ is as follows. First we measure the correlation time $\tauca$ as described in Section~\ref{sec:meas_corr_time}. Then we calculate the time-delayed auto-correlation function (\ref{correlation_def}) and multiply it by the correction factor $\exp(\dt^2/\tauca^2)$, which accounts for the fact that the amplitude of the turbulence decays in time, so that the resulting corrected auto-correlation function only includes information about the decay due to the poloidal correlation length. Finally, the following function is fitted to this corrected auto-correlation function (see discussion in~\ref{sec:poloidal_corr_lengths} and Section~\ref{sec:autocorrelation}):
\begin{eqnarray}\label{time_delay_shape}
C_\mathrm{auto}(\dt) &=& \exp\left[ - \frac{\vz^2\dt^2\tan^2\alpha}{\ellza^2} \right] \cos\left( \kza \vz \dt \tan\alpha \right), \\
&=&  \exp\left[ - \frac{\dt^2}{\ellza'^2} \right] \cos\left( \kza' \dt \right),\label{eff_time_delay_shape}
\end{eqnarray}
where the primed quantities are the fitting parameters, which have the property that $\kza' \ellza' = \kza \ellza$ and, therefore, the requirement to know $\vz \tan\alpha$ is eliminated. This method is tested successfully in \ref{sec:poloidal_corr_lengths} using our model of a fluctuating density field (described in Section~\ref{sec:toymodel} and \ref{sec:toymodel_appendix}).

\subsubsection{Spatial correlation parameters from the MAST BES.}\label{sec:example_spatial_corr_fun}
\begin{figure}
\centering
\includegraphics[width=\textwidth]{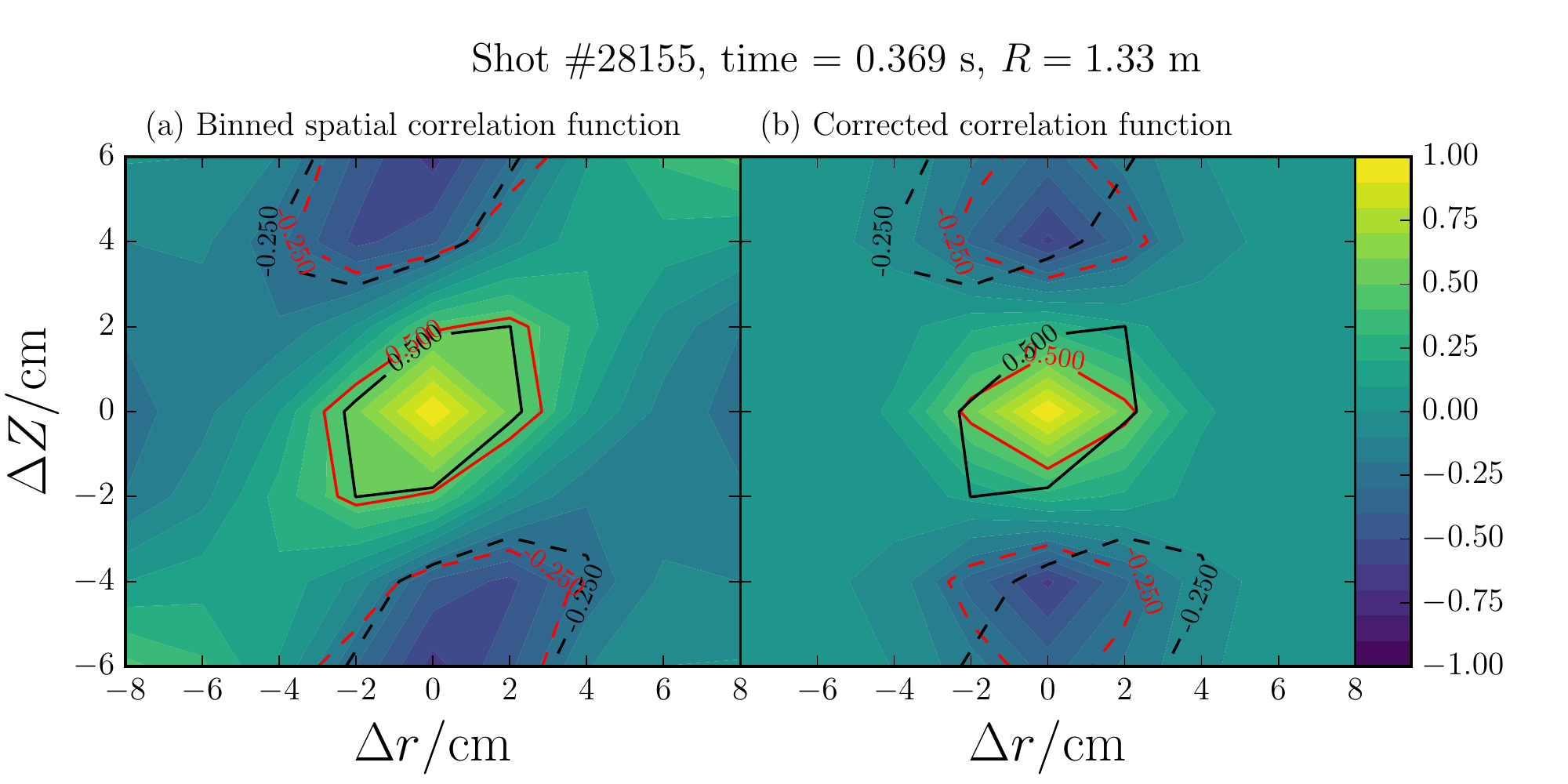}
\caption[Example binned and corrected spatial correlation function]{(a) The zero-time-delay binned spatial correlation function (see Section~\ref{sec:example_spatial_corr_fun}) of the BES intensity signal in MAST shot \#28155 at $t=0.369\ \mathrm{s}$ plotted using the outer set of channels. The red contours show the fit (\ref{fit_fun}) to the data. (b) The corrected correlation function using (\ref{an_ellrt_inv}-\ref{an_tantheta_inv}) and the PSF parameters $L_1=3.1\ \mathrm{cm}, L_2=1.5\ \mathrm{cm}$, and $\apsf=-24^\circ$ (defined in Section~\ref{sec:MeasuringrealPSFs}). These PSF parameters are similar to those of Case 2, described in Section~\ref{sec:PSFsInMAST} and shown in Figure~\ref{fig:psf_alpha}. In (b) the black contour line is the same as that in (a), the red contour line is for the corrected parameters. The fitted and corrected parameter values are given in Table~\ref{tab:example_data}. \label{fig:example_data} }
\end{figure}
\begin{table}
\centering
\begin{tabular}{r | r r r r r}% Taken from note on 21/09/2016 - spike filtered analysis
	Parameter				& $\ell_r/\mathrm{cm}$& $\ell_Z/\mathrm{cm}$ & $k_r/\mathrm{cm}^{-1}$ & $k_Z/\mathrm{cm}^{-1}$ & $\Theta/\mathrm{deg}$  \\ \hline
	(a) BES fit value 			& $3.7\pm0.5$ & $8.6\pm0.7$ & $-0.20\pm0.07$ & $0.53\pm0.01$ & $21\pm7$ \\ 
	(b) Corrected value (lab) 	& $2.6\pm0.7$ & $7.7\pm0.8$ &  $0.03\pm0.20$  & $0.66\pm0.03$ & $-2\pm17$

\end{tabular}
\caption[Table of example spatial fit parameters for BES data in Figure~\ref{fig:example_data}.]{Fitting parameters, using (\ref{fit_fun}), for Figure~\ref{fig:example_data}.\label{tab:example_data}}
\end{table}
The BES system on MAST has a set of 32 channels arranged into an $8\times 4$ radial-poloidal array, with spacing between channels of approximately $2\ \mathrm{cm}$. As this BES array covers almost a quarter of the minor radius of the tokamak, it is not expected that the turbulence is precisely homogeneous across the entire array. Therefore, we split it into two radial-poloidal sub-arrays of $5\times4$ channels each, referred to here as ``inner" and ``outer" arrays.

A complete set of spatial correlation functions (\ref{correlation_def}) for a single sub-array can be represented by a $20\times20$ matrix. In order to visualise this matrix, it is necessary to associate spatial coordinates with each matrix element. As the spacing between BES channels is uniform, and we are assuming that the turbulence is homogeneous, a single pair of relative coordinates $(\dr,\dz)$ can correspond to multiple values in the spatial-correlation-function matrix. To resolve this problem, a binned spatial correlation function is constructed by averaging over all values of the spatial correlation function matrix that have the same relative coordinates. An example binned correlation function is shown in Figure~\ref{fig:example_data}(a). The slight difference between the binned correlation function contours and the contours of the fit (\ref{fit_fun}), shown in red, is due to the fact that this fit, along with all others in this work, has been made to the complete set of values of the spatial correlation function, with no binning.

% ----------------------------------------------------------------------------------------------------------
% ----------------------------------------------------------------------------------------------------------
% ----------------------------------------------------------------------------------------------------------
\section{Correlation parameters of the physical density field}\label{sec:physical_fields}
The physical field of interest is the ion number density. The correlation function of this field is most easily described in the plasma frame, which will be defined in Section~\ref{sec:plasma_frame}. In Section~\ref{sec:lab_frame}, the transformation of correlation parameters from the plasma frame to the laboratory frame (that of the experimental observer) is made. The correlation parameters of the density field in the laboratory frame would be equivalent to those of the intensity field measured by a BES system if the PSFs in (\ref{PSF_def}) were delta functions. However, as we will show, making this assumption is rarely justified, and, therefore, the distinction between the laboratory-frame and BES measurements is emphasised here: the ``laboratory frame" refers specifically to the correlation properties of the density field, whilst BES measurements are of the correlation properties of the intensity field. 

\subsection{Plasma frame}\label{sec:plasma_frame}
The plasma frame is a coordinate system aligned with the magnetic field and moving with the mean velocity of the plasma, i.e., rotating with the toroidal $\vz$ and poloidal $\vZ$ plasma velocities. The coordinates $(x, y, z)$ describe the spatial position of a plasma element in this frame in the radial $(\vec{\hat{x}})$, binormal $(\vec{\hat{y}})$ and parallel to the magnetic field $(\vec{\hat{z}})$ directions. We assume that the correlation function of the turbulent density field in these coordinates can be fitted by\footnote{We assume there is no oscillatory structure in the parallel direction: $k_{z|P}=0$; and note that the inclusion of a finite $k_{z|P}$ should not considerably alter our results, as it will be ordered (see Section~\ref{sec:asymtotics}) the same size as $\ellzp^{-1}$, which is small compared to the lengths $\ellxp^{-1}$ and $\ellyp^{-1}$.}
\begin{equation}\label{plasma_frame_correlation_function}
\fl C_P(\Delta x, \Delta y, \Delta z, \Delta t) =  \exp\left( - \frac{\Delta x^2}{\ellxp^2} - \frac{\Delta y^2}{\ellyp^2} - \frac{\Delta z^2}{\ellzp^2} - \frac{\Delta t^2}{\taucp^2} \right) \cos \left( \kxp \Delta x + \kyp \Delta y\right),
\end{equation}
where $\ellxp$ is the radial correlation length, $\ellyp$ is the binormal correlation length, $\ellzp$ is the parallel correlation length, $\taucp$ is the correlation time, $\kxp$ is the radial wavenumber and $\kyp$ is the binormal wavenumber. We have used the subscript $P$ to label correlation parameters defined in the plasma frame.
% ----------------------------------------------------------------------------------------------------------
% ----------------------------------------------------------------------------------------------------------
% ----------------------------------------------------------------------------------------------------------

\subsection{Laboratory frame}\label{sec:lab_frame}
In the laboratory frame, we measure a fluctuating time series in a radial-poloidal cross-section described by the coordinates $(r, Z)$. Therefore, as well as transforming out of the plasma frame, we also have to consider the projection onto this two-dimensional plane. In~\ref{sec:lab_frame_corr}, the full expression (\ref{lab_corr_function_full}) for the laboratory-frame correlation function $C_L(\dr,\dz,\dt)$ is given. For clarity, here we consider the spatial and temporal laboratory-frame correlation functions separately, in order to relate measurable parameters in the laboratory frame to the plasma-frame parameters.

\subsubsection{Spatial correlation function.}
The spatial laboratory-frame correlation function defined from (\ref{lab_corr_function_full}) as $C_{\mathrm{spatial}|L}(\dr, \dz) \equiv C_L(\dr,\dz,\dt = 0)$, can be written in the form
\begin{equation}\label{lab_frame_correlation_function}
C_{\mathrm{spatial}|L}(\dr, \dz) = \exp \left( -\frac{\dr^2}{\ellrt^2} - \frac{\dz^2}{\ellzt^2} \right)\cos\left( \krt \dr+ \kzt \dz \right),
\end{equation}
where the relationships of the laboratory-frame correlation parameters to the plasma-frame correlation parameters of (\ref{plasma_frame_correlation_function}) are
\begin{eqnarray}
\ellrt &=& \ellxp, \label{ellrt_ellxp} \\
\ellzt &=&  \left(\frac{\cos^2\alpha }{ \ellyp^2} + \frac{ \sin^2\alpha}{\ellzp^2}\right)^{-1/2}, \label{lab_pol_length} \\
\krt &=& \kxp, \label{krt_kxp} \\
\kzt &=& \kyp  \cos\alpha, \label{kzt_kyp}
\end{eqnarray}
and we have introduced the subscript label $L$ to identify the laboratory-frame parameters. The correlation function (\ref{lab_frame_correlation_function}) has the same functional form as the fit function (\ref{fit_fun}), used to extract the BES correlation parameters (with $p=0$).

\subsubsection{Temporal correlation function.}\label{sec:autocorrelation}
The time-delayed, single-point auto-correlation function defined from (\ref{lab_corr_function_full}) as $C_{\mathrm{temporal}|L}(\dt) \equiv C_L(\dr=0,\dz=0,\dt)$, is 
\begin{equation}\label{lab_temporal_corr_func}
C_{\mathrm{temporal}|L}(\dt) =  \exp \left( -  \frac{\dt^2}{\tau_\mathrm{auto}^2} \right)\cos\left(  \kyp\vz \dt \sin\alpha     + \kyp\vZ\dt\cos\alpha   \right),
\end{equation}
where
\begin{equation}\label{tauauto}
\tau_\mathrm{auto} \equiv \left[\frac{ (\vz \sin\alpha + \vZ\cos\alpha)^2}{ \ellyp^2 } + \frac{ (\vz\cos\alpha + \vZ\sin\alpha)^2}{ \ellzp^2} + \frac{1}{ \taucp^2}\right]^{-1/2},
\end{equation}
is a function of the spatial correlation parameters as well as of the plasma-frame correlation time $\taucp$. The functional form of $C_{\mathrm{temporal}|L}$ above is different from the $C_\mathrm{auto}$ in (\ref{time_delay_shape}), because in (\ref{time_delay_shape}) we have only kept the lowest-order (see Section~\ref{sec:asymtotics}) contributions to the full temporal correlation function (\ref{lab_temporal_corr_func}).

The CCTD method (see Section~\ref{sec:meas_corr_time}) attempts to remove the spatial terms in (\ref{tauauto}) by assuming that the spatial properties of the turbulence remain constant in time and, therefore, the difference in the peak amplitudes of the time-delayed cross correlations is only due to the temporal decorrelation. The correlation time resulting from the CCTD method is $\tauct$. To interpret this correlation time, we have to find the relationship between $\tauct$ and the plasma-frame correlation time $\taucp$. To do this, we can simply find the maximum of the time-delayed cross-correlation function  (\ref{lab_corr_function_full}) at a fixed poloidal displacement $\dz$ and with $\dr=0$. However, it is informative, and will become useful for later discussions in Section~\ref{sec:gs2temporal}, to approach the problem by using an asymptotic expansion based on the typical time and spatial scales of the problem.

\subsubsection{Asymptotic ordering for CCTD method.}\label{sec:asymtotics}
We start by assuming that the ion gyroradius $\rho_i$ is small compared to the minor radius of the tokamak $a$, providing us with the small parameter 
\begin{equation}\label{epsilon_def}
\epsilon\equiv\frac{\rho_i}{a},
\end{equation}
which is just the gyrokinetic ordering~\cite{Abel2013}. The perpendicular correlation lengths of the turbulence have been measured to be typically of order a few ion gyroradii (see Figure~\ref{fig:example_data}, where $\rho_i=1.1\ \mathrm{cm}$), i.e., $\ellxp,\ellyp\sim\rho_i$, while the parallel correlation length is significantly longer $\ellzp\sim a$~\cite{Liewer1985,Ritz1988,Koch2007}. Therefore, we order the perpendicular spatial correlation lengths to be shorter than the parallel correlation length: $\ellxp, \ellyp \sim \epsilon \ellzp$. The poloidal velocity is ordered to be smaller than the toroidal velocity $\vZ \sim \epsilon\vz$~\cite{Hinton1985, Catto1987,Field2009}. The toroidal velocity is taken to be the same order of magnitude as the thermal velocity, $\vth = \sqrt{2 T_i/m_i}$, where $T_i$ is the ion temperature, and $m_i$ is the ion mass. This then leaves space for a subsidiary expansion in low-Mach number $M \equiv \vz/\vth$, which will be performed in Section~\ref{sec:lowmachnumber}. 

We relate the lengths, times and velocities to each other by noting that the poloidal displacement of a perturbation $\dz$ scales with the toroidal velocity and laboratory-frame time delay of the peak of the cross-correlation function $\dtpeakt$: $\dz \sim \vz \dtpeakt$, and that this poloidal displacement is typical of the perpendicular correlation lengths $\ellxp$ and $\ellyp$. In addition to this, we use the critical balance conjecture~\cite{Barnes2011,Ghim2013,Schekochihin2016} to relate the parallel correlation length to the thermal velocity and the plasma-frame correlation time $\ellzp \sim \vth \taucp$. Critical balance is an assumption that the length of a correlated structure parallel to the magnetic field is determined by the time taken for the structure to decorrelate in the perpendicular plane, $\taucp$, being comparable to the time taken to transmit information along the magnetic field line, $\ellzp/\vth$.

Combining all these relations, we arrive at the following asymptotic ordering:
\begin{eqnarray}\label{fullordering}
\vz \taucp, \vth \taucp \sim \ellzp &\sim& \mathcal{O}(R),\\
\vz\dtpeakt \sim \dz, \dr, \ellxp, \ellyp \quad\quad &\sim& \mathcal{O}(\epsilon R),  \\
\vZ \dtpeakt &\sim&\mathcal{O}(\epsilon^2R), \label{endfullordering}
\end{eqnarray}
where the parameter $R$ is the tokamak major radius, representing the scale of the longest structures in the system. From (\ref{fullordering}-\ref{endfullordering}), we see that the time delay of the peak of the cross-correlation function is ordered small compared to the correlation time of the turbulence $\dtpeakt \sim \epsilon \taucp$. Inspection of the time-delayed cross-correlation functions measured from experiment, in Figure~\ref{fig:example_time_delay}, shows that this approximation may be considered reasonable, as all $\dtpeak \lesssim \frac{1}{2}\tauca$.

\subsubsection{Laboratory-frame correlation time}\label{sec:lab_frame_corr_time}
The full details of the calculation of the laboratory-frame correlation time, $\tauct$, are given in \ref{sec:toy_CCTD}. The outline of the procedure is as follows: the exponent of the correlation function (\ref{lab_corr_function_full}) is expanded order by order in $\epsilon$, and at each order the global maximum of the time-delayed cross-correlation function found. The time-delay envelope, described by the peaks of the time-delayed cross-correlation functions, $\dtpeakt(\dz)$, is then calculated as a function of spacing between poloidal channels, $\dz$. It is found that there is no decay in the peak amplitude of the time-delayed cross-correlation function until second order in the expansion. At second order, the time-delay envelope has a decay time given by
\begin{equation}\label{lab_corr_time}
\frac{1}{\tauct^2} \equiv \frac{1}{\taucp^2} + \frac{\vz^2}{\ellzp^2\cos^2\alpha},
\end{equation}
thus defining the laboratory-frame correlation time measured using the CCTD method. 

\subsubsection{Low-Mach-number ordering.}\label{sec:lowmachnumber}
The expression for the laboratory-frame correlation time (\ref{lab_corr_time}) can be further simplified if we adopt, as an ordering subsidiary to (\ref{fullordering}-\ref{endfullordering}), the assumption that the Mach number is small $M = \vz/\vth \sim \vz \taucp/\ellzp \ll 1$. As a result, the second term in (\ref{lab_corr_time}) can be neglected and we find that the laboratory-frame correlation time coincides with the plasma-frame correlation time.

\subsubsection{Laboratory-frame poloidal velocity}\label{sec:lab_frame_pol_vel}
Using $\dtpeakt(\dz)$ and $\dz$, the laboratory-frame apparent poloidal velocity is calculated as $v_{\mathrm{pol}|L}=\dz/\dtpeakt$, analogous to the BES measurement of $v_{\mathrm{pol}|B}$, described in Section~\ref{sec:meas_app_pol_vel}. The resulting apparent poloidal velocity is
\begin{equation}
v_{\mathrm{pol}|L} \equiv \vz\tan\alpha + \vZ.
\end{equation}
To lowest order in the $\epsilon$ expansion (\ref{fullordering}-\ref{endfullordering}), the apparent poloidal velocity is purely the projection of the toroidal velocity on to the poloidal plane
\begin{equation}\label{vpol_lab}
v_{\mathrm{pol}|L} = \vz\tan\alpha +  \mathcal{O}(\epsilon \vz).
\end{equation}
Another corollary of the $\epsilon$ expansion is that the lowest-order expression for the laboratory-frame poloidal correlation length (\ref{lab_pol_length}) is 
\begin{equation}\label{lab_ellZ_approx}
\ellzt = \ellyp/\cos\alpha + \mathcal{O}(\epsilon \ellyp),
\end{equation}
which can be used to extract the plasma-frame binormal correlation length, without knowledge of the parallel correlation length.

% ----------------------------------------------------------------------------------------------------------
% ----------------------------------------------------------------------------------------------------------
% ----------------------------------------------------------------------------------------------------------

\section{Point-Spread Functions}\label{sec:psfs}
As described in Section~\ref{sec:introduction}, the BES indirectly measures the density field via the intensity of light emitted from excited neutral-beam atoms. The transformation between these two fields is dictated by the Point-Spread Functions (PSFs) of the instrument, see (\ref{PSF_def}). In this section, we introduce the PSFs and explain how we model their structure.

To understand how the PSFs affect the BES measurements of turbulent quantities, it is necessary to have a good description of the properties of the PSFs. The complicated dependence of the PSFs on the plasma equilibrium, neutral-beam profile, and atomic physics (see variation of the PSFs in Figure~\ref{fig:psf_alpha}) means that a phenomenological approach is highly beneficial. In Section~\ref{sec:MeasuringrealPSFs}, we describe how the shape of the PSFs is characterised, in terms of only three parameters. Then, in Section~\ref{sec:PSFsInMAST}, we describe how these parameters vary in MAST. We then introduce, in Section~\ref{sec:GaussianPSFs}, an idealised form of a PSF as a tilted Gaussian function, using the three derived parameters and the peak amplitude of the PSF. In order to calculate analytically the effect of PSFs on the measured turbulence parameters, we further assume that all the PSFs have the same shape. Therefore, we approximate the PSFs for a specific time and set of channels (inner or outer --- see Section~\ref{sec:example_spatial_corr_fun}) by using the mean of the PSF parameters. The accuracy of these approximations is tested in Section~\ref{sec:PSFapproxTest}, where we introduce a measure to quantify the difference between the real PSFs and the Gaussian-model PSFs.

% ------------------------------------------------------------------------------------------------------------------------------------------
% ------------------------------------------------------------------------------------------------------------------------------------------
% ------------------------------------------------------------------------------------------------------------------------------------------

\subsection{Calculation of PSFs for BES systems}\label{sec:PSFcalc}
A full description of how the PSFs for MAST are calculated is available in~\cite{Ghim2010}, however, given that the structure of the PSFs is important for the current work, we review the most important aspects here. The PSFs are calculated by considering the emission from the $n=3\rightarrow2$ excited state of the primary-beam atom (Deuterium) on a series of two-dimensional planes aligned perpendicular to the line of sight (LoS) of a detector channel. The emission from each of the two-dimensional planes is integrated on to the focal plane of the optical system by using the fact that the density fluctuations are extended along the parallel direction of the magnetic field~\cite{Liewer1985,Ritz1988,Koch2007}. This is equivalent to interpolating a two-dimensional field of density fluctuations along the magnetic field line, to construct a three-dimensional field, and then integrating along the LoS, through this density field, to determine the line-integrated emissivity. Therefore, any misalignment of the LoS and the magnetic field line in the sampling volume would cause smearing of the image. This is the main factor affecting the shape of the PSFs in MAST, as will be discussed in Section~\ref{sec:PSFsInMAST}.

Additionally, the radial motion of the beam atoms within the finite lifetime ($3-10 \mathrm{ns}$) of the excited state causes the radial width of the PSFs to be in the range of $0.5-1.5\mathrm{cm}$, with the exact width depending on the velocity of the beam and the plasma density~\cite{Hutchinson2002}.

The amplitude of the PSFs is proportional to both the local density of the plasma and the beam density. The beam density decreases as the atoms in the neutral beam penetrate the plasma and are ionised by collisions, primarily by charge exchange with the plasma ions. Therefore, the emissivity decreases with decreasing major radius and this is reflected in a decrease in the amplitude of the PSFs. In this work, we normalise each set of PSFs by the maximum amplitude in this set of PSFs, so that we only consider differences between the PSFs.

% ------------------------------------------------------------------------------------------------------------------------------------------
% ------------------------------------------------------------------------------------------------------------------------------------------
% ------------------------------------------------------------------------------------------------------------------------------------------

\subsection{Characterisation of the PSFs}\label{sec:MeasuringrealPSFs}
\begin{figure}
	\centering
	\includegraphics[width=0.33\textwidth]{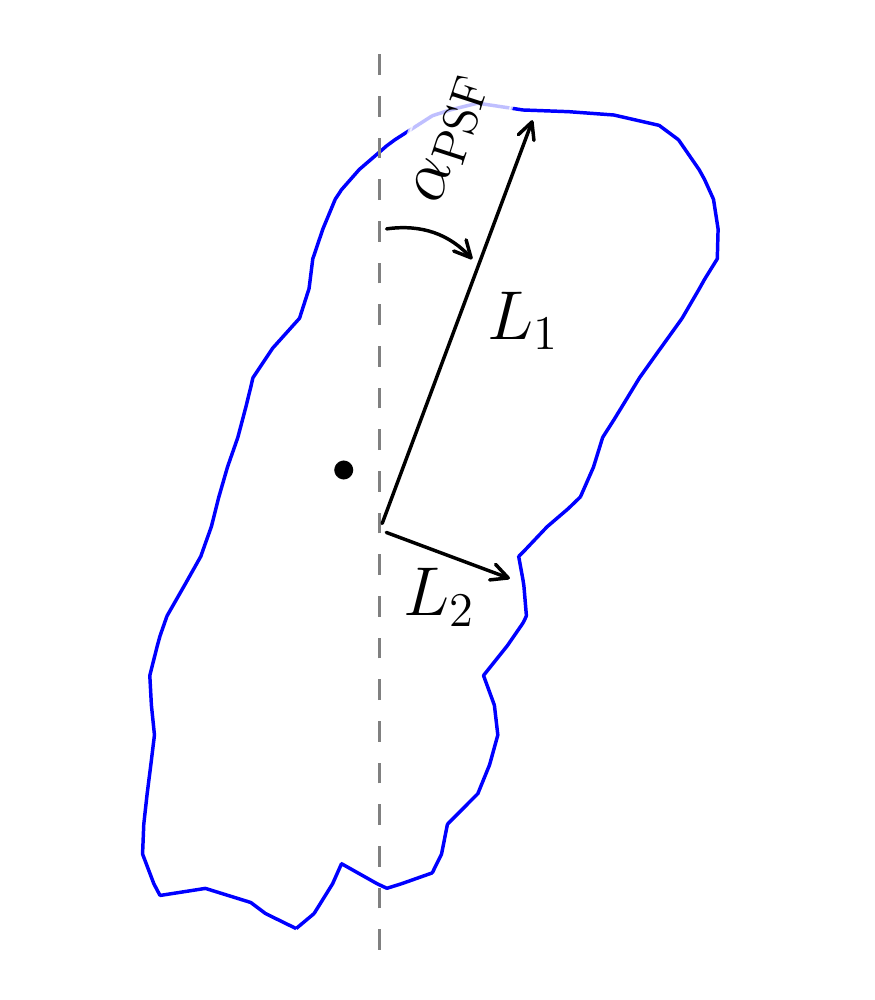}
	\caption[Characterisation of PSFs]{A schematic to illustrate the PSF angle $\alpha_{\mathrm{PSF}}$, the principal component $L_1$, and the secondary component $L_2$. The characterisation of the PSFs in terms of these three parameters is described in Section~\ref{sec:MeasuringrealPSFs}. \label{fig:psf_alpha_charac}}
\end{figure}
The shape of the PSFs can be characterised by two orthogonal vectors, $\vec{p}_1$ and $\vec{p}_2$, using principal-component analysis~\cite{Jolliffe2002}.  For this purpose, each PSF is normalised to produce a probability distribution from which a sample of points is generated. The principal-component analysis can then be applied to this distribution of points to find the direction of the principal vector $\vec{p}_1 = (p_{1r}, p_{1Z})$ in the radial-poloidal plane. The tilt of the PSF is given by the angle
\begin{equation}
\apsf = \mathrm{arctan}\left(\frac{p_{1r}}{p_{1Z}}\right)
\end{equation}
from the positive $Z$-axis in the anti-clockwise sense (this is defined in the same sense as the tilt of the correlation function, $\Theta$, see (\ref{theta_B_def}), and means that all $\alpha_\mathrm{PSF} < 0$).  The secondary component is defined to be perpendicular to the principal component. We can determine the `size' of a PSF by constructing a line with the gradient of the principal component that passes through the peak amplitude of the PSF. The segment of this line bounded by the points of intersection with the $e^{-1}$-amplitude contour of the PSF then defines the length $2 L_1$. Following the same procedure with the secondary component defines the length $2 L_2$. A schematic of this analysis is given in Figure~\ref{fig:psf_alpha_charac}.

The three parameters $L_1, L_2$ and $\apsf$ cannot be used to reconstruct precisely the contour of the real PSF from which they have been calculated, because they describe the shape of a tilted ellipse. As can be seen in Figure~\ref{fig:psf_alpha}, the PSFs in MAST are clearly not elliptical. However, the characterisation in terms of the three PSF parameters can be used in a simplified model (Section~\ref{sec:GaussianPSFs}) that allows us to construct the correlation function of the intensity field analytically (Section~\ref{sec:PSF_analytic}). We will later check how important the neglected effects are, by comparing with numerical calculations that use the real PSFs (Section~\ref{sec:numeric}).

% ------------------------------------------------------------------------------------------------------------------------------------------
% ------------------------------------------------------------------------------------------------------------------------------------------
% ------------------------------------------------------------------------------------------------------------------------------------------

\subsection{PSFs in MAST: 5 representative cases}\label{sec:PSFsInMAST}
\begin{table}[h]
\centering
\begin{tabular}{l|r|r|r|r|r|}
Case								& 1 & 2 & 3 & 4 & 5 \\ \hline
Shot 								& 28155 & 28155 & 28155 & 27268 & 29891 \\ 
Sub-array 							& in & in & out  & in &  out \\ 
Time/ms 							& 125 & 250 &  125 & 250 & 375 \\
$L_1/\mathrm{cm}$			        		& 2.0  & 3.1  &  2.2 & 2.8  &   2.4 \\
$L_2/\mathrm{cm}$			        		& 1.1  & 1.4  &   1.3 & 1.3  & 1.5  \\
$\alpha_\mathrm{PSF}/\mathrm{deg}$ 	&-70  & -33  &   -59 & -46 & -144 \\
\end{tabular}
\caption[Mean PSF characteristics for the five cases.]{The mean characteristics of the PSFs from the 5 cases that we are considering (see Figure~\ref{fig:psf_alpha}). The PSF parameter values in the table are averages over all the PSF parameter values from each channel in the respective sub-array, which have been calculated using principal-component analysis (Section~\ref{sec:MeasuringrealPSFs})}
\label{tab:psf_parameters}
\end{table}
In Spherical Tokamaks (STs), such as MAST, because of their tight aspect ratio, the magnetic pitch angle varies significantly over the radial extent of the BES system, which causes, through the misalignment of the LoS and the magnetic field line (see Section~\ref{sec:PSFcalc}), the shape of the PSFs to vary from channel to channel. As examples we consider the five cases shown in Figure~\ref{fig:psf_alpha}, with their measured parameters given in Table~\ref{tab:psf_parameters}. Cases 1-4 are taken from Double-Null Divertor (DND)  discharges, where the BES views on the mid-plane of the plasma. Case 5 is from a Lower-Single-Null Divertor (LSND) configuration, where the BES views above the magnetic axis. In the DND cases (especially Case 4), it is clear that the pitch angle of the magnetic field increases with radius, causing the poloidal extent of the PSFs to increase. In Case 5, the different viewing geometry means the magnetic field causes the PSFs to be tilted in the opposite sense to the DND cases. We include Case 5 because of this feature, as it will be useful to see how the different tilts of the PSFs affect the correlation parameters of the intensity field (see Section~\ref{sec:tiltcorrfun}).

The time evolution of the $q$ profile in MAST causes the PSFs also to vary in time, as can be seen by comparing Case 1 and Case 2, taken from the same shot at two different times. The temporal evolution is mainly due to the increase in the poloidal component of the magnetic field, $B_Z$, during the shot, causing $\alpha_\mathrm{PSF}$ to increase. This variation of $B_Z$ in space and time is typical for the majority of DND shots on MAST.

The four DND PSF cases (Cases 1-4) cover most of the possible variation in the PSF parameters for DND discharges, as will be shown in Section~\ref{sec:PSFapproxTest}. We focus on DND discharges because the symmetry around the mid-plane implies that magnetic-shear effects on the turbulence do not need to be taken into account when measuring the turbulence parameters. However, we have included Case 5 as an example of the PSFs for LSND discharges because the PSF shapes are significantly different, highlighting the importance of properly understanding and accounting for the PSF effects.

% ------------------------------------------------------------------------------------------------------------------------------------------
% ------------------------------------------------------------------------------------------------------------------------------------------
% ------------------------------------------------------------------------------------------------------------------------------------------

\subsection{Gaussian-model PSFs}\label{sec:GaussianPSFs}
Gaussian-model PSFs are constructed from the measured characteristics of the real PSFs from each BES channel $i$: the principal components $L_{1i}, L_{2i}$, the tilt angle $\alpha_{\mathrm{PSF}i}$, and the peak amplitude $A_{\mathrm{PSF}i}$. These are given by
\begin{equation}\label{GaussianPSF}
P_i^\mathrm{Gauss}(\vec{r}-\vec{r}_i, A_{\mathrm{PSF}i}, L_{1i}, L_{2i}, \apsfi) = A_{\mathrm{PSF}i}\mathrm{exp}\left( - \frac{\Delta Z'^2 }{L_{1i}^2} -\frac{\Delta r'^2 }{L_{2i}^2} \right),
\end{equation}
\begin{eqnarray}
\mathrm{with} \quad  \Delta r' &=& \Delta r \cos(\alpha_{\mathrm{PSF}i})+\Delta Z \sin(\alpha_{\mathrm{PSF}i}), \nonumber \\
\mathrm{and} \quad \Delta Z' &=& \Delta Z \cos(\alpha_{\mathrm{PSF}i})- \Delta r \sin(\alpha_{\mathrm{PSF}i}), \nonumber
\end{eqnarray} 
where $\vec{r}-\vec{r}_i =  (\Delta r, \Delta Z) = (r-r_i, Z-Z_i)$, and $r_i, Z_i$ are the positions of the peak amplitudes of the PSFs. This paper will make much use of these Gaussian-model PSFs, because of their analytic tractability. From here onwards, unless otherwise stated, we assume that each set of the Gaussian-model PSFs have the same values of $L_1$, $L_2$, and $\apsf$, and these are taken to be the mean of the $L_{1i}$, $L_{2i}$, and $\alpha_{\mathrm{PSF}i}$, respectively, over the inner or outer set of channels. However, we keep the amplitude dependence in $P_i^\mathrm{Gauss}$ to show, in Section~\ref{sec:PSF_an_2D}, that the BES covariance function (\ref{Cbes3}) is independent of the PSF amplitude. The requirement for using the mean values is also specified in Section~\ref{sec:PSF_an_2D}.  Before moving on, let us investigate how good an approximation these Gaussian-model PSFs are to the real PSFs. 

% ----------------------------------------------------------------------------------------------------------
% ----------------------------------------------------------------------------------------------------------
% ----------------------------------------------------------------------------------------------------------

\subsection{Difference between real and Gaussian-model PSFs}\label{sec:PSFapproxTest}
\begin{figure}[h]
\centering
\includegraphics[width=0.65\textwidth]{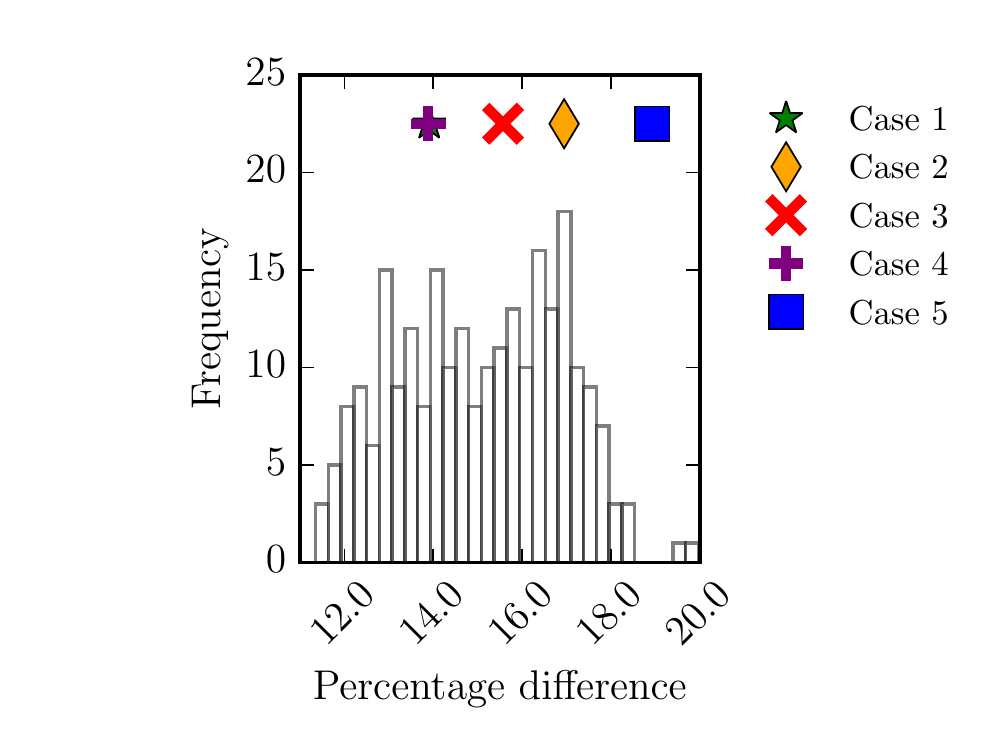}
\caption[Histogram of percentage differences between real and Gaussian-model PSFs.]{Difference between the real PSFs and the Gaussian-model PSFs using the difference measure (\ref{deltaP0def}) for 245 sets of PSFs calculated at various times in 20 MAST shots in DND configuration.  The values for each of the cases presented in Figure~\ref{fig:psf_alpha} are indicated with the coloured symbols. Case 5 is not included in the calculation of the histogram, yet is labelled for comparison. Cases 1 and 4 have almost the same value of (\ref{deltaP0def}).\label{fig:comp_full_gauss_psf_simple} }
\end{figure}
We define a measure of the difference between a set of the real PSFs, $P_i^\mathrm{real}$, and the Gaussian-model PSFs, $P_i^\mathrm{Gauss}$, in order to determine the quality of the approximation (\ref{GaussianPSF}). The measure is the average over the set of real PSFs of
\begin{equation}\label{deltaP0def}
\Delta P_0 =  \frac{1}{A_{\mathrm{PSF}i}}\mathrm{RMS}\left[P_i^\mathrm{real}(\vec{r}-\vec{r}_i) - P_i^\mathrm{Gauss}(\vec{r}-\vec{r}_i,A_{\mathrm{PSF}i},L_1, L_2, \apsf) \right]_\vec{r} ,
\end{equation}
i.e., the root-mean-square (RMS) difference between a real PSF and the Gaussian model (\ref{GaussianPSF}) of the set of PSFs, relative to the peak amplitude of the real PSF. The central position $\vec{r}_i$ of the real PSF is defined in \ref{sec:app_diff_gaus_psf}. The measure (\ref{deltaP0def}) is plotted in Figure~\ref{fig:comp_full_gauss_psf_simple} for 245 sets of PSFs taken from a database of MAST shots~\cite{Ghim2013}, and shows that the difference between the real and Gaussian-model PSFs lies in the range 12-18\%. There are two key factors that contribute to this difference: the use of the average values for the PSF parameters (ignoring the spatial variation across the sub-array) and the assumption of a Gaussian shape. In \ref{sec:psf_testing}, we show that the main contribution to this difference comes from the assumption of a Gaussian shape for the PSFs rather than from their spatial variation. We note that the percentage differences given in Figure~\ref{fig:comp_full_gauss_psf_simple} do not represent the error in the measured correlation parameters associated with using the Gaussian-model PSFs in place of the real PSFs when applying these to a fluctuating density field, which will be discussed in Section~\ref{sec:validity_of_Gaussian}.

\begin{table}[h]
	\centering
	\begin{tabular}{l|r|r|r|r|r}
		% See note on 27/04/2016 for source of simple case data only (includes all effects in one number)
		& Case 1 	& Case 2 	&Case 3 	& Case 4  	& Case 5 \\ \hline
		Difference from Gaussian/\%  	&    14 	&    17 	&  16 	& 14		&  19         \\   % This is the simple case
		RMS / deg.        				&    2.4 	&    3.9 	&  3.6 	& 3.1 	&  5.4         \\             
	\end{tabular}
	\caption[Differences from Gaussian for five cases and difference between real and Gaussian model PSF effects on tilt angle.]{Table of values for the difference measure (\ref{deltaP0def}) defined in Section~\ref{sec:PSFapproxTest} for each of the five cases described in Section~\ref{sec:PSFsInMAST}. Additionally, values are given for the root-mean-square (RMS) of the differences between the tilt angle $\Theta_B$ measured from synthetic-BES data generated using real and Gaussian-model PSFs. The mean is taken over the difference in values for each PSF case in Figure~\ref{fig:blobby_plots}(d), see Section~\ref{sec:validity_of_Gaussian} for the relevant discussion.}
	\label{tab:rms_real_gauss}
\end{table}

On the histogram plot of Figure~\ref{fig:comp_full_gauss_psf_simple}, we also indicate the values of the difference measure (\ref{deltaP0def}) for the five PSF cases introduced in Section~\ref{sec:PSFsInMAST}. The numerical values are given in Table~\ref{tab:rms_real_gauss}. The example cases have been chosen so that they span as large a range of the difference measure as possible and can, therefore, be used to quantify how the difference between the effect of real and Gaussian-model PSFs on correlation parameters (see Section~\ref{sec:validity_of_Gaussian}) depends on how well the Gaussian-model PSFs agree with the real PSFs.

% ------------------------------------------------------------------------------------------------------------------------------------------
% ------------------------------------------------------------------------------------------------------------------------------------------
% ------------------------------------------------------------------------------------------------------------------------------------------

\section{Analytic calculations of the effect of PSFs}\label{sec:PSF_analytic}

\subsection{Effect of PSFs on the 2D spatial structure of the correlation function}\label{sec:PSF_an_2D}
In this section, we calculate analytically the relationship between the laboratory-frame correlation lengths and wavenumbers to those measured by the BES system by taking account of PSF effects. Using (\ref{PSF_def}) and (\ref{covfun}), we write the covariance function between two BES detector channels, located at positions $\vec{r}_i$ and $\vec{r}_j$, as
\begin{eqnarray}
\fl C^\mathrm{cov}_B(\vec{r}_i,\vec{r}_j) &=& \frac{ \left\langle \int P_i(\vec{r} - \vec{r}_i) \delta n(\vec{r},t) \mathrm{d}^2\vec{r} \int P_j(\vec{r'} - \vec{r}_j) \delta n(\vec{r',t}) \mathrm{d}^2\vec{r'} \right\rangle}{\left\langle \int \mathrm{d}^2\vec{r} P_i(\vec{r}-\vec{r}_i) \meann \int \mathrm{d}^2\vec{r'} P_j(\vec{r'}-\vec{r}_j) \meann \right\rangle}, \nonumber \\
&=&  \frac{1}{\hat{P}_i\hat{P}_j} \int \int\mathrm{d}^2\vec{r}  \mathrm{d}^2\vec{r'} P_i(\vec{r} - \vec{r}_i)  P_j(\vec{r'} - \vec{r}_j)  \left\langle \frac{\delta n(\vec{r},t) \delta n(\vec{r'},t)}{\meann^2} \right\rangle , \label{Cbes2} \\
&=&  \int\mathrm{d}^2 \Delta\vec{r} \tilde{P}_{ij}(\Delta\vec{r}, \Delta\vec{r}_{ij}) C^\mathrm{cov}_L(\Delta \vec{r}), \label{Cbes3}
\end{eqnarray}
where $n(\vec{r},t)=\meann + \delta n(\vec{r},t)$ is the total density field, which is split into a mean part independent of time and constant in space $\meann$, and a fluctuating part $\delta n(\vec{r},t)$, analogously to the intensity field (\ref{intensity_decomp}) in Section~\ref{sec:measurement}. In (\ref{Cbes2}), we assumed that the PSFs do not vary over the ensemble (time) average and have suppressed the dependence on time delay $\dt$ in the covariance functions $C^\mathrm{cov}_L$ and $C^\mathrm{cov}_B$. The denominator terms are calculated explicitly using (\ref{GaussianPSF}):
\begin{equation}
\hat{P}_i \equiv \int \mathrm{d}^2\vec{r} P_i(\vec{r}-\vec{r}_i) = A_{\mathrm{PSF}i} \pi L_{1i} L_{2i}.
\end{equation}
In (\ref{Cbes3}), we introduced the quantity
\begin{equation}
\tilde{P}_{ij}(\Delta\vec{r}, \Delta\vec{r}_{ij}) \equiv \frac{1}{\hat{P}_i\hat{P}_j}\int\mathrm{d}^2 \Delta \vec{a} P_i(\Delta\vec{r} + \Delta\vec{a})  P_j(\Delta\vec{a} + \Delta\vec{r}_{ij}), \label{PSFintegral}
\end{equation}
which is independent of the PSF amplitudes, and, therefore, so is $C^\mathrm{cov}_B$. We have also assumed that the covariance function (\ref{fulltimedelay}), $C^\mathrm{cov}_L(\Delta \vec{r}) =\langle\delta n(\vec{r}) \delta n(\vec{r'}) / \meann^2 \rangle$, is only a function of the relative position, $\Delta \vec{r} = \vec{r} - \vec{r'}$, i.e., the turbulence is spatially homogeneous, consistent with our assumption in Section~\ref{sec:ztd_measurement}. We have changed the integration variables from ($\vec{r}, \vec{r'}$) to ($\Delta \vec{r}, \Delta \vec{a} = \vec{r'} - \vec{r}_i$) so that the BES covariance function becomes a function of the difference $\Delta\vec{r}_{ij} = \vec{r}_i - \vec{r}_j$. However, this does not mean that the BES covariance function is independent of the absolute measurement position, as the $\Delta\vec{r}_{ij}$ are still dependent on the PSF indices. In order for the BES covariance function to be dependent only on relative position, we also have to require that the PSFs not vary between channels, i.e., that $L_{1i}, L_{2i}$ and $\alpha_{\mathrm{PSF}i}$ should all be the same, as postulated in Section~\ref{sec:GaussianPSFs}. The correlation function is then simply
\begin{equation}
C_B(\Delta\vec{r}_{ij}) = \frac{C^\mathrm{cov}_B(\Delta\vec{r}_{ij})}{C^\mathrm{cov}_B(0)} \label{Cbes4},
\end{equation}
where we assume that the correlation function is independent of the channel indices, but retain them to distinguish the discrete BES channel spacing from the continuous density field in the plasma.

In order to proceed, we use the Gaussian-model PSFs given by (\ref{GaussianPSF}). This allows us to compute (\ref{PSFintegral}) explicitly:
\begin{eqnarray}
\fl\tilde{P}(\Delta\vec{r}, \Delta\vec{r}_{ij}) &=& \frac{1}{2 \pi L_1L_2} \exp\left\{-\frac{ \left[(\Delta r-\Delta r_{ij})\sin\alpha_\mathrm{PSF} -  (\Delta Z-\Delta Z_{ij})\cos\alpha_\mathrm{PSF}\right]^2  }{2 L_1^2}\right\} \nonumber \\
&\quad& \times \exp\left\{-\frac{\left[(\Delta r-\Delta r_{ij})\cos\alpha_\mathrm{PSF} +  (\Delta Z-\Delta Z_{ij})\sin\alpha_\mathrm{PSF}\right]^2}{ 2L_2^2 }\right\}. \label{Ptwsimple}
\end{eqnarray}
In order to evaluate the integral in (\ref{Cbes3}), we use the laboratory-frame covariance function (\ref{fulltimedelay}), with time delay $\dt=0$. We then compute (\ref{Cbes4}) to find the spatial correlation function that the BES would measure\footnote{It is also possible to get to (\ref{Cbesfinal}) by using (\ref{lab_frame_correlation_function}) in (\ref{Cbes3}), as the result is normalised in (\ref{Cbes4}).}:
\begin{equation}
\fl C_{\mathrm{spatial}|B}(\Delta\vec{r}_{ij}) =  \exp\left(-\frac{\Delta r_{ij}^2}{\ellra^2} - \frac{\Delta Z_{ij}^2}{\ellza^2} - \frac{\Delta r_{ij}\Delta Z_{ij}}{\ell_{rZ|B}^2}\right)\cos\left(\kra \Delta r_{ij} + \kza \Delta Z_{ij}\right), \label{Cbesfinal}
\end{equation}
where 
\begin{eqnarray}
\ellra^2 &=& \ellrt^2 + \frac{4 L_1^2 L_2^2 + 2 \ellzt^2(L_1^2 \sin^2\apsf + L_2^2\cos^2\apsf)}{ \ellzt^2 + 2 (L_1^2 \cos^2\apsf + L_2^2\sin^2\apsf)}, \label{an_ellra} \\
\ellza^2 &=& \ellzt^2 + \frac{4 L_1^2 L_2^2 + 2 \ellrt^2(L_1^2 \cos^2\apsf + L_2^2\sin^2\apsf)}{ \ellrt^2 + 2 (L_1^2 \sin^2\apsf + L_2^2\cos^2\apsf)}, \label{an_ellza} \\
\ell_{rZ|B}^2 &=& \frac{D^4}{2(L_1^2-L_2^2)\sin2\apsf}, \label{an_ellrza}\\
\kra &=& \frac{1}{D^4} \left[ \krt \ellrt^2\ellzt^2 + 2L_1^2\cos\apsf(\krt\ellrt^2\cos\apsf + \kzt\ellzt^2\sin\apsf) \right. \nonumber \\
&\quad& \left. + 2L_2^2\sin\apsf(\krt\ellrt^2\sin\apsf - \kzt\ellzt^2\cos\apsf)\right],  \label{an_kra}\\
\kza &=& \frac{1}{D^4} \left[ \kzt \ellrt^2\ellzt^2 + 2L_1^2\sin\apsf(\krt\ellrt^2\cos\apsf + \kzt\ellzt^2\sin\apsf) \right. \nonumber \\
&\quad& \left. - 2L_2^2\cos\apsf(\krt\ellrt^2\sin\apsf - \kzt\ellzt^2\cos\apsf)\right],\label{an_kza}
\end{eqnarray}
and
\begin{eqnarray}\label{Dterm}
D^4 &=& 4L_1^2L_2^2 + \ellrt^2\ellzt^2 + 2L_2^2(\ellrt^2\sin^2\apsf + \ellzt^2\cos^2\apsf)\nonumber\\ &\quad& +2L_1^2(\ellrt^2\cos^2\apsf + \ellzt^2\sin^2\apsf).
\end{eqnarray}
Equations (\ref{an_ellra}-\ref{an_kza}) can be easily inverted (see Section~\ref{sec:correctingPSFs}) to express the laboratory-frame parameters as functions of the measured BES parameters and the PSF parameters only. It is clear from the expressions for the BES correlation lengths  (\ref{an_ellra}) and (\ref{an_ellza}) that in the limit of small PSFs, i.e., when $L_1, L_2 \ll \ellrt, \ellzt$, the measured BES and laboratory-frame values are equal.  Furthermore, we see that the second terms on the right-hand sides of (\ref{an_ellra}) and (\ref{an_ellza}) are both positive definite, therefore, the effect of PSFs is always to increase the radial and poloidal correlation lengths above the laboratory-frame values. Consequently, if we take the limit of the radial and poloidal laboratory-frame correlation lengths becoming small compared to the PSF lengths, $\ellrt, \ellzt \ll L_1, L_2$, then the BES-measured correlation lengths will be functions only of the PSF parameters.

The way in which the PSFs affect the wavenumbers is more complicated: from (\ref{an_kra}) and (\ref{an_kza}), we see that the BES-measured wavenumbers depend on the sign of the laboratory-frame wavenumbers, as well as on the sign of the tilt angle of the PSFs. This relationship will be further elucidated in Section~\ref{sec:tiltcorrfun}.

\begin{figure}
\centering
\includegraphics[width=0.85\textwidth]{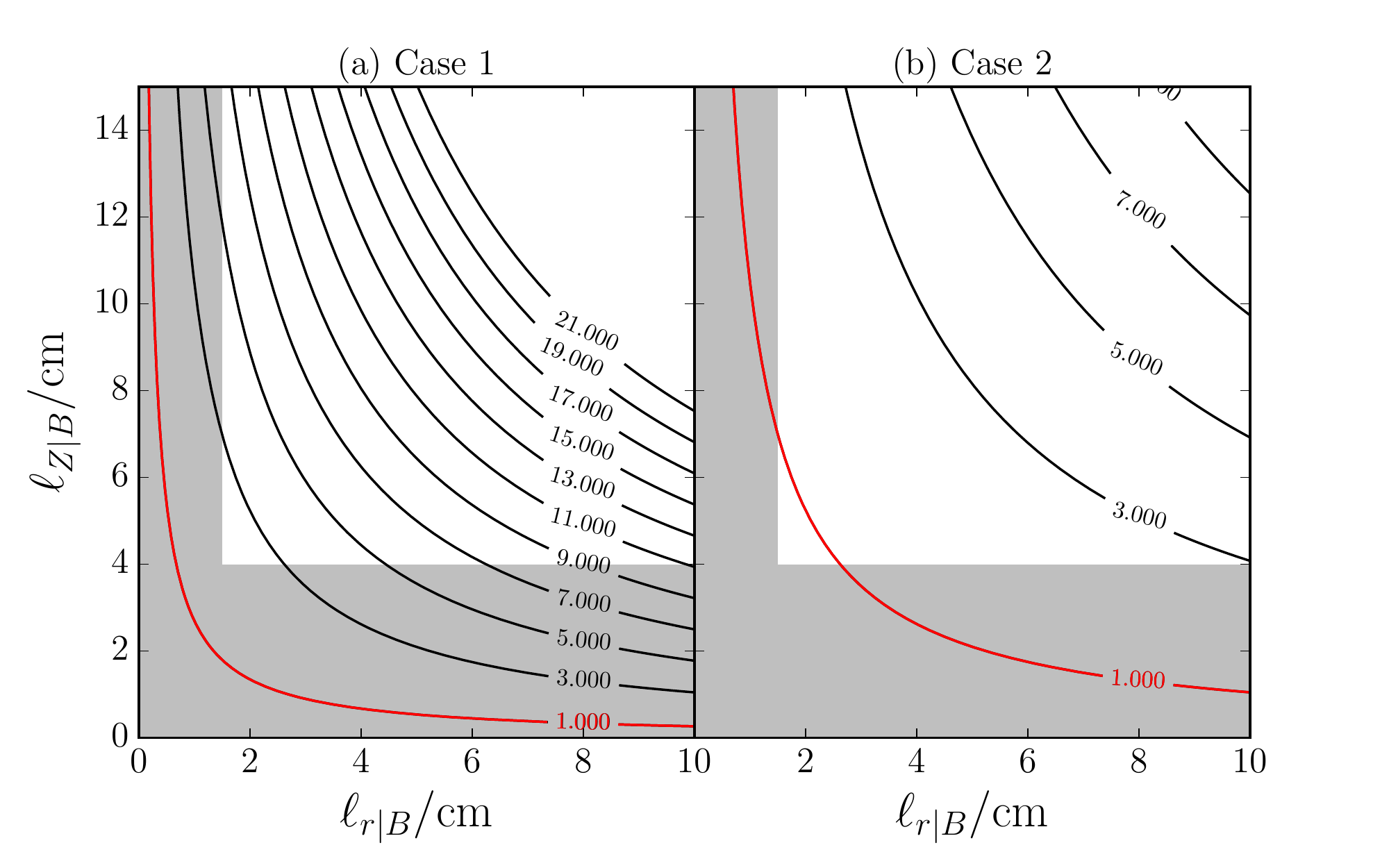}
\caption[Contour lines of $\Omega$]{Contour lines of $\Omega$, defined by (\ref{conditionOnSmallCrossTerm}), for (a) Case-1 PSFs and (b) Case-2 PSFs (see Table~\ref{tab:psf_parameters}). The regions outside the grey areas indicate the range of typical correlation lengths measured by the MAST BES system. The red contour line shows $\Omega=1$, which indicates the boundary of the region where cross-term (\ref{an_ellrza}) becomes non-negligible. \label{fig:smallcrossterm} }
\end{figure}

The appearance of a cross-term (\ref{an_ellrza}) between $\Delta r_{ij}$ and $\Delta Z_{ij}$, parametrised by the scale $\ellrza$, is a result purely of the PSFs, as no such term is present in the laboratory-frame correlation function (\ref{lab_frame_correlation_function}). Operationally, including the cross-term as an independent fitting parameter causes the fit of (\ref{Cbesfinal}) to (\ref{correlation_def}) to become under-constrained and, therefore, reduces our ability to extract the parameters of interest. For this reason, it is opportune to neglect the cross-term in the fitting function. Provided that $\ell_{rZ|B}^2 \gg \ellra\ellza$, it is reasonable to do so. This requirement can be written in terms of the BES-measured and the PSF parameters as follows
\begin{equation}
\Omega \equiv \frac{\ellza\ellra\left(   1 + \sqrt{ 1 + (2/\ellra\ellza)^2 (L_1^2-L_2^2)^2 \sin^2(2 \apsf) }  \right) }{4|(L_1^2-L_2^2)\sin2\apsf|} \gg 1. \label{conditionOnSmallCrossTerm}
\end{equation}
This is trivially satisfied when $\apsf = m\pi/2, m\in\mathbb{Z}$.  Figure~\ref{fig:smallcrossterm} shows that for most measured values of $\ellra$ and $\ellza$, the requirement (\ref{conditionOnSmallCrossTerm}) is satisfied. Generally speaking, we can always use the fitting function (\ref{fit_fun}) without the cross-term and then calculate $\Omega$ for fitted BES-measured correlation parameters to check that it is sufficiently large. However, this post-fitting test does not guarantee that the cross-term could have been safely neglected, because the inputs into the evaluation of $\Omega$ are calculated from fitting (\ref{fit_fun}) and not from fitting (\ref{Cbesfinal}) to the correlation function (\ref{correlation_def}).

% ------------------------------------------------------------------------------------------------------------------------------------------
% ------------------------------------------------------------------------------------------------------------------------------------------
% ------------------------------------------------------------------------------------------------------------------------------------------

\subsection{Effect of PSFs on the fluctuation amplitude}\label{sec:PSF_an_amp}
The effect of the PSFs on the mean-square fluctuation amplitude can be quantified by considering the auto-covariance function, setting $\drij=0$ in (\ref{Cbes3}):
\begin{equation}\label{sigma_amp_B}
\sigma_{\mathrm{amp}|B}^2 \equiv \int\mathrm{d}^2 \Delta\vec{r} \tilde{P}(\Delta\vec{r}, 0) C^\mathrm{cov}_L(\Delta \vec{r}),
\end{equation}
where we have introduced the notation $\sigma_{\mathrm{amp}|B}$, to distinguish the amplitude calculated using Gaussian-model PSFs from the equivalent quantity $\overline{\delta I/I}$  measured in experiment and numerical simulations, see (\ref{flucAmpDef}), as these are only guaranteed to be the same if the assumptions of Section~\ref{sec:PSF_an_2D} are satisfied, i.e., that the turbulence is homogeneous and all PSFs are the same. Then, using (\ref{fit_fun}) and (\ref{Ptwsimple}) to complete the integral in (\ref{sigma_amp_B}), we find
\begin{eqnarray}
 \frac{\sigma_{\mathrm{amp}|B}^2}{\sigma_{\mathrm{amp}|L}^2} = \frac{ \ellrt\ellzt}{ D^2} &\exp&\left\{ -\frac{1}{2 D^4} \bigg[ 2 L_1^2L_2^2(\krt^2\ellrt^2 + \kzt^2\ellzt^2)  \right.  \nonumber \\
&\quad& + \ellrt^2\ellzt^2 L_1^2(\krt\sin\apsf-\kzt\cos\apsf)^2   \nonumber \\
&\quad& \left.+  \ellrt^2\ellzt^2 L_2^2(\krt\cos\apsf+\kzt\sin\apsf)^2 \bigg] \vphantom{\frac{1}{1}}\right\}, \label{PSFflucamp}
\end{eqnarray}
where $\sigma_{\mathrm{amp}|L}^2 \equiv C_L^\mathrm{cov}(0)$ is the mean-square fluctuation amplitude in the laboratory frame (\ref{fluc_amp_lab}). The quantity $\sigma_{\mathrm{amp}|L}$ is the analytic equivalent of the fluctuation amplitude of the density field $\overline{\delta n/n}$, defined analogously to $\overline{\delta I/I}$ in (\ref{flucAmpDef}) replacing the intensity field with the density field.

Generally, we see that the BES-measured mean-square fluctuation amplitude $\sigma_{\mathrm{amp}|B}^2$ is a linear function of the laboratory-frame mean-square fluctuation amplitude $\sigma_{\mathrm{amp}|L}^2$, which means that inverting the relationship (\ref{PSFflucamp}) is easy (see Section~\ref{sec:inversion}). As both $D^2 \ge \ellrt\ellzt$ and the exponent in (\ref{PSFflucamp}) is always negative, the PSFs always cause the fluctuation amplitude of the measured signal to be lower than the amplitude of the laboratory-frame density field. For most standard values of turbulence parameters (see Section~\ref{sec:numeric}), the exponent in (\ref{PSFflucamp}) is small and the dominant effect comes from the coefficient $\ellrt\ellzt/D^2$. 

% ------------------------------------------------------------------------------------------------------------------------------------------
% ------------------------------------------------------------------------------------------------------------------------------------------
% ------------------------------------------------------------------------------------------------------------------------------------------

\subsection{Effect of PSFs on the correlation time and poloidal velocity}\label{sec:PSF_an_corr_time}
Following the CCTD method described in Section~\ref{sec:meas_corr_time}, the correlation time is computed by finding the envelope of the peaks of the time-delayed cross-correlation functions. We adopt the  asymptotic ordering introduced in Section~\ref{sec:asymtotics}, with the additional ordering of the PSF lengths $L_1, L_2$ as the same order as the radial, $\ellxp$, and poloidal, $\ellyp$, correlation lengths. The derivation, including PSF effects, of the correlation time and apparent poloidal velocity using the CCTD method is given in \ref{sec:appendix_CCTD}; here we simply present the results.

The apparent poloidal velocity is unaffected by PSFs:
\begin{equation}
v_{\mathrm{pol}|B} =  \vz \tan\alpha + \vZ  +  \mathcal{O}(\epsilon^2 \vz) \label{vpolPSF},
\end{equation}
where the second term is $\mathcal{O}(\epsilon^1)$ smaller than the first term and, therefore, under the assumptions of Section~\ref{sec:asymtotics} any measurement will be dominated by the toroidal rotation. 

The correlation time is also independent of the PSFs:
\begin{equation}\label{BEScorrtime}
\frac{1}{\tauca^2} = \frac{1}{\tauct^2} = \frac{1}{\taucp^2}  + \frac{\vz^2}{\ellzp^2 \cos^2\alpha}  + \mathcal{O}(\epsilon ),
\end{equation}
where the term containing $\vz$ can be neglected if, in addition to (\ref{fullordering}-\ref{endfullordering}), we assume a subsidiary ordering in small Mach number, as was done in Section~\ref{sec:lowmachnumber}. Then the correlation time measured using the CCTD method on the intensity field, $\tauca$, is the same as the plasma-frame correlation time.

% ----------------------------------------------------------------------------------------------------------
% ----------------------------------------------------------------------------------------------------------
% ----------------------------------------------------------------------------------------------------------

\section{Characterising the effect of PSFs on correlation parameters}\label{sec:numeric}

We wish to understand how PSFs affect the turbulent parameters and also whether, in determining this, spatially invariant Gaussian-model PSFs (\ref{GaussianPSF}) are a good approximation for the real PSFs. In order to determine the effects of real PSFs, a fluctuating density field must be generated to which the real PSFs can be applied. Our model for such a fluctuating density field is described in Section~\ref{sec:toymodel}. Then, in Section~\ref{sec:effects_real_PSFs}, we apply the real PSFs (Cases 1-5) to this numerically generated model field using (\ref{PSF_def}), measure the resulting correlation parameters as described in Section~\ref{sec:measurement}, and discuss how the correlation parameters of this synthetic-BES data differ from the correlation parameters in the laboratory frame of the model fluctuating field (i.e., before applying the real PSFs). In Section~\ref{sec:validity_of_Gaussian}, we use the analytically derived laboratory-frame correlation parameters of the model field (\ref{sec:toymodel_appendix}), and the measured PSF parameters for the five MAST representative cases (Table~\ref{tab:psf_parameters}), to calculate the analytic BES correlation parameters using (\ref{an_ellra}-\ref{an_kza}) and (\ref{sigma_amp_B}). We then compare these  with the correlation parameters measured from the synthetic-BES data of Section~\ref{sec:effects_real_PSFs} and quantify how well the Gaussian-model PSFs reproduce the effects of the real PSFs. In Section~\ref{sec:summary_sec_6}, we summarise the results.

% ------------------------------------------------------------------------------------------------------------------------------------------
% ------------------------------------------------------------------------------------------------------------------------------------------
% ------------------------------------------------------------------------------------------------------------------------------------------

\subsection{Model fluctuating density field}\label{sec:toymodel}

We will use a model fluctuating density field designed so that the correlation function of its time series is exactly the plasma-frame correlation function (\ref{plasma_frame_correlation_function}) or the laboratory-frame correlation function (\ref{lab_frame_correlation_function}) and (\ref{lab_temporal_corr_func}), depending on which frame it is calculated in. This is shown analytically in \ref{sec:toymodel_appendix}, where the details of our model field are presented. We stress that this model is purely phenomenological and contains no physical prescription for the formation of turbulent structures, and is thus similar to the models used in other tests of measurement techniques~\cite{Ghim2010,Zoletnik2005,Tal2011}.

A time series of density fluctuations is generated by constructing a signal from the sum of $N$ localised perturbations. The initial locations at which the perturbations are formed are randomly chosen in space and time (uniformly distributed within the domain). Each individual perturbation has a functional form similar to the plasma-frame correlation function (\ref{plasma_frame_correlation_function}), but with an amplitude taken from a Gaussian distribution with zero mean and standard deviation $\sigma_\mathrm{amp}$ and a phase taken from a uniform random distribution in the range $[0, 2\pi]$.

Each perturbation is allowed to grow and then decay over a finite number of correlation times. The radial wavenumber evolves with time as $k_x = k_{x0} + k_y S t $, where $k_{x0}$ is the radial wavenumber at peak amplitude, $S$ is the flow shear, and $t$ is time. The flow shear does not appear in any of the following expressions because it is absorbed into the definition of the plasma-frame radial correlation length and correlation time, see (\ref{ellxp}) and (\ref{taucp_from_toy}), is kept constant throughout this work, and has only a minor effect on the measured quantities.

The perturbations are created in the plasma frame. To transform them into the laboratory frame, the toroidal velocity $\vz$, poloidal velocity $\vZ$, and the pitch angle of the magnetic field, must be specified. In this work, these are assumed to be constants in position and time. The pitch angle is set to $\alpha = 30^\circ$, which is representative of the value in the outer-core of MAST plasmas. 

The output of the model is a two-dimensional radial-poloidal fluctuating density field in the laboratory frame, to which the real PSFs can be applied by evaluating the integral (\ref{PSF_def}). This then gives a radial-poloidal fluctuating intensity field (a synthetic ``BES measurement'') that can be analysed using the same methods that are applied to the experimental BES data, as described in Section~\ref{sec:measurement}.

% ------------------------------------------------------------------------------------------------------------------------------------------
% ------------------------------------------------------------------------------------------------------------------------------------------
% ------------------------------------------------------------------------------------------------------------------------------------------

\subsection{The effects of real PSFs}\label{sec:effects_real_PSFs}
\begin{figure}
\centering
\begin{minipage}{0.45\textwidth}
\includegraphics[width=\textwidth]{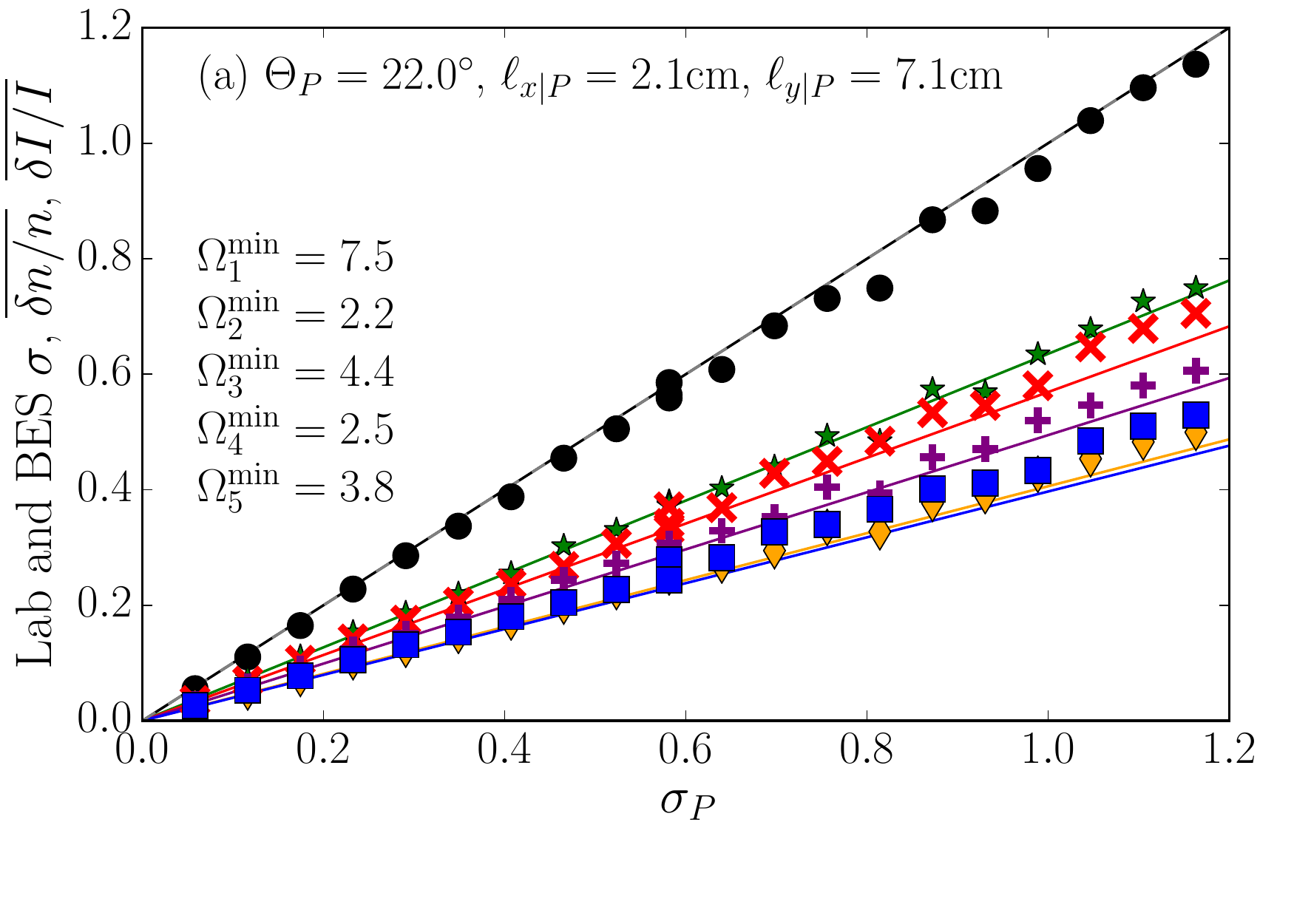}
\includegraphics[width=\textwidth]{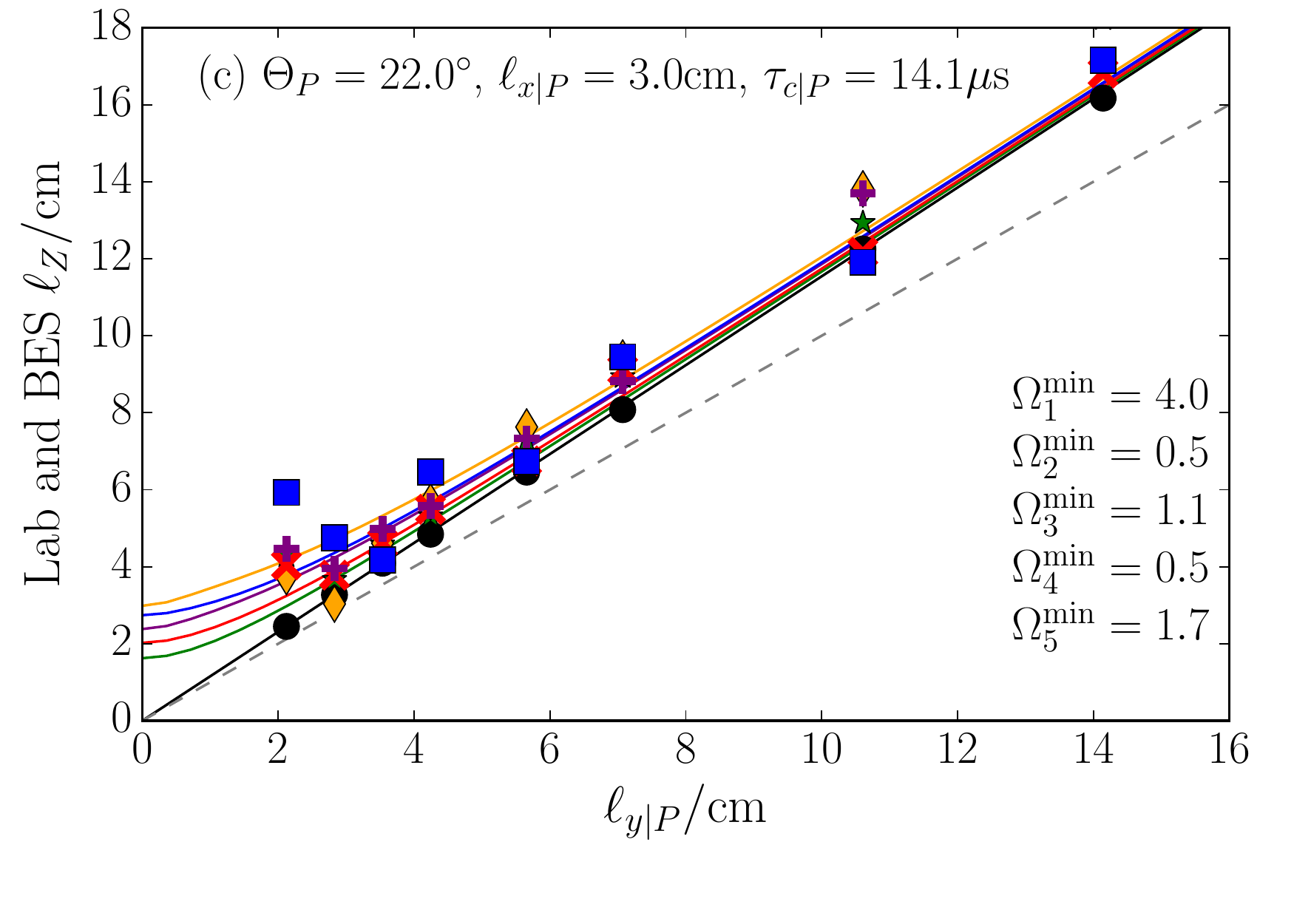}
\includegraphics[width=\textwidth]{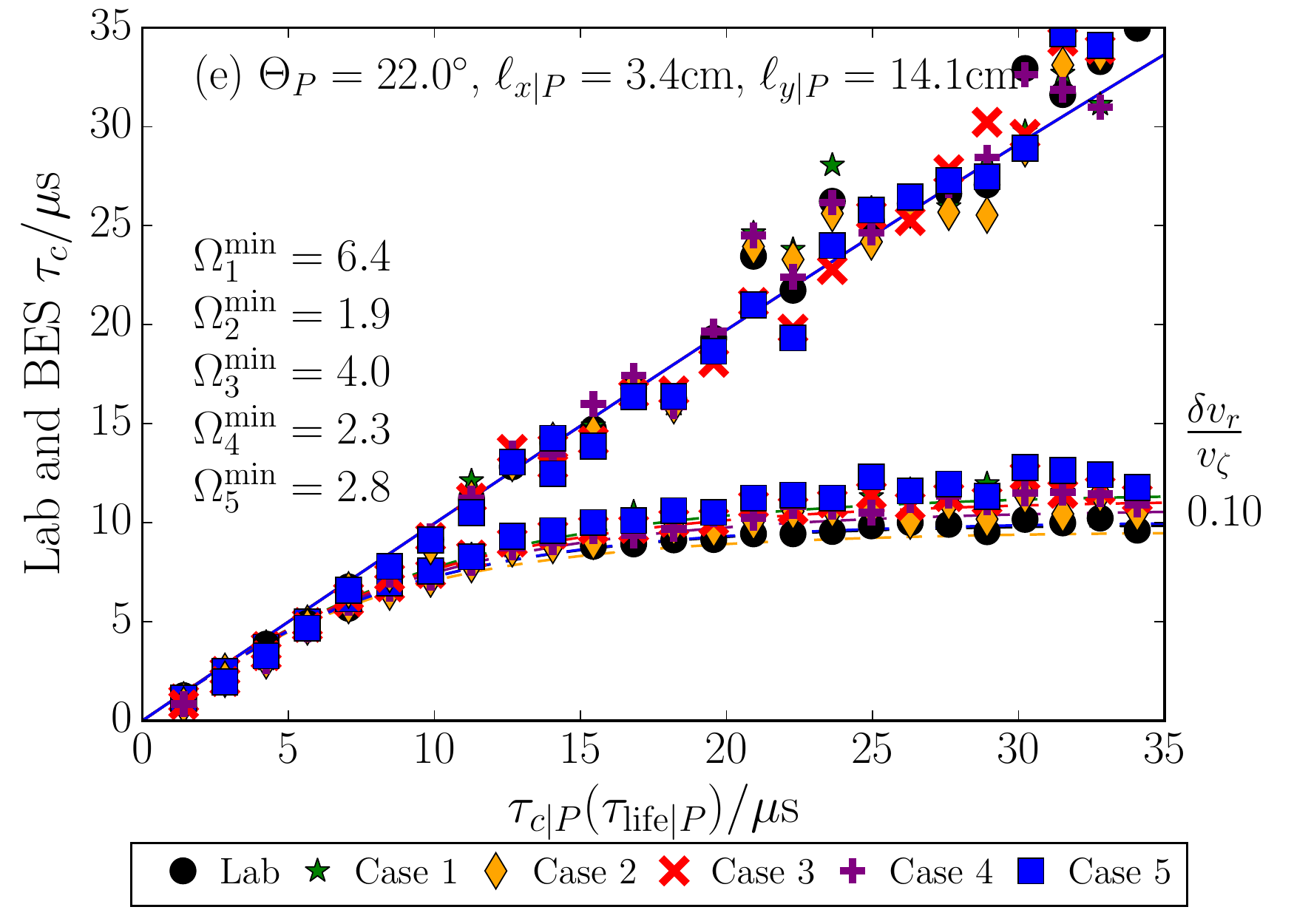}
\end{minipage}
\begin{minipage}{0.45\textwidth}
\includegraphics[width=\textwidth]{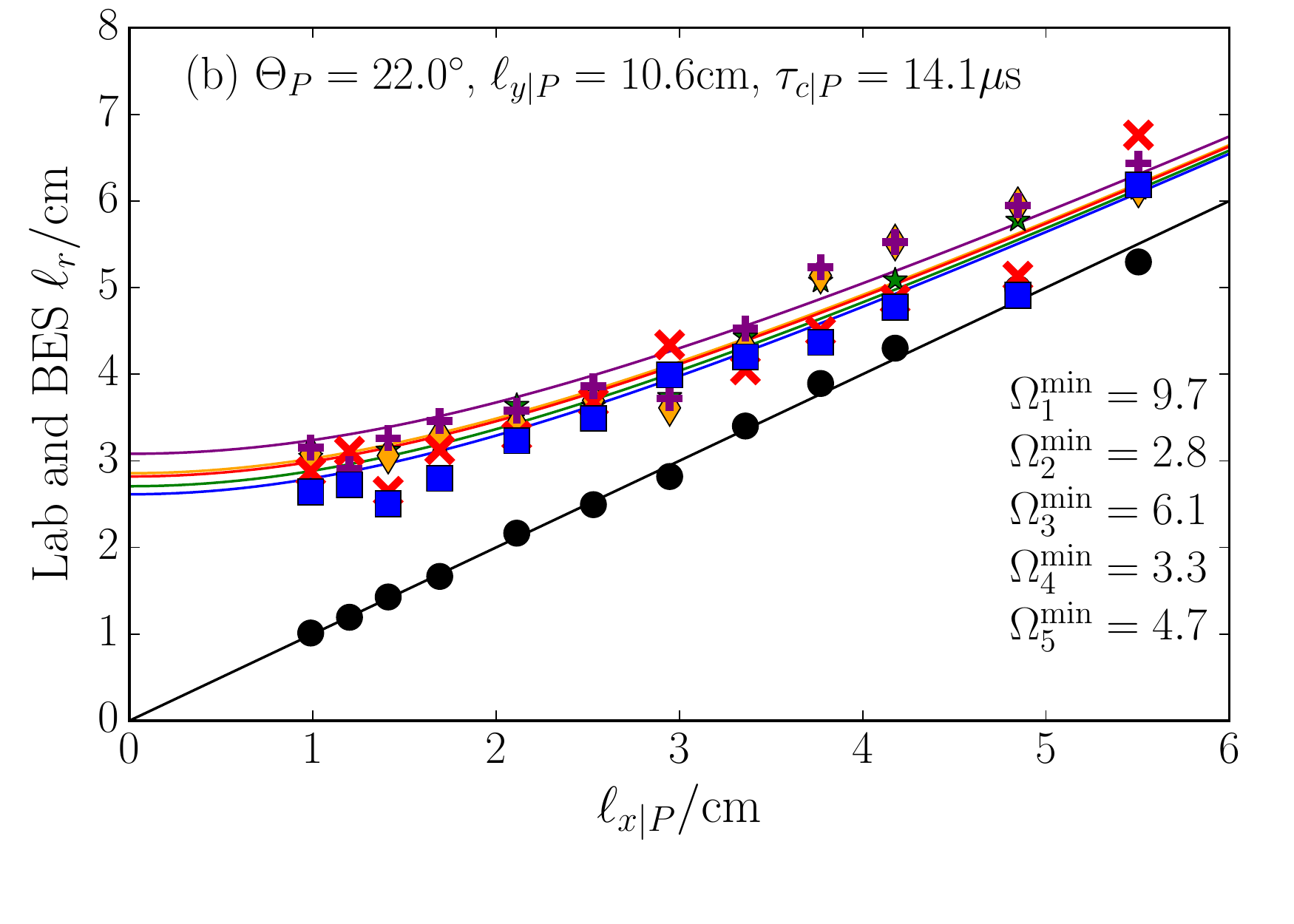}
\includegraphics[width=\textwidth]{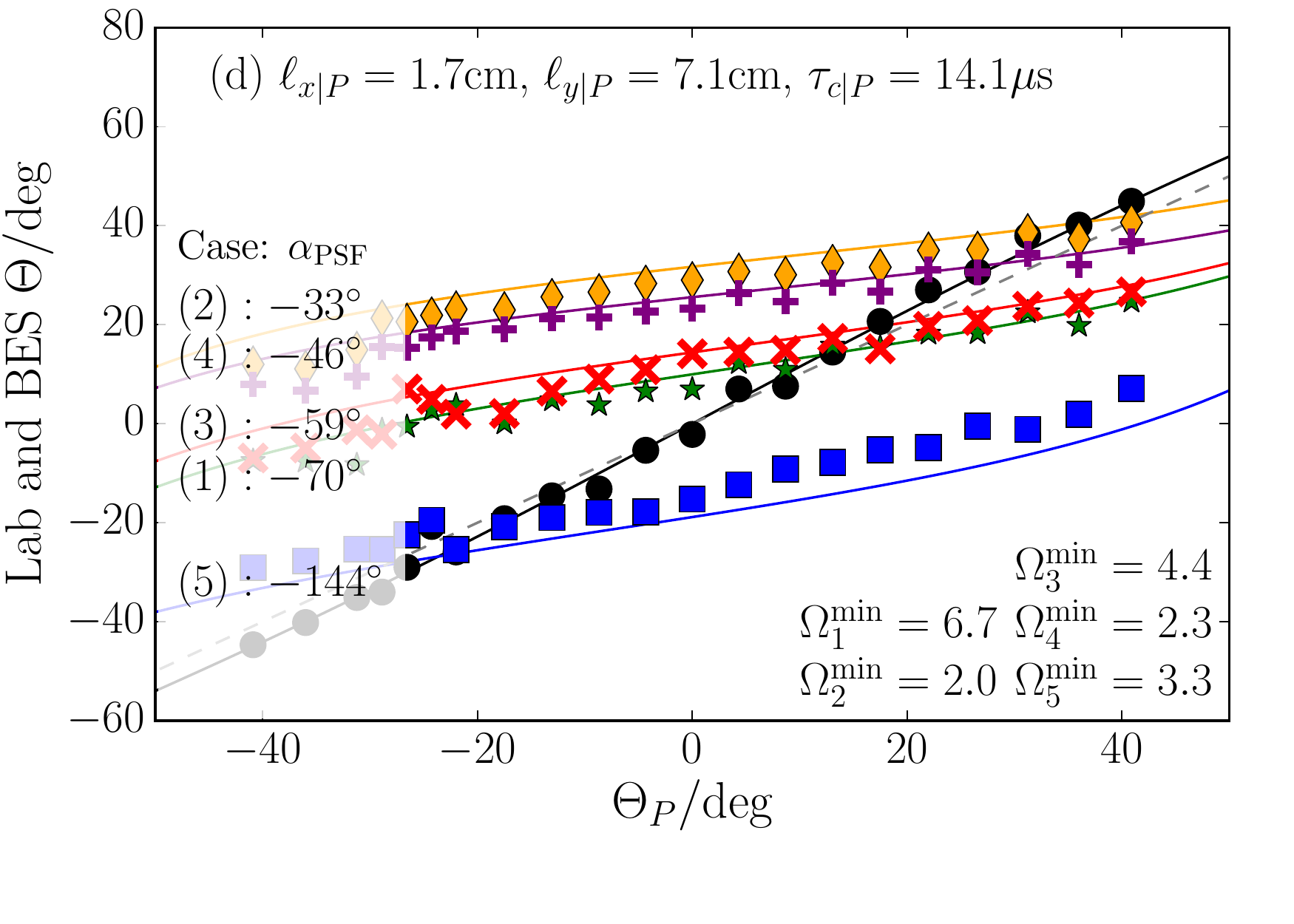}
\includegraphics[width=\textwidth]{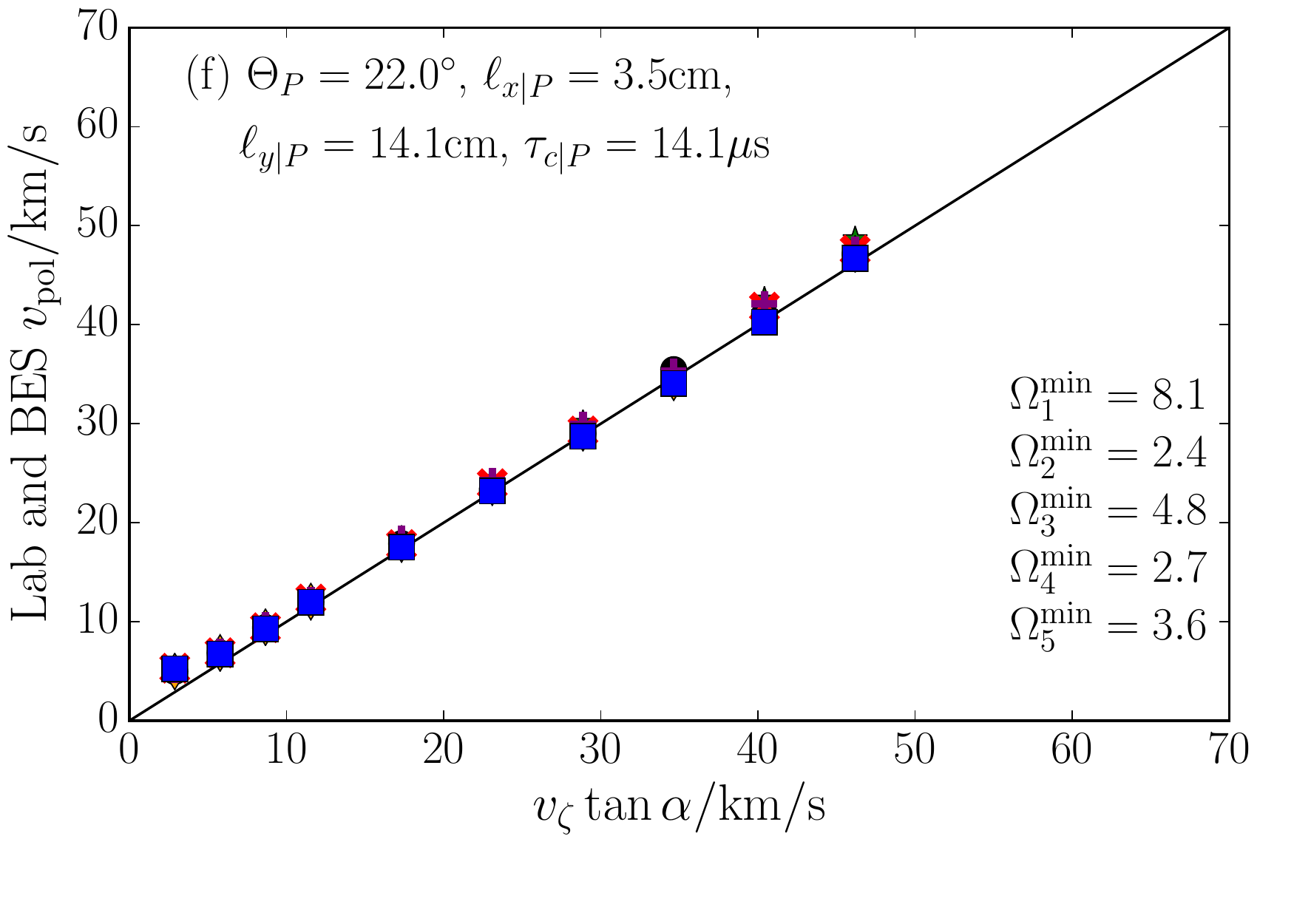}
\end{minipage}
\caption[Real and Gaussian-model PSF effects on the turbulence correlation parameters]{(a) Fluctuation amplitude, (b) radial correlation length, (c) poloidal correlation length, (d) tilt angle; (e) correlation time, and (f) apparent poloidal velocity in the laboratory-frame with and without PSF effects, plotted against their plasma-frame values. The markers correspond to numerical calculations using the model of fluctuating fields described in Section~\ref{sec:toymodel}  and are discussed in Section~\ref{sec:effects_real_PSFs}. The black markers correspond to measurements of the correlation parameters made in the laboratory frame, whilst the different coloured markers correspond to measurements including the effects of the five PSF cases, which have been introduced in Section~\ref{sec:PSFsInMAST} and are illustrated in Figure~\ref{fig:psf_alpha}. The black solid lines correspond to the laboratory-frame relations (\ref{fluc_amp_lab}), (\ref{ellrt_ellxp}), (\ref{lab_ellZ_approx}), (\ref{krt_kxp}-\ref{kzt_kyp}), (\ref{lab_corr_time}), and (\ref{vpol_lab}) in each of the panels, respectively. The coloured lines correspond to the Gaussian-model PSF calculations (\ref{PSFflucamp}), (\ref{an_ellra}), (\ref{an_ellza}), (\ref{an_kra}-\ref{an_kza}), (\ref{BEScorrtime}), and (\ref{vpolPSF}), respectively, which are discussed in Section~\ref{sec:validity_of_Gaussian}. The lowest numerical values of $\Omega$, from (\ref{conditionOnSmallCrossTerm}), are given in each panel for each PSF case. The lower set of curves in (e), labelled with values of $\dvr/\vz$, are discussed in Section~\ref{sec:gs2temporal}. \label{fig:blobby_plots} }
\end{figure}

In this section, we consider what effects the real PSFs have on the measurement of turbulence parameters. We consider each of the plasma-frame parameters in turn: the fluctuation amplitude, radial correlation length, binormal correlation length, tilt angle (the ratio of radial and binormal wavenumbers), correlation time, and apparent poloidal velocity. Proceeding through the panels in Figure~\ref{fig:blobby_plots}(a-f), each of these plasma-frame correlation parameters is varied, keeping all other quantities constant. To start with, we are only concerned with the discrete data points (the solid lines will be discussed in Section~\ref{sec:validity_of_Gaussian}). The laboratory-frame correlation parameters measured from the model fluctuating density field, using the methods of Section~\ref{sec:measurement}, are marked with filled black circles. The equivalent correlation parameters of the synthetic-BES intensity fields, generated using (\ref{PSF_def}) with the five PSF cases (Figure~\ref{fig:psf_alpha} and Section~\ref{sec:PSFsInMAST}), are marked with coloured shapes. By comparing the correlation parameters of these five PSF cases to the laboratory-frame correlation parameters, we see what effect the real PSFs have.

\subsubsection{Fluctuation amplitude.}\label{sec:modelflucamp}
In Figure~\ref{fig:blobby_plots}(a), we see that the real PSFs cause the BES-measured fluctuation amplitude $\dII$ to decrease compared to the laboratory-frame density-fluctuation amplitude $\dnn$. The extent of this decrease depends on the PSF case, and, by using the PSF parameters from Table~\ref{tab:psf_parameters}, can be seen to be approximately proportional to the area, $A\simeq L_1L_2$, of the PSFs. It is also evident that, for a given set of real PSFs, there is a linear relationship between the laboratory-frame and synthetic-BES fluctuation amplitude.

\subsubsection{Radial correlation length.}\label{sec:real_radialcorrlength}
The radial correlation length measured from the synthetic-BES data is longer than the laboratory-frame radial correlation length, as can be seen in Figure~\ref{fig:blobby_plots}(b). The increase in radial correlation length due to the PSFs is greater for shorter laboratory-frame radial correlation lengths. This occurs because the laboratory-frame radial correlation length becomes shorter than the size of the PSFs ($\simeq L_1$) and, therefore, the radial width of the spatial correlation function becomes dominated by the blurring due to the PSFs as they average over the small-scale perturbations. 

The difference between the effect of the different real-PSF cases is small ($\simeq 0.5\ \mathrm{cm}$) for most laboratory-frame radial correlation lengths, suggesting that the detailed shape of the PSFs is not very important in determining to what extent the radial correlation length is increased by the PSFs.

\subsubsection{Poloidal correlation length.}\label{sec:real_poloidalcorrlength}
In Figure~\ref{fig:blobby_plots}(c), we see that the effect of the real PSFs on the laboratory-frame poloidal correlation length follows similar trends to the radial correlation length discussed in Section~\ref{sec:real_radialcorrlength}. However, at very small values, the poloidal correlation length measured from the synthetic-BES data can increase even as the laboratory-frame poloidal correlation length decreases. This occurs when both the laboratory-frame poloidal and radial correlation lengths are similar to the PSF size. As a result, when fitting (\ref{fit_fun}) to extract the spatial correlation parameters, the parameters cannot be well constrained. This can be seen by considering the values of $\Omega$ (\ref{conditionOnSmallCrossTerm}), which are smallest at low $\ellyp$, and are given for each PSF case by $\Omega^\mathrm{min}\equiv\min\Omega$ in Figure~\ref{fig:blobby_plots}(c). The values of $\Omega^\mathrm{min} \leq 1$, clearly do not satisfy the requirement (\ref{conditionOnSmallCrossTerm}), $\Omega \gg 1$, and, therefore, suggest that (\ref{fit_fun}) is not the appropriate fitting function to use. Nevertheless, in most experimental measurements of turbulence in MAST, the poloidal correlation length is about three times longer than the radial correlation length~\cite{Ghim2013,FieldPPCF2014}, and thus this non-monotonic regime is unlikely to be relevant.

\subsubsection{Tilt angle of correlation function.}\label{sec:tiltcorrfun}
The ratio of the radial, $\kxp$, to binormal, $\kyp$, wavenumbers of turbulence is of particular interest in theories of suppression of turbulence by flow shear~\cite{Fedorczak2012a,Schekochihin2012}. Therefore, we consider the effect of PSFs on this tilt angle, rather than on the two wavenumbers separately. Thus, in addition to (\ref{theta_B_def}), we define the tilt angle in the laboratory frame to be
\begin{equation}
\Theta_L = -\arctan\left(\frac{\krt}{\kzt}\right),
\end{equation}
and the plasma-frame tilt angle to be
\begin{equation}
\Theta_P = -\arctan\left(\frac{\kxp}{\kyp}\right),
\end{equation}
where the wavenumbers in the laboratory frame are related to those in the plasma frame through (\ref{krt_kxp}) and (\ref{kzt_kyp}).

In Figure~\ref{fig:blobby_plots}(d), the tilt angles $\Theta_B$ of the correlation function of the synthetic-BES data generated using the five representative real-PSF cases are significantly different from the laboratory-frame tilt angle. Generally, the effect of the real PSFs is to decrease the range of possible values of tilt angle that can be observed. For example, consider Case 2, where the tilt angle ranges between $\simeq 10^\circ$ and $45^\circ$, despite the laboratory-frame tilt angle ranging between $-45^\circ$ and $45^\circ$. The reduction in the range of measurable tilt angles due to PSF effects can be understood by first realising that the radial and poloidal wavenumbers can be determined by the position closest to the peak of the correlation function where the correlation function changes sign. The effect of the PSFs is to change this zero-crossing position, by integrating (unevenly) over the positive and negative regions of the correlation function. Thus, when the tilt angle of the laboratory-frame correlation function is aligned with the PSF angle $\Theta_L \simeq \apsf \Mod{ 180^\circ}$, large parts of both negative regions of the correlation function are integrated over (provided the sizes of the PSF and of the correlation function are similar), resulting in the correlation function of the synthetic-BES data having a significantly different zero-crossing position, and, therefore, a different tilt angle. Conversely, the synthetic-BES tilt angles are closest to the laboratory-frame tilt angles when the laboratory-frame tilt angle and the PSF angle are misaligned, i.e., $\Theta_L \simeq \apsf + 90^\circ \Mod{ 180^\circ}$.

The tilt angle also differs significantly between the five real-PSF cases. In Figure~\ref{fig:blobby_plots}(d), we have labelled each curve by the PSF angle from Table~\ref{tab:psf_parameters}, which shows a clear correlation between the PSF angle and the systematic angular shift between the tilt angle of the correlation function for each of the PSF cases. The reason for these differences can be understood by following the same argument as presented in the previous paragraph. Therefore, we see that the alignment between $\Theta_L$ and $\apsf$ can have a significant effect on the measured tilt angle $\Theta_B$.

\subsubsection{Correlation time.}\label{sec:corrtimemodel}
In this section, we only consider the upper set of data points in Figure~\ref{fig:blobby_plots}(e); the lower set of data points will be discussed in Section~\ref{sec:gs2temporal}. There is almost no difference between the laboratory-frame correlation time, $\tauct$, and the correlation time measured from the synthetic-BES data, $\tauca$, for each of the five PSF cases. 

At correlation times above $\simeq 15\ \mu\mathrm{s}$, the agreement between the laboratory-frame and synthetic-BES values worsens, but not in a systematic manner. This is because, for longer correlation times, the perturbations decay less quickly as they pass the poloidal detector channels, and, therefore, the change in amplitude that needs to be measured in order to calculate the correlation time becomes smaller. At sufficiently small changes in amplitude, statistical noise from the model fluctuating field can start to affect the measurement of the change in amplitude.

\subsubsection{Apparent Poloidal velocity.}\label{sec:real_polvelocity}
The apparent poloidal velocity measured in the laboratory frame and from the synthetic-BES data with the PSFs of all five cases shows considerable agreement, as manifested in Figure~\ref{fig:blobby_plots}(f). Therefore, the real PSFs have almost no effect on the poloidal-velocity measurement, under the assumptions of our model of fluctuating fields.

\subsection{Validity of using Gaussian-model-PSFs}\label{sec:validity_of_Gaussian}
As described in Section~\ref{sec:GaussianPSFs}, using the Gaussian-model PSFs relies on the following modelling assumptions:
\begin{enumerate}
	\item the shape of each PSF is well described by a Gaussian, parameterised by $L_1, L_2$ and $\apsf$;
	\item all the PSFs in the sub-array (inner/outer) of the BES being considered have the same Gaussian parameters;
\end{enumerate}
It is the aim of this section to demonstrate that the above approximations are reasonable, by comparing the effects of the Gaussian-model PSFs, using the PSF parameters in Table~\ref{tab:psf_parameters}, with the effects of the real PSFs on the turbulence correlation parameters that were discussed in Section~\ref{sec:effects_real_PSFs}.

\subsubsection{Fluctuation amplitude.}\label{sec:gauss_fluc_amp}
In Figure~\ref{fig:blobby_plots}(a), for each of the PSF cases introduced in Section~\ref{sec:PSFsInMAST}, the effect of the Gaussian-model PSFs given by (\ref{sigma_amp_B}) reproduce the observed effect of the real PSFs on the laboratory-frame fluctuation amplitude. Details of how this comparison is made are provided in \ref{sec:toyflucamp}. The most significant difference between the fluctuation amplitudes calculated using the Gaussian-model PSFs and the real PSFs is for Case 5. Indeed, real PSFs of Case 5 show the largest difference from the Gaussian-model PSFs (Figure~\ref{fig:comp_full_gauss_psf_simple}). The larger reduction in the fluctuation amplitude determined using the Gaussian model for Case 5 can be explained by the real PSFs for this case being more peaked than a Gaussian, and so behaving more like delta functions under the integral (\ref{PSF_def}), reducing the effective area over which the PSFs average.

\subsubsection{Radial correlation length.}\label{sec:gauss_radial_corr_leng}
The radial correlation length calculated using the Gaussian-model PSFs (\ref{an_ellra}) shows good qualitative agreement with the radial correlation length measured from the synthetic-BES data, as can be seen in Figure~\ref{fig:blobby_plots}(b). The radial correlation length calculated using the Gaussian-model PSFs tends towards a non-zero constant, dependent on $\ellzt$, $L_1$, $L_2$, and $\apsf$, as the laboratory-frame radial correlation length decreases below the PSF size. The fact that (\ref{an_ellra}) does not go to zero as the laboratory-frame radial correlation length goes to zero means that, if an experimentally measured radial correlation length is lower than the smallest value of (\ref{an_ellra}), the corresponding laboratory-frame value cannot be recovered. Therefore, there is a resolution limit of the BES, which is formalised in Section~\ref{sec:resolutionlimit}.

\subsubsection{Poloidal correlation length.}
The expression (\ref{an_ellza}) for the poloidal correlation length with Gaussian-model PSFs is the same as that for the radial correlation length (\ref{an_ellra}) when the labels $r$ and $Z$ are interchanged $r\leftrightarrow Z$ and the PSF lengths are interchanged $L_1\leftrightarrow L_2$. Hence, the poloidal correlation length given by (\ref{an_ellza}) has similar features to the radial correlation length (\ref{an_ellra}) discussed in Section~\ref{sec:gauss_radial_corr_leng}. However, in real experiments, the poloidal correlation length is longer than both the radial correlation length and the PSF lengths. In such a parameter regime, the laboratory-frame poloidal correlation length is similar to the BES poloidal correlation length, which can be seen in Figure~\ref{fig:blobby_plots}(c), where (\ref{an_ellza}) is plotted for each of the five PSF cases. We also see, in this figure, that the poloidal correlation length given by (\ref{an_ellza}) is qualitatively the same as the poloidal correlation length measured from the synthetic-BES data, which has been discussed in Section~\ref{sec:real_poloidalcorrlength}. 

\subsubsection{Tilt angle of the correlation function.}\label{sec:gauss_tilt_angle}
The tilt angle $\Theta_B$ defined in (\ref{theta_B_def}) is calculated using the radial (\ref{an_kra}) and poloidal (\ref{an_kza}) wavenumbers. In Figure~\ref{fig:blobby_plots}(d), this analytic calculation shows good qualitative agreement with the tilt angles measured from the synthetic-BES data. The root-mean-square (RMS) differences between these two tilt angles, for each of the five PSF cases, are given in Table~\ref{tab:rms_real_gauss}. By comparing these RMS values with the difference measure (\ref{deltaP0def}) between the shapes of the real and Gaussian-model PSFs, also given in Table~\ref{tab:rms_real_gauss}, we see that there is a clear correlation between the two quantities. This suggests that the observed difference is due to the imperfect validity of the assumptions that we have listed at the beginning of this section. The largest RMS difference between the tilt angles is $5.4^\circ$, which corresponds to an approximate error of $10\%$ due to using the Gaussian-model PSFs compared to using the real-PSFs.

\subsubsection{Correlation time and apparent poloidal velocity.}
The correlation time given by (\ref{BEScorrtime}) for Gaussian-model PSFs is exactly the same as the laboratory-frame correlation time (\ref{lab_corr_time}). Indeed, in Figure~\ref{fig:blobby_plots}(e), we see that (\ref{BEScorrtime}) agrees well with the synthetic-BES correlation time for all five representative PSF cases.

The analytic calculation of the apparent poloidal velocity including Gaussian-model PSF effects (\ref{vpolPSF}) shows that there is no difference in this quantity from the laboratory frame. As the real-PSFs also have no effect on the measurement of the apparent poloidal velocity (see Section~\ref{sec:real_polvelocity}), there is, trivially, good agreement between the effects of the Gaussian-model and real PSFs.

\subsection{Summary}\label{sec:summary_sec_6}
As the evidence presented in the above sections shows good agreement between the real- and Gaussian-model-PSF effects for all measured turbulence parameters, it seems reasonable to conclude that the Gaussian-model PSFs describe the effects of the real PSFs on the measurement of the turbulent parameters well. More strongly, we have seen that the error in using Gaussian-model PSFs instead of the real PSFs (that was estimated in Section~\ref{sec:gauss_tilt_angle} to be no more than $10\%$) is smaller than the changes to the measured correlation parameters that are caused by the finite size of the PSFs (see Figure~\ref{fig:blobby_plots}(a),(b), and (d) for changes over $100\%$). Therefore, it is worthwhile to use the Gaussian-model PSFs to assess, and, in Section~\ref{sec:correctingPSFs}, correct for, the PSF effects, even if they are not exactly the same as the real PSFs.

% ----------------------------------------------------------------------------------------------------------
% ----------------------------------------------------------------------------------------------------------
% ----------------------------------------------------------------------------------------------------------

\section{Correcting for PSF effects}\label{sec:inversion}

\subsection{Determination of laboratory-frame spatial correlation parameters and fluctuation amplitude}\label{sec:correctingPSFs}
The equations (\ref{an_ellra}-\ref{an_kza}) for the spatial properties of the turbulence can easily be inverted to find the laboratory-frame parameters as functions of the measured BES parameters:
\begin{eqnarray}
\fl\ellrt^2 &=& \frac{1}{2}\left[\vphantom{\sqrt{()}}   \ellra^2 - 4 ( L_2^2 \cos^2\apsf + L_1^2 \sin^2\apsf ) \right. \nonumber \\
&\quad& \left. + \sqrt{ \ellra^4 + 4 (\ellra/\ellza)^2 (L_1^2-L_2^2)^2 \sin^2(2 \apsf) } \label{an_ellrt_inv}\right], \\
\fl\ellzt^2 &=& \frac{1}{2}\left[\vphantom{\sqrt{()}}  \ellza^2 - 4 ( L_1^2 \cos^2\apsf + L_2^2\sin^2\apsf )  \right. \nonumber \\
&\quad& \left. + \sqrt{ \ellza^4 + 4 (\ellza/\ellra)^2 (L_1^2-L_2^2)^2 \sin^2(2 \apsf) } \label{an_ellzt_inv} \right],
\end{eqnarray}
where the positive square roots are the only physically relevant choice, and
\begin{eqnarray}
\fl\krt &=& \left[1 + \frac{2(L_1^2\sin^2\apsf + L_2^2\cos^2\apsf)}{\ellrt^2}\right]\kra - \left[\frac{(L_1^2-L_2^2)\sin(2\apsf)}{\ellrt^2}\right]\kza \label{krt_inv},\\ 
\fl\kzt &=& \left[1 + \frac{2(L_1^2\cos^2\apsf + L_2^2\sin^2\apsf)}{\ellzt^2}\right]\kza - \left[\frac{(L_1^2-L_2^2)\sin(2\apsf)}{\ellzt^2}\right]\kra. \label{kzt_inv}
\end{eqnarray}
The wavenumbers (\ref{krt_inv}) and (\ref{kzt_inv}) can be combined to find the laboratory-frame tilt angle:
\begin{eqnarray}
\fl\tan\Theta_L = 
\frac{\ellzt^2 \left[ \tan\Theta_B \sec2\apsf (L_1^2 + L_2^2 + \ellrt^2) - (L_1^2 - L_2^2) (\tan\Theta_B + \tan2 \apsf)  \right]}{\ellrt^2 \left[ (L_1^2 + L_2^2 + \ellzt^2)\sec2\apsf + (L_1^2 - L_2^2) (1 - \tan\Theta_B \tan2 \apsf)  \right]}, \label{an_tantheta_inv}
\end{eqnarray}
which for brevity has been written in terms of the laboratory-frame correlation lengths $\ellrt$ and $\ellzt$, given by (\ref{an_ellrt_inv}) and (\ref{an_ellzt_inv}), respectively. 

The laboratory-frame fluctuation amplitude can easily be determined from (\ref{PSFflucamp}) once the laboratory-frame spatial parameters (\ref{an_ellrt_inv}-\ref{kzt_inv}) have been calculated. One of the advantages of the above explicit expressions for the laboratory-frame parameters is that the uncertainty in the resulting quantities can be derived from the uncertainties on the measurements of the parameters that characterise the BES turbulence and the PSFs.

\subsection{Determination of plasma-frame spatial correlation parameters and fluctuation amplitude}
The transformations of the laboratory-frame spatial correlation parameters into plasma-frame spatial correlation parameters are described by (\ref{ellrt_ellxp}-\ref{kzt_kyp}), which require the pitch angle $\alpha$ of the magnetic field  to be known. The only complication is in the reconstruction of the plasma-frame binormal correlation length (\ref{lab_pol_length}), as we have no measurement of the plasma-frame parallel correlation length $\ellzp$. However, by using the asymptotic ordering (\ref{fullordering}-\ref{endfullordering}), this parallel correlation length can be neglected and the plasma-frame binormal correlation length can be calculated from (\ref{lab_ellZ_approx}). As discussed in \ref{sec:toyflucamp}, the fluctuation amplitude in the plasma frame is the same as that in the laboratory frame.

\subsection{Resolution limit of BES}\label{sec:resolutionlimit}
\begin{figure}[h]
\centering
\includegraphics[width=1.0\textwidth]{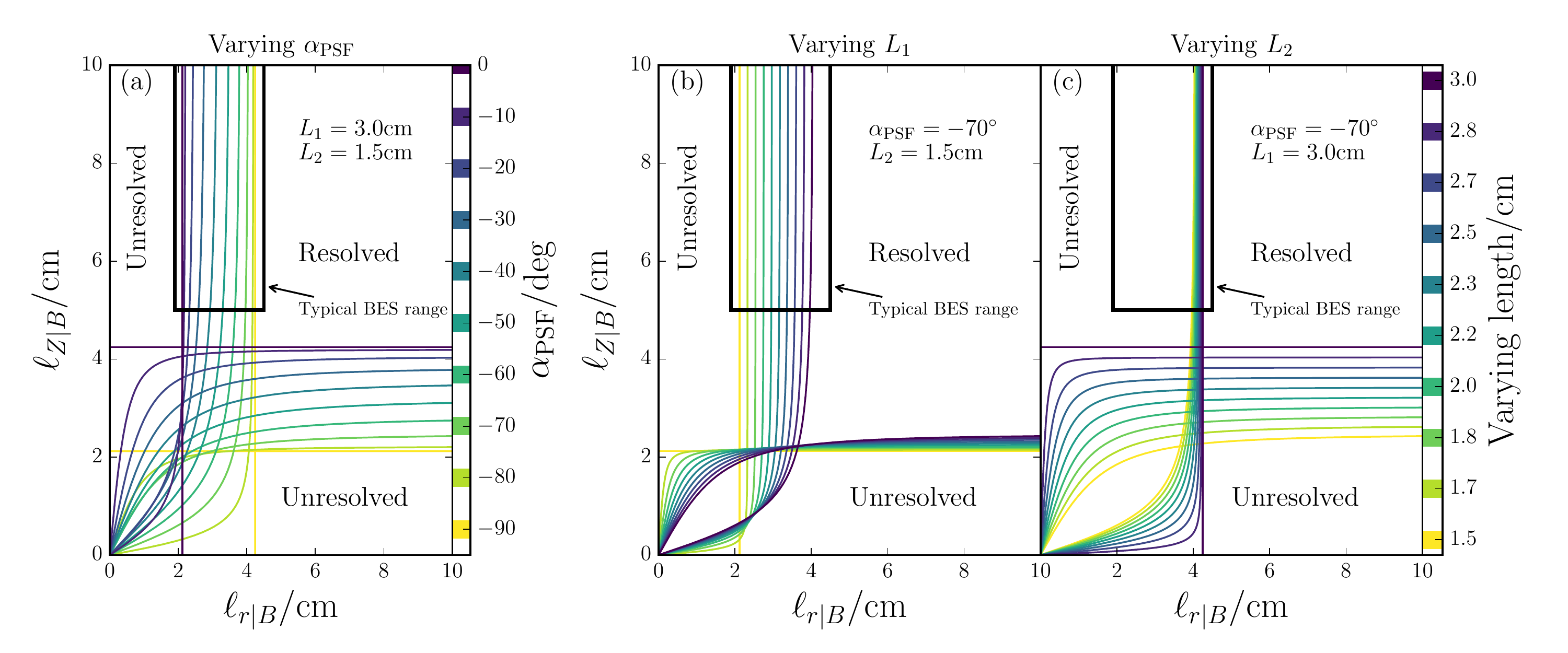}
    \caption[Resolution limits of BES measurements]{Plots of the resolution limits (\ref{ellra_boundary}) and (\ref{ellza_boundary}) showing: (a) the effect of the PSF angle $\apsf$, (b) the effect of the length of the principal PSF component $L_1$, (c) the effect of the length of the secondary PSF component $L_2$, on the resolution limit of the BES system. The black box indicates the typical range of measured correlation lengths using the MAST BES system~\cite{Ghim2013}.\label{fig:analyticboundaries}}
\end{figure}

In all of the above calculations (Section~\ref{sec:PSF_analytic} to Section~\ref{sec:correctingPSFs}), provided the PSF parameters are known, it is always possible to convert between real-valued laboratory-frame parameters and BES parameters, however, this is not necessarily true for the data taken from the real BES diagnostic. It is possible for the laboratory-frame parameters deduced from this inversion to be imaginary-valued, if the expressions in the right-hand-side of (\ref{an_ellrt_inv}) and (\ref{an_ellzt_inv}) are evaluated to be negative. This occurs when the BES-measured correlation length (either $\ellra$ or $\ellza$) has a value that is smaller than the minimum of value of the same length determined using Gaussian-model PSFs, as can be seen in Figures~\ref{fig:blobby_plots}(b) and (c). We can, therefore, define the resolution limits as the minimum values of $\ellra$ and $\ellza$ that can be obtained from (\ref{an_ellrt_inv}) and (\ref{an_ellzt_inv}), which we will refer to as $\ellra^\mathrm{res}$ and  $\ellza^\mathrm{res}$, respectively. These minima occur when the laboratory-frame values are zero ($\ellrt = 0$ or $\ellzt = 0$). Equations (\ref{an_ellrt_inv}) and (\ref{an_ellzt_inv}) can then be solved for  $\ellra^\mathrm{res}$ and $\ellza^\mathrm{res}$, as functions of the measured BES spatial correlation lengths and the PSF parameters. The radial resolution limit is then
\begin{equation}
\ellra^\mathrm{res} = \frac{2 \ellza (L_1^2\sin^2\apsf + L_2^2\cos^2\apsf)}{\sqrt{2 \ellza^2(L_1^2\sin^2\apsf + L_2^2\cos^2\apsf) + (L_1^2 - L_2^2)^2 \sin^2 2\apsf}}, \label{ellra_boundary}
\end{equation} 
and the poloidal resolution limit is
\begin{equation}
\ellza^\mathrm{res} = \frac{2 \ellra (L_1^2\cos^2\apsf + L_2^2\sin^2\apsf)}{\sqrt{2 \ellra^2(L_1^2\cos^2\apsf + L_2^2\sin^2\apsf) + (L_1^2 - L_2^2)^2 \sin^2 2\apsf}}.\label{ellza_boundary}
\end{equation} 
As the laboratory-frame correlation lengths are required to calculate the laboratory-frame wavenumbers, these expressions also determine when the wavenumbers, $\kra$ and $\kza$, are resolved.

Possible reasons why the experimentally measured BES correlation lengths can be below the resolution limits (\ref{ellra_boundary}) and (\ref{ellza_boundary}) include uncertainties in the measurement from the BES, such as background $D_\alpha$ emission and electronic noise in the detector, as well as uncertainties in the equilibrium profiles that are used to calculate the PSFs. 

Even measurements above, but near, the resolution boundaries may be unreliable. For example, when the laboratory-frame radial correlation length is calculated using (\ref{an_ellrt_inv}) any small uncertainty in the BES-measured radial correlation length causes a large uncertainty in the laboratory-frame radial correlation length, see Figure~\ref{fig:blobby_plots}(b). Therefore, the resolution limits (\ref{ellra_boundary}) and (\ref{ellza_boundary}) must be considered as the absolute lower boundaries of the diagnostic. In order to be confident in the reliability of a reconstruction, an additional constraint on the values of the reconstructed laboratory-frame correlation lengths may be used, such that the reconstructed values are above a certain threshold, e.g., $\ellrt\ge \ellra^\mathrm{res}$.

In Figure~\ref{fig:analyticboundaries}, expressions (\ref{ellra_boundary}) and (\ref{ellza_boundary}) are plotted for different combinations of PSF parameters. The region in the upper right side of each plot can be considered resolved. Figure~\ref{fig:analyticboundaries}(a) demonstrates that the tilting of the PSFs can cause a significant change to the position of the resolution limits in the range of $2\ \mathrm{cm} < \ellra < 4 \ \mathrm{cm}$. As most of the MAST BES-measured radial correlation lengths lie in this interval~\cite{Ghim2013}, it is clear that properly accounting for PSF effects is important. The closeness of the resolution boundary to the MAST BES-measured radial correlation lengths also means that a small error, in either of these two quantities, can result in the measurement being essentially unresolved.

In Figures~\ref{fig:analyticboundaries}(b) and \ref{fig:analyticboundaries}(c), the PSF angle $\apsf$ is fixed and the lengths $L_1$ and $L_2$ are varied. From Table~\ref{tab:psf_parameters}, we see that the principal length $L_1$ can increase from $2.0 \ \mathrm{cm}$ to $3.1 \ \mathrm{cm}$ during a shot (Case 1 to Case 2). When $\apsf$ is constant, this means that the radial resolution limit can lay anywhere in the entire range of typical MAST BES measurements of the radial correlation length. Fortunately, typical measured values of the poloidal correlation length tend to be greater than the poloidal resolution limit.

% ----------------------------------------------------------------------------------------------------------
% ----------------------------------------------------------------------------------------------------------
% ----------------------------------------------------------------------------------------------------------

\section{Testing the inversion method using gyrokinetic simulations}\label{sec:testing}
In this section, we discuss only the spatial correlation parameters and fluctuation amplitude, as the model of fluctuating fields that we have been using (Section~\ref{sec:toymodel}) has shown that the PSFs have no effect on the temporal correlation parameters. Later, in Section~\ref{sec:gs2temporal}, however, we find that PSFs do affect the correlation-time measurement, and attempt to extend our model to explain this.

\subsection{Assumptions of our method}
We have derived, in Section~\ref{sec:PSF_analytic}, the effect of PSFs on a specified correlation function and have tested numerically the validity of these analytic calculations by applying real PSFs to a time series generated by a model fluctuating density field in Section~\ref{sec:numeric}. Our conclusions about the effect of the PSFs on the turbulent correlation parameters rely on the functional form (\ref{plasma_frame_correlation_function}) that we have assumed for the correlation function. In order to test this assumption, we generate a density-fluctuation time series (in the laboratory frame) that is more physically motivated than that generated by our artificial model of fluctuating fields by using the turbulence data obtained in numerical simulations of a MAST-relevant plasma~\cite{vanWyk2016}, with the local, nonlinear, gyrokinetic flux-tube code GS2~\cite{Kotschenreuther1995}.

The correlation parameters for this numerical data are calculated, as in Section~\ref{sec:measurement}, both from the laboratory-frame data and from synthetic-BES data created by applying real PSFs using (\ref{PSF_def}). We then use the analytic relations of Section~\ref{sec:inversion} and the Gaussian model of the PSFs to estimate the spatial laboratory-frame correlation parameters (corrected parameters) from the correlation parameters of the synthetic-BES data. By comparing the corrected parameters with the correlation parameters measured from the raw density field calculated by GS2, we have a direct measure of the quality of the assumptions that underpin our procedure.

\subsection{Gyrokinetic simulations}\label{sec:gksim}
The full details of the gyrokinetic simulation that we use can be found in~\cite{vanWyk2016}. The equilibrium used for the simulation is taken from MAST shot $\# 27268$ at $t=0.250\ \mathrm{s}$ and at a major radius of $R=1.32\ \mathrm{m}$. The ion-temperature gradient is $a/L_{T_i} = 4.9$, where $a$ is the minor radius of the last closed flux surface, and  $L_{T_i}^{-1} = T_i^{-1}\partial{T_i}/\partial{r}$ is the ion (Deuterium) temperature gradient length, with $T_i$ the ion temperature and $r$ the radial coordinate used by GS2, as described in~\cite{vanWyk2016}. This simulation was performed with an experimentally relevant equilibrium flow shear $\gamma_E = 0.16 \ \vth/a$, where $\vth = \sqrt{2 T_i/m_i}$ is the ion thermal velocity, and $m_i$ the ion mass. The ion and electron species are both treated kinetically with the true ion-electron mass ratio. Artificial damping is applied to separate the ion and electron spatial scales in order to reduce computation time. 

The output of the simulation that we use is a two-dimensional field of density fluctuations spanning $40\ \mathrm{cm}\times 80\ \mathrm{cm}$ in the radial-poloidal plane\footnote{The GS2 output for a single flux tube is generated using a single set of equilibrium parameters, however, we note that, when the PSFs are applied to the density field, the spatial variation of these PSFs is caused by radial variations in the equilibrium, thus introducing some inconsistency into the analysis. This would be of concern if we were comparing these simulations directly with experiment rather than simply using them to validate our data-reconstruction technique.}. The implementation of flow shear in GS2 introduces spurious aliasing of the density field, which increases with radial distance from the centre of the simulation domain. Therefore, we only analyse data from the central region ($\pm 5\ \mathrm{cm}$) of the simulation output. This aliasing effect will be illustrated and discussed in more detail in Section~\ref{sec:gs2temporal}.

\subsection{Spatial correlation parameters}\label{sec:gs2spatial}
\begin{figure}[h]
\centering
\includegraphics[width=\textwidth]{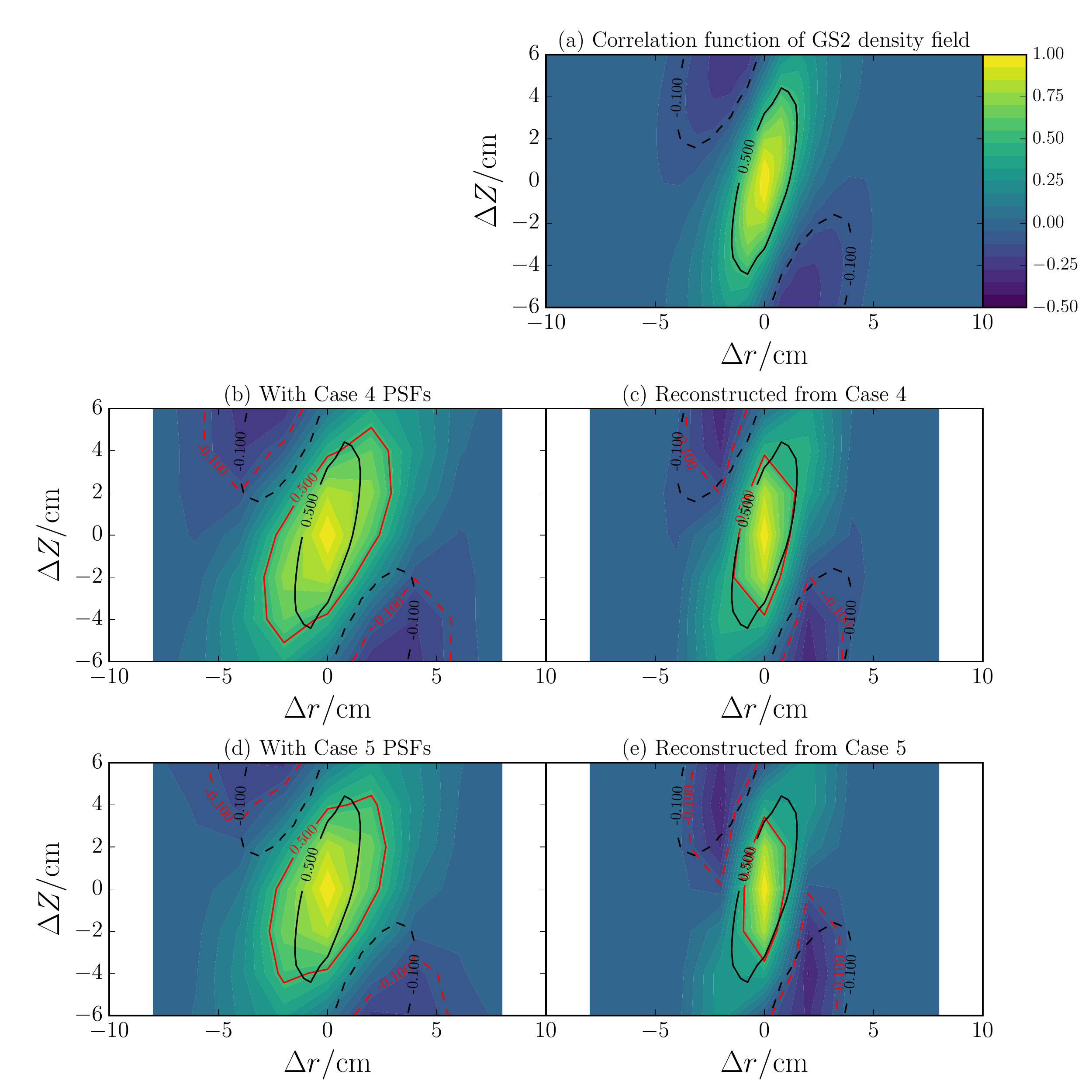}
\caption[Testing the reconstruction of spatial correlation parameters using gyrokinetic simulation.]{Binned spatial correlation functions from a GS2 simulation of MAST~\cite{vanWyk2016}: (a) raw GS2 laboratory-frame binned correlation function, also given as black contours in all other panels for comparison; (b) and (d) the correlation function after application, using (\ref{PSF_def}), of Case 4 and Case 5 PSFs, respectively (see Section~\ref{sec:PSFsInMAST}); (c) and (e) the corrected correlation function after the analytic reconstruction (Section~\ref{sec:correctingPSFs}) is applied to (b) and (d), respectively. Values for the fitted correlation parameters using (\ref{fit_fun}) are given in Table~\ref{tab:GS2_table}. \label{fig:GS2psfCheck} }
\end{figure}

\begin{table}[h]
\centering
\begin{tabular}{l|g|r g|r g|}
% see Evernote note 01/03/2016 for new third radial data values 
Parameter                        	& GS2 value 	& Case 4 	$\rightarrow$& Corrected 	& Case 5  $\rightarrow$	& Corrected \\ \hline
Fluc. amp. $[\%] $             	& $2.2$		& $1.4$  	&             	  $2.1$		&        	$1.4$		&      $2.6$     		\\
$\ell_r/\mathrm{cm}$        	& $2.16$           & $3.94$       	& $2.49$       	& $3.43$      		& $2.23$          \\
$\ell_Z/\mathrm{cm}$ 	& $8.53$     	& $10.1$       	& $9.78$       	& $10.1$      		& $9.67$          \\
$k_r/\mathrm{cm}^{-1}$	& $-0.633$     	& $-0.341$       	& $-0.639$    	& $-0.299$    		& $-0.873$         \\
$k_Z/\mathrm{cm}^{-1}$  	& $0.280$      	& $0.246$       	& $0.248$     	& $0.244$      		& $0.278$          \\
$\tau_c/\mu\mathrm{s}$  	& $7.8\pm3.3$ &  $5.7\pm1.4$    &  $ - $        		&  $8.0\pm3.4$     			& $ - $           \\
$\vpol/km/s$			& $42.8\pm9.3$ & $41.9\pm8.2$   &         $ - $       			&      $40.1\pm8.2$  			&      $ - $        \\
\end{tabular}
\caption[Spatial correlation parameters of Figure~\ref{fig:GS2psfCheck}.]{Correlation parameters measured from GS2 simulations with and without PSFs, and the corrected correlation parameters after using the reconstruction method described in Section~\ref{sec:correctingPSFs}. The correlation functions from which the spatial parameters are extracted are plotted in Figure~\ref{fig:GS2psfCheck}. The temporal correlation properties are plotted in Figure~\ref{fig:GS2velocities}.}
\label{tab:GS2_table}
\end{table}

The laboratory-frame binned spatial correlation function (see discussion in Section~\ref{sec:example_spatial_corr_fun}) of the GS2 data calculated from (\ref{correlation_def}) is plotted in Figure~\ref{fig:GS2psfCheck}(a) and the corresponding fitting parameters measured using (\ref{fit_fun}) are given in Table~\ref{tab:GS2_table}. We consider the application of two sets of PSFs to the GS2 data, Cases 4 and 5, as typical examples of DND and LSND PSFs, which have been introduced previously in Section~\ref{sec:PSFsInMAST}. The correlation functions of the resulting synthetic-BES data for these two cases are shown in Figure~\ref{fig:GS2psfCheck}(b) and (d), respectively. Both cases show nearly a factor of two increase of the radial correlation length in the synthetic-BES data, compared to the laboratory-frame correlation length in Figure~\ref{fig:GS2psfCheck}(a). The broadening of the correlation function due to the PSFs is more pronounced in the radial direction than the poloidal direction because the radial correlation length is nearer the PSF size, $L_1$, than the poloidal correlation length. 

The radial wavenumber is a factor of two smaller in the synthetic-BES correlation function compared to the laboratory-frame correlation function. We note that Case-5 PSFs cause a slightly greater reduction of the radial wavenumber, because their principal component is more aligned with the tilt axis of the correlation function than that of the Case-4 PSFs (whose principal component is perpendicular to the tilt axis), see Figure~\ref{fig:blobby_plots}(d). This is because Case-5 PSFs have a greater ``averaging'' effect over the oscillatory structure of the correlation function, as discussed in Section~\ref{sec:tiltcorrfun}.

The measured properties of the PSFs (Table~\ref{tab:psf_parameters}) are used with the inversion equations of Section~\ref{sec:correctingPSFs} to correct for the PSF effects. The results of this procedure are plotted in Figures~\ref{fig:GS2psfCheck}(c) and (e). We see that these corrected correlation functions match the raw GS2 laboratory-frame correlation function, in Figure~\ref{fig:GS2psfCheck}(a), much better than the uncorrected correlation functions of the synthetic-BES data. A closer inspection of the numerical values in Table~\ref{tab:GS2_table} shows that, for both PSF cases, the corrected values of radial and poloidal correlation lengths are near, but overestimate slightly, the raw GS2 correlation lengths. 

For Case-4 PSFs, the correction procedure works well for the radial wavenumber ($<\!\!\! 1\%$ difference), but not as well for the poloidal wavenumber ($11\%$ difference). Conversely, for Case-5 PSFs, the corrected radial wavenumber shows a significant mismatch with the raw-GS2 value ($38\%$ difference), whilst the poloidal wavenumber shows good agreement ($<\!\! 1\%$ difference). The large magnitude of some of these differences between the corrected and raw-GS2 wavenumbers is due to the fact that the radial correlation length is approximately the same size as the principal component of the PSFs and, therefore, any small errors in the fitting of the correlation function, combined with the differences between the Gaussian-model PSFs and the real PSFs, are amplified. The worse performance of the correction procedure for Case 5 compared to Case 4 is also related to the fact that the Gaussian model is a better approximation to the Case-4 PSFs than to the Case-5 PSFs (see Section~\ref{sec:PSFapproxTest}).

\subsection{Fluctuation amplitude}
The fluctuation amplitude is calculated using (\ref{flucAmpDef}) and is given in the first row of Table~\ref{tab:GS2_table}. As expected from Section~\ref{sec:modelflucamp}, the PSFs reduce the observed fluctuation amplitude. Correcting for the PSF effects using (\ref{PSFflucamp}) works well in the analysis of Case 4. In contrast, the correction for Case-5 PSFs overestimates the fluctuation amplitude, which is consistent with (\ref{PSFflucamp}) overestimating the reduction in the fluctuation amplitude when using Gaussian-model PSFs for this case, compared to the numerical evaluation using the real PSFs, as seen in Figure~\ref{fig:blobby_plots}(a).

% ----------------------------------------------------------------------------------------------------------
% ----------------------------------------------------------------------------------------------------------
% ----------------------------------------------------------------------------------------------------------

% ----------------------------------------------------------------------------------------------------------
% ----------------------------------------------------------------------------------------------------------
% ----------------------------------------------------------------------------------------------------------

\section{Temporal correlation parameters: further refinements}\label{sec:gs2temporal}
\begin{figure}
\centering
\includegraphics[width=0.9\textwidth]{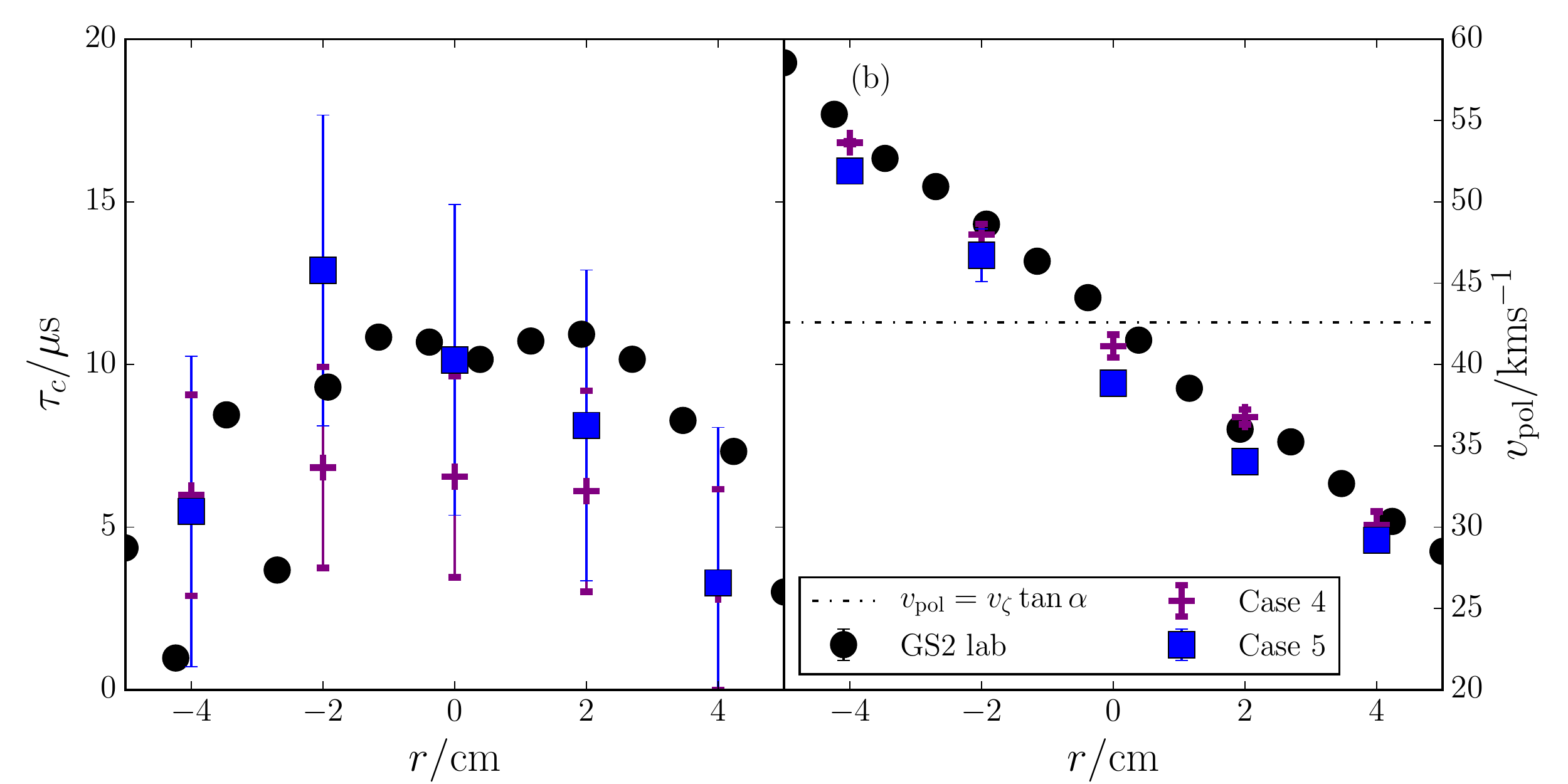}
\caption[PSF effects on the correlation time measured from gyrokinetic simulations]{The effect of PSFs on the correlation time measured from the GS2 simulation of MAST shot \#27268: (a) Correlation time calculated using the CCTD technique (Section~\ref{sec:meas_corr_time}) in the laboratory-frame (black circles) and from synthetic-BES data with the Case 4 (purple crosses) and the Case 5 (blue squares) PSFs, against radial position relative to the centre of the GS2 simulation domain. The errors on the PSF data are calculated from (\ref{errontau}). (b) Apparent poloidal velocity measured from the same simulation. The nominal toroidal rotation projected onto the poloidal plane $\vpol=\vz\tan\alpha=42.6\mathrm{km/s}$ is shown by the horizontal dot-dashed line. \label{fig:GS2velocities} }
\end{figure}

In Table~\ref{tab:GS2_table}, the mean correlation times measured with and without the PSFs of Case 4 are significantly different from each other. This appears to contradict the calculation of Section~\ref{sec:PSF_an_corr_time}, where the PSFs were shown not to affect the measurement of the correlation time. 

The measurements of the correlation time and apparent poloidal velocity are made at each radially separated array of poloidal channels in the 2D BES array, and similarly at each radially distinct set of poloidal grid points in the GS2 numerical domain. Therefore, it is possible to plot radial profiles of $\tauca$ and $v_{\mathrm{pol}|B}$ resulting from these measurements, which is done in Figure~\ref{fig:GS2velocities}. In Figure~\ref{fig:GS2velocities}(a), we see that the GS2 laboratory-frame correlation time decreases when $|r| \gtrsim 2\ \mathrm{cm}$. This has been identified as a numerical effect due to aliasing caused by the algorithm used to implement flow shear in GS2. Therefore, we will not consider data points outside of this radial range in the following discussion.

As we see in Figure~\ref{fig:GS2velocities}(a), the correlation times calculated with the Case-5 PSFs are similar to the laboratory-frame correlation times, which is possibly because these PSFs have a narrower shape, closer to a delta function, as discussed in Section~\ref{sec:gauss_fluc_amp}. However, the correlation times calculated using the Case-4 PSFs have values that are approximately half of the true laboratory-frame correlation times. Additionally, in Figure~\ref{fig:GS2velocities}(b), we see that both Case-4 and Case-5 PSFs produce apparent poloidal velocities that have a small, but noticeable, difference from the laboratory-frame apparent poloidal velocities. In this section, we attempt to explain these differences.

\subsection{Physical interpretation of the correlation time}\label{sec:physical_corr_time}
The correlation time in the plasma frame is defined in (\ref{plasma_frame_correlation_function}) to be $\taucp$, which can be `measured' by calculating the time-delay auto-correlation function of the density field (by replacing $\delta I$ in (\ref{correlation_def}) with $\delta n$)  at a fixed spatial point in the plasma frame, and fitting this function by (\ref{plasma_frame_correlation_function}) with $\Delta x=\Delta y=\Delta z=0$. The value of the correlation time $\taucp$ is determined by two effects: (1) the Lagrangian decay of a moving perturbation with time, and (2) the Eulerian decorrelation of a perturbation as it is advected past the measurement location by the turbulent velocity field. These two effects are related, as the turbulent velocity field is determined from the electrostatic potential related to the perturbed density field, and, therefore, the velocity field will decorrelate at a similar rate to the density perturbations. Therefore, the correlation time $\taucp$ can be defined to be the fundamental decorrelation timescale associated with the turbulence.

In our model of fluctuating fields (see Section~\ref{sec:toymodel}), which has a correlation function that is exactly the plasma-frame correlation function (\ref{plasma_frame_correlation_function}), we did not give the individual perturbations (\ref{3dfun}) any small random velocities, and, therefore, it was assumed that there was no Eulerian contribution to the correlation time. In the following section we describe how we refine our model to include such velocities, and discuss the consequences of doing so for the functional form of the correlation function.

\subsection{Introducing a fluctuating radial velocity into the model of fluctuating fields.}\label{sec:fluc_rad_velocity}
In \ref{sec:basic_equations}, we describe how to include the effect of a fluctuating radial velocity $\vr$ on the motion of our model perturbations; in \ref{sec:plasma_frame_corr}, the plasma-frame correlation function (\ref{plasma_frame_correlation_function_with_dvr}) is calculated assuming that this radial velocity is a Gaussian-distributed random variable with zero mean and standard deviation $\dvr$. We do not include a fluctuating binormal velocity, because, for typical binormal correlation lengths, the effect of PSFs on the binormal correlation length has been shown to be small (Section~\ref{sec:real_poloidalcorrlength}).

In \ref{sec:plasma_frame_corr}, the plasma-frame correlation time of our model is found to be
\begin{equation}\label{plasma_frame_corr_time_vr}
\frac{1}{\taucp^2} = \frac{1}{\taucpl^2} + \frac{1}{\tautp^2},
\end{equation}
where $\taucpl$ is the Lagrangian lifetime of a perturbation defined by (\ref{tau_life}) and 
\begin{equation}\label{turnovertime}
\tautp = \frac{\ellxp}{\dvr}\frac{\sqrt{2}}{\sqrt{2 + \kxp^2\ellxp^2}},
\end{equation}
is the eddy turn-over-time. In order to calculate $\taucp$, we assume, in addition to the ordering (\ref{fullordering}-\ref{endfullordering}), that the fluctuating radial velocity is small, $\dvr/\vz=\mathcal{O}(\epsilon^1 = \rho_i/a)$, compared to the toroidal velocity. This assumption is reasonable, as the fluctuating radial velocity in the gyrokinetic simulation used in Section~\ref{sec:testing} is found to be $2\%$ of the toroidal velocity.

In \ref{sec:toy_CCTD}, the laboratory-frame correlation time $\tauct$ is calculated and found to be the same as the plasma-frame correlation time $\taucp$ in the low-Mach number limit (Section~\ref{sec:lowmachnumber}), just as we found before in (\ref{lab_corr_time}). The appearance of both the lifetime and the eddy-turn-over time in (\ref{plasma_frame_corr_time_vr}) means that the measured correlation time in the laboratory frame $\tauct$ will be dominated by the smaller of $\taucpl$ or $\tautp$. We demonstrate this effect in Figure~\ref{fig:blobby_plots}(e), where the laboratory-frame correlation time $\tauct$ is plotted against the lifetime $\taucpl$. We see that, as the fluctuating radial velocity increases relative to the toroidal velocity, the correlation time $\tauct$ becomes less sensitive to changes in the lifetime.

In \ref{sec:PSFcorrtime}, the PSF effects on the temporal parameters are calculated from the laboratory-frame correlation function (\ref{lab_corr_function_full}), whilst retaining the radial-velocity effects. The resulting BES-measured correlation time is then (\ref{corrtimewithPSFs})
\begin{equation}\label{simpcorrtimewithPSFs}
\frac{1}{\tauca^2} = \frac{1}{\taucpl^2}  + \frac{1}{\tauta^2},
\end{equation}
where the BES eddy-turn-over time is
\begin{equation}\label{tautadef}
\tauta \equiv  
\left[(2+  \krt^2 \ellrt^2)\frac{\dvr^2}{2 \ellra^2} + \left(\frac{\kzt^2 \ellzt^2}{2 \ellra^2} - \frac{\kzt^2 \ellzt^4}{2 D^4}  \right) \dvr^2 \right]^{-1/2},
\end{equation}
where $D$ is given by (\ref{Dterm}), and we have written (\ref{tautadef}) in terms of both laboratory-frame and BES-measured correlation parameters, in order to be concise (the relationships between these two sets of parameters are given by (\ref{an_ellrt_inv}-\ref{kzt_inv})). 

The reason why the BES eddy-turn-over time (\ref{tautadef}) is different from the plasma- and laboratory-frame eddy-turn-over times (\ref{turnovertime}) is because, by introducing the fluctuating radial velocity into our model of fluctuating fields, the plasma-frame correlation function (\ref{plasma_frame_correlation_function_with_dvr}) of the individual perturbations is no longer equal to (\ref{plasma_frame_correlation_function}). The difference between these two correlation functions is that in (\ref{plasma_frame_correlation_function_with_dvr}) cross terms (e.g, $\Delta x \dt$) between spatial and temporal relative coordinates are introduced by the fluctuating radial velocity. Then, when the PSF integral (\ref{PSF_def}) is calculated in (\ref{Cbes2}), mixing occurs between the spatial and temporal correlation parameters. 

As the fluctuating component of the radial velocity is not measured experimentally, quantifying how large the difference between the laboratory-frame and BES-measured correlation times is difficult. However, we may estimate the maximum size of the PSF effects if we assume that the plasma-frame correlation time is dominated by the eddy-turn-over time. Then the fractional difference between the laboratory-frame (\ref{plasma_frame_corr_time_vr}) and BES-measured (\ref{simpcorrtimewithPSFs}) correlation times is independent of $\dvr$, and given by
\begin{equation}
\tauerr \equiv \frac{|\tauca-\tauct|}{\tauca}
\simeq \Bigg|1-\frac{\tautt}{\tauta}\Bigg| \simeq \Bigg|1 - \frac{\ellrt}{\ellra}\Bigg|, \label{errontau}
\end{equation}
where the final expression is reached by assuming that the poloidal correlation length $\ellzt\gg \ellrt, L_1, L_2$, which is reasonable based on experimental measurements (see Table~\ref{tab:example_data} and \cite{Ghim2013}). The error calculated using (\ref{errontau}) gives an order-of-magnitude estimate of the error associated with the PSF effects (we discuss the sign of the error in Section~\ref{sec:real_Gauss_corr_time}). The errors on the correlation times of the synthetic-BES data plotted in Figure~\ref{fig:GS2velocities}(a), have been calculated using (\ref{errontau}), which shows that these are the correct size to account for the differences between the synthetic-BES and the laboratory-frame correlation times. This error estimate can be considered to be independent of our model, as the result (\ref{errontau}) can also be reached by simply arguing that the typical size of PSF effects can be given by the ratio $\ellrt/\ellra$.

\subsection{Limitations of our model}\label{sec:real_Gauss_corr_time}
\begin{figure}
\centering
\includegraphics[width=0.6\textwidth]{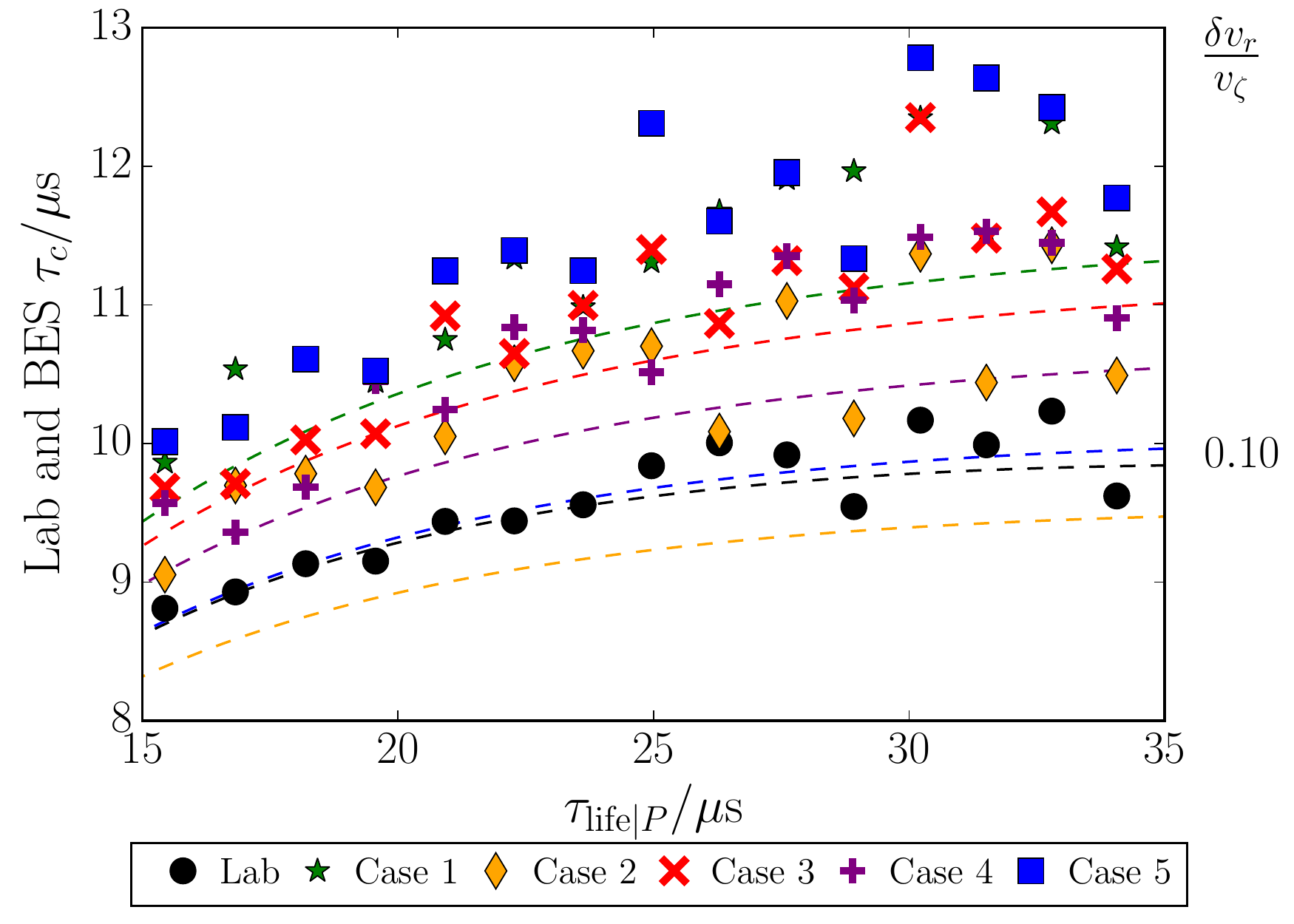}
\caption[Enlargement of Figure~\ref{fig:blobby_plots}(e)]{Enlargement of Figure~\ref{fig:blobby_plots}(e) around the  $\dvr/\vz = 10\%$ curve, to illustrate the difference between the Gaussian-model PSF calculation (\ref{simpcorrtimewithPSFs}) of the effects of PSFs on the correlation-time measurement (coloured dashed lines) and the effect of the real PSFs (coloured markers). The length of the simulation was increased from $T = 2\ \mathrm{ms}$ to $T = 20\ \mathrm{ms}$ to calculate these data points in order to reduce the statistical noise in the measurement.  \label{fig:zoomtau} }
\end{figure}

Unfortunately, the laboratory-frame correlation time cannot be reconstructed from the synthetic-BES correlation time measured from the gyrokinetic simulations using (\ref{tautadef}), even if the value of the fluctuating radial velocity $\dvr$ is known. There are two reasons why this is the case. The first is our assumption of Gaussian-model PSFs, and the second is our assumption that the fluctuating velocity field follows a Gaussian distribution.

We demonstrate the discrepancy that arises due to using Gaussian-model PSFs by applying the real PSFs to the density field generated by our model of fluctuating fields, just as we did in Section~\ref{sec:effects_real_PSFs}. From Figure~\ref{fig:zoomtau}, which is an enlargement of Figure~\ref{fig:blobby_plots}(e), we see that the correlation times calculated with the real PSFs and with the Gaussian-model PSFs (\ref{simpcorrtimewithPSFs}) do not agree (most notably for Case 5). The difference between the Gaussian-model and the real PSFs plays a more important role in the correlation-time measurement than in the measurements of spatial correlations in Section~\ref{sec:testing} because the change in amplitude of the correlation function required to measure the correlation time is a second-order effect (in $\epsilon=\rho_i/a$, see Section~\ref{sec:asymtotics}). 

Furthermore, the modelling assumption of a Gaussian-distributed $\vr$ that we have used is too limiting. This can be seen by comparing the change in the correlation time caused by the real PSFs for Case 4 in Figure~\ref{fig:zoomtau}, where the assumption of a Gaussian-distributed $\vr$ has been used, to the change in the correlation time for the same real PSFs on the GS2 density field in Figure~\ref{fig:GS2velocities}(a). In Figure~\ref{fig:zoomtau}, the synthetic-BES correlation time is longer than the laboratory-frame correlation time, whereas in Figure~\ref{fig:GS2velocities}(a), the synthetic-BES correlation time is shorter than the laboratory-frame correlation time. As the exact same set of real PSFs has been used to generate both sets of synthetic data, the difference in behaviour between the correlation times of the two sets of synthetic data must be due to the different models used to generate the fluctuating fields. This is not surprising, given that our assumption of a Gaussian-distributed $\vr$ with no interactions between the individual perturbations is a gross simplification of a truly turbulent velocity field.

\begin{figure}
	\centering
	\includegraphics[width=0.6\textwidth]{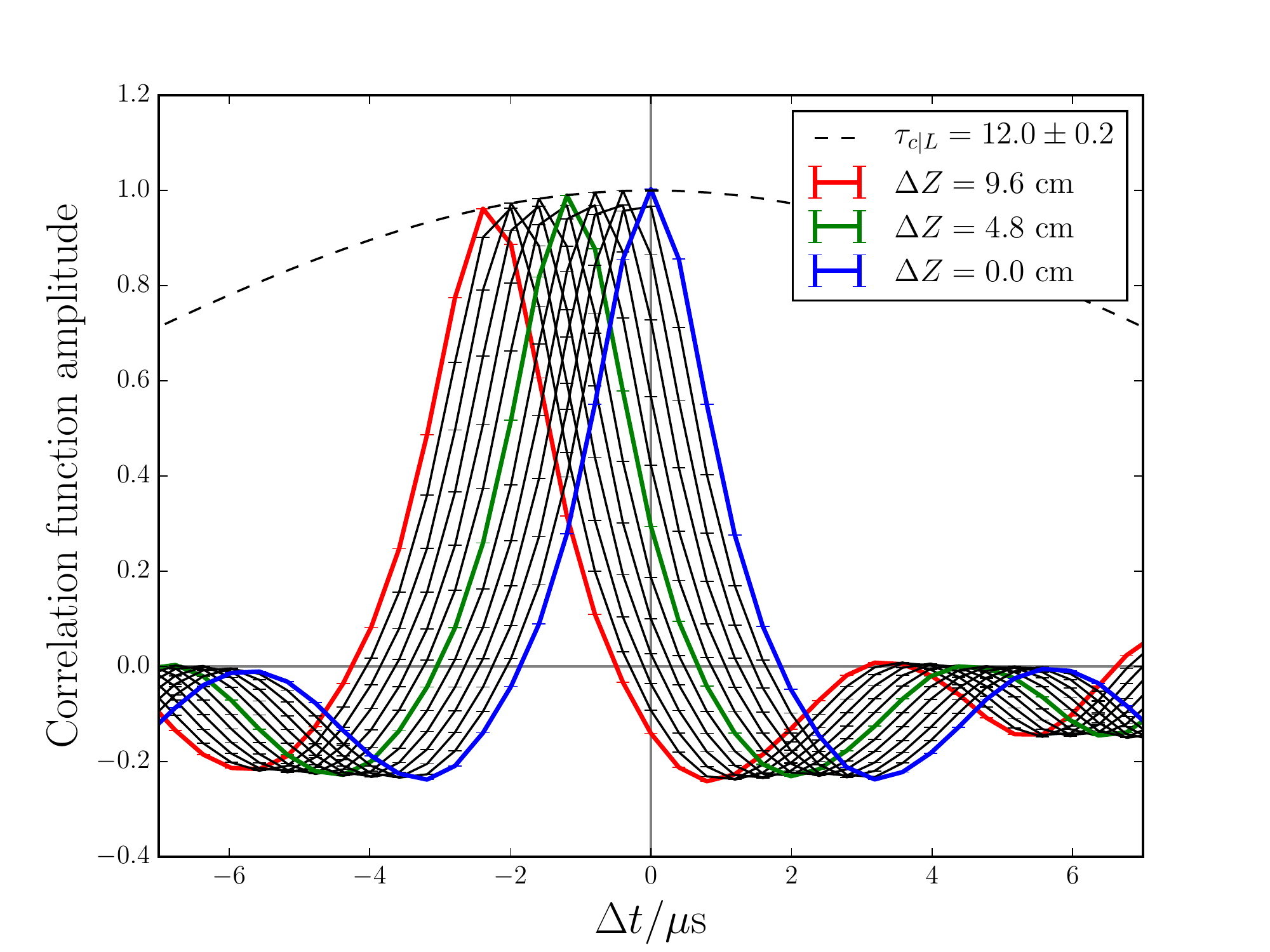}
	\caption[Time-delayed correlation functions from GS2]{Time-delayed correlation functions from the gyrokinetic simulation introduced in Section~\ref{sec:gksim}, at $r = 0.35\ \mathrm{cm}$ from the centre of the simulation domain, for a range of poloidal displacements $\dz$. The range of $\dz$ used to generate this plot has been increased compared to that used to calculate the data in Table~\ref{tab:GS2_table}, in order to emphasise the quality of the Gaussian fit (\ref{correlation_timedelay_fit}) to the peaks of the curves for each $\dz$. This should be compared to the time-delay correlation functions of the experimentally measured intensity field from the MAST BES data in Figure~\ref{fig:example_time_delay}.  \label{fig:gs2_timedelay} }
\end{figure}
Finally, we note that one of the key assumptions in this paper is that the functional form for the time decay of the correlation function is a Gaussian; see (\ref{plasma_frame_correlation_function}). In previous work~\cite{Ghim2013}, an exponential decaying function, $f(t) = \exp(-|t|/\tauca)$, was used to measure the correlation time $\tauca$ from the MAST BES signals. Indeed, this might in fact be a better fit to the experimental correlation function in Figure~\ref{fig:example_time_delay}. Therefore, it could be argued that the failure of our procedure to account for PSF effects on the correlation-time measurement was due to the failure of our underlying assumption on the structure of the time-delay correlation function. However, the peaks of the time-delay correlation function from the gyrokinetic simulations (shown in Figure~\ref{fig:gs2_timedelay}) that were used in this section to identify these issues are, in fact, well fit by the Gaussian function (\ref{correlation_timedelay_fit}).

\subsection{Apparent poloidal velocity.}\label{sec:apparent_pol_vel}
We see in Figure~\ref{fig:GS2velocities}(b) that the apparent poloidal velocity varies in the range of $30\--60\ \mathrm{km/s}$ across the radial extent of the analysis region. This is due to the flow shear in the simulated density field adding to the bulk toroidal rotation. The apparent poloidal velocities measured from the synthetic-BES data for the two PSF cases are also plotted in Figure~\ref{fig:GS2velocities}(b) and show reasonably good agreement with the laboratory-frame apparent poloidal velocity, deviating, at most, by $5\ \mathrm{km/s}$. Evidently, the effect of the PSFs on this measurement is smaller than on the correlation-time measurement, which is because the toroidal velocity is significantly greater than all other velocities in the GS2 simulation. 

Nevertheless, the real PSFs do have a small systematic effect on the apparent-poloidal-velocity measurement, which is not included in our analytic expression (\ref{vpolPSF}) using Gaussian-model PSFs, even when one accounts for a fluctuating radial velocity field (see~\ref{sec:PSFcorrtime}).  Therefore, we see, again, that our model is not able to account fully for the effects of PSFs on the measurement of the temporal properties of the turbulence. We can, however, be reassured that measurements of apparent poloidal velocities can be accurate to within $10\%$ of the laboratory-frame value, without having to be corrected for PSF effects.

% ----------------------------------------------------------------------------------------------------------
% ----------------------------------------------------------------------------------------------------------
% ----------------------------------------------------------------------------------------------------------

\section{Conclusions}\label{sec:conclusion}
In the Introduction to this work, in Section~\ref{sec:introduction}, we posed two questions. The first of these was what effect PSFs have on the measurement of the correlation parameters of plasma turbulence (fluctuation amplitude, $\dII$, radial, $\ellra$, and poloidal, $\ellza$, correlation lengths and wavenumbers, $\kra, \ \kza$, correlation time, $\tauca$, and apparent poloidal velocity, $v_{\mathrm{pol}|B}$) using BES systems. The culmination of the analysis to answer this question occurs in Section~\ref{sec:numeric}, where it was found that:
\begin{enumerate}
	\item The measured fluctuation amplitude (Section~\ref{sec:meas_fluctuationamp}) of the intensity field differs from the fluctuation amplitude of the density field by a linear factor that is dependent on the area of the PSFs.
	\item The measured radial correlation length (Section~\ref{sec:ztd_measurement}) of the intensity field can be significantly longer than the true radial correlation length of the density field, because the latter was often similar to the size of the PSFs.
	\item As the poloidal correlation lengths are measured to be longer than the PSF length the effects of the PSFs on them is less significant than on the radial correlation lengths.
	\item The tilt angle of the correlation function changes significantly due to PSF effects, and can even change sign.
	\item The CCTD method for measuring the correlation time and apparent poloidal velocity (Section~\ref{sec:meas_corr_time}) can also be affected by PSF effects, as has been shown in Section~\ref{sec:gs2temporal}.
\end{enumerate}
In the process of investigating how to quantify these effects, we have developed 
\begin{enumerate}
	\item a new method for measuring the poloidal correlation length using information from a fit to the time-delayed auto-correlation function, which is subsequently used to help constrain the fit to the zero-time-delay spatial correlation function (see Section~\ref{sec:ztd_measurement}); and
	\item a model of fluctuating fields (Section~\ref{sec:toymodel} and \ref{sec:toymodel_appendix}) that uses randomly distributed perturbations to generate time series that real PSFs can be applied to by numerically evaluating the integral (\ref{PSF_def}); this model of fluctuating fields could be used to test instrument function effects on fluctuating fields of various kinds, including density, temperature, potential, etc. on other devices and aid the development of new turbulence diagnostics.
\end{enumerate}

The second question posed in the Introduction was whether it was possible to correct for the PSF effects identified above. Below we enumerate the findings of this work in response to this question. 
\begin{enumerate}
	\item A simple, analytic method has been developed, designed to reconstruct the fluctuation amplitude (\ref{PSFflucamp}) and the spatial correlation parameters (\ref{an_ellrt_inv}-\ref{an_tantheta_inv}) of the density field from measurements of the correlation parameters of the intensity field and the principal-component lengths and inclination angle of the PSFs; assuming a Gaussian model shape for the PSFs (Section~\ref{sec:GaussianPSFs}). This is described in Section~\ref{sec:correctingPSFs} and has been tested successfully on gyrokinetic simulations of turbulence in Section~\ref{sec:testing}. 
	\item Our reconstruction method also provides a definition of the spatial resolution limits (\ref{ellra_boundary}) and (\ref{ellza_boundary}), of a BES system in terms of the principal components of the PSFs; see Section~\ref{sec:resolutionlimit}. In addition to post-hoc testing of the validity of measurements, these equations can also be used to optimise the design of new BES systems.
	\item In Section~\ref{sec:PSF_an_corr_time}, we found that the CCTD method measures exactly the plasma-frame correlation time of the turbulence and is unaffected by PSF effects, under the assumptions that the Lagrangian lifetime of individual perturbations is of order $a/\rho_i$ longer than the time taken for a perturbation to be advected past the detector array by the bulk toroidal velocity, that the Mach number associated with the toroidal rotation is small, and that any fluctuating radial velocity is smaller than the bulk toroidal velocity by at least $\rho_i^2/a^2$.
	\item The fluctuating radial velocity in gyrokinetic simulations of plasma turbulence in MAST is only $\rho_i/a$ smaller than the bulk toroidal velocity. In Section~\ref{sec:fluc_rad_velocity}, we showed that, in this case, the CCTD method measured a correlation time that was a combination of both Lagrangian and Eulerian times. The (Eulerian) eddy-turn-over time, in addition to  depending on the fluctuating radial velocity, depends on the spatial scales of the turbulence and, therefore, provides a means through which PSFs can affect the measurement of the correlation time.
	\item Due to the limitations of our model of fluctuating fields and our Gaussian-model PSFs, which are discussed in Section~\ref{sec:real_Gauss_corr_time}, it is not possible to reconstruct the laboratory-frame correlation time precisely. Nevertheless, in Section~\ref{sec:fluc_rad_velocity}, we have provided an estimate of the error on the measured correlation time that is associated with PSF effects.
	\item The poloidal-velocity measurement is dominated by the bulk toroidal velocity. In Section~\ref{sec:apparent_pol_vel}, we saw that while it was affected by the real PSFs, these effects were, at least, $\rho_i/a$ smaller than the bulk toroidal velocity. Therefore, within an accuracy of $10\%$, no corrections are required for measurements of the toroidal velocity.
\end{enumerate}

To summarise the above, the main results of this investigation can be separated into two parts: the first is the elucidation of the assumptions behind the measurement methods used to extract statistical quantities from time series of turbulence data (most notably the correlation-time measurement), while the second is the development of a method for the reconstruction of the fluctuation amplitude and spatial correlation parameters of a turbulent density field from BES measurements. Thus, given a two-point, two-time correlation function of the BES-measured intensity field, we provide an algorithm for extracting the corresponding correlation function for the true density field. This method can now be applied to turbulence measurements, as has been done, for example, in Figure~\ref{fig:example_data}(b), where the corrected spatial correlation function is plotted for real BES measurements from MAST shot \#28155 and shows that both the tilt angle of the correlation function and the radial correlation length undergo substantial correction from the raw measured values. The first use of this methodology to extract new physics from BES measurements on MAST is \cite{Fox2016a}.

We finish by noting that these results are not restricted in applicability just to MAST or just to spherical tokamaks. For example, the PSFs for the BES system on the conventional-aspect-ratio device DIII-D~\cite{Luxon2002} can be of a similar size to the measured spatial correlation functions~\cite{Shafer2006,Shafer2012}, just as is the case for MAST turbulence and PSFs discussed here. Therefore, as discussed in~\cite{Shafer2012}, accounting for PSFs is also important for such devices. Indeed, using our method with the DIII-D BES data would probably produce better reconstructions of the correlation parameters of the turbulent density field than for the MAST BES, because the spatial variation of the PSFs across the BES array is smaller in DIII-D, due to the smaller variation in the pitch angle of the magnetic field, and, therefore, the assumption that all the PSFs are the same is better satisfied.

\ack
We would like to thank Daniel Dunai, Jon Hillesheim, Tim Horbury, Gerg\H{o} Pokol and Anne White for many useful discussions. This work has been carried out within the framework of the EUROfusion Consortium and has received funding from the Euratom research and training programme 2014-2018 under grant agreement No 633053 and from  the RCUK Energy Programme [grant number EP/I501045]. The views and opinions expressed herein do not necessarily reflect those of the European Commission. Financial support was also provided to MFJF by Merton College, Oxford. YcG was supported by the National R\&D Program through the National Research Foundation of Korea (NRF) funded by the Ministry of Science, ICT and Future Planning (Grant No.\ 2014M1A7A1A01029835) and by the KUSTAR-KAIST Institute, KAIST, Korea. The work of AAS was supported in part by grants from UK STFC and EPSRC.

\appendix
% ------------------------------------------------------------------------------------------------------------------------------------------------------
% ------------------------------------------------------------------------------------------------------------------------------------------------------
% ------------------------------------------------------------------------------------------------------------------------------------------------------

% ------------------------------------------------------------------------------------------------------------------------------------------------------
% ------------------- APPENDIX A: POLOIDAL CORRELATION LENGTHS FITTING       ----------------------------------------
% ------------------------------------------------------------------------------------------------------------------------------------------------------

% ------------------------------------------------------------------------------------------------------------------------------------------
% ------------------------------------------------------------------------------------------------------------------------------------------
% ------------------------------------------------------------------------------------------------------------------------------------------

\section{Measuring the product of the poloidal wavenumber and poloidal correlation length}\label{sec:poloidal_corr_lengths}

\begin{figure}[h]
	\centering
	\includegraphics[width=0.85\textwidth]{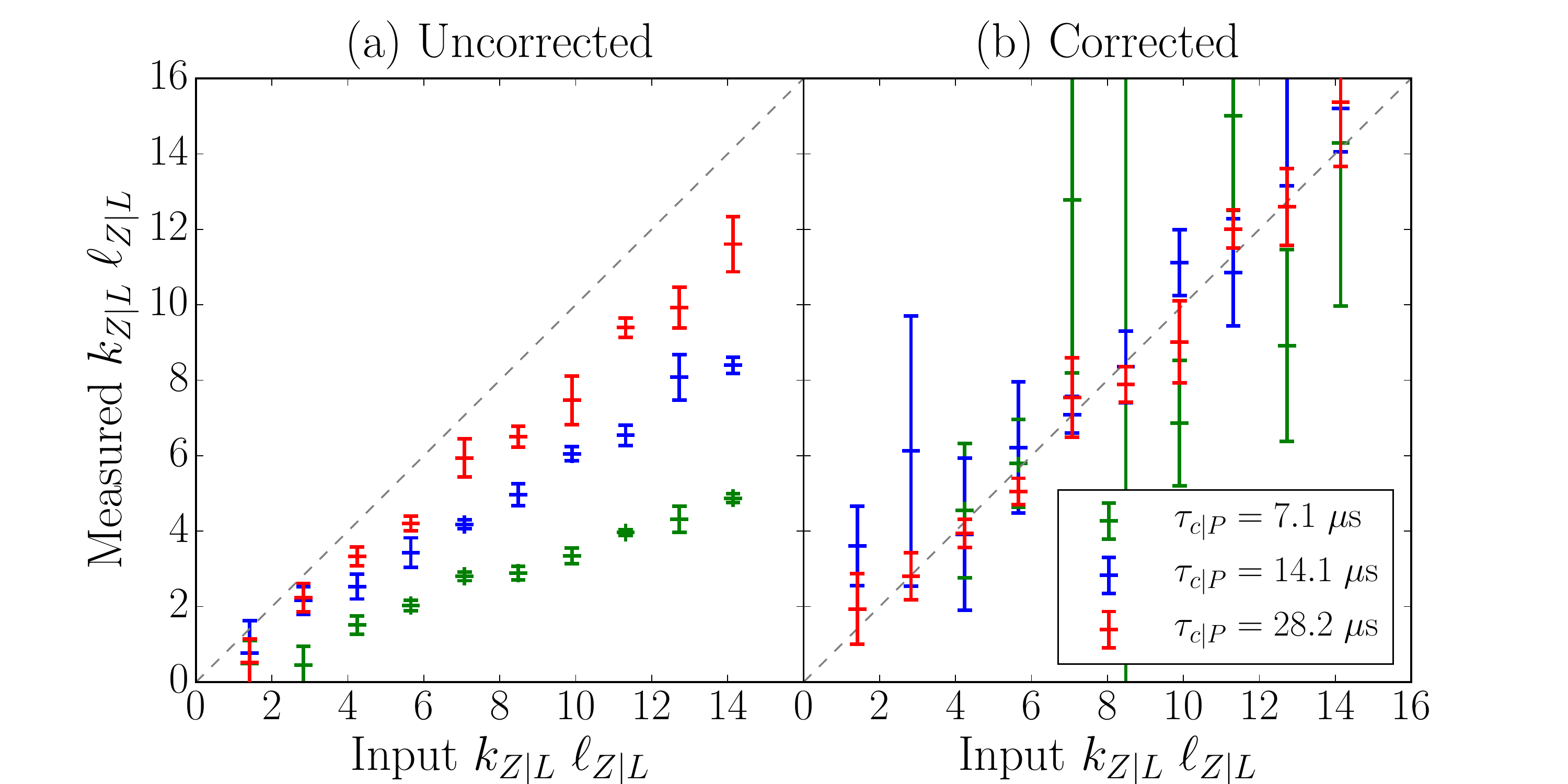}
	\caption[Testing the method of reconstructing $\kzt\ellzt$]{A scan in $\kzt\ellzt$ has been performed using our model of fluctuating fields (see Section~\ref{sec:toymodel}) by varying $\kzt$ and keeping constant $\ellzt = 16.3\ \mathrm{cm}$, for three different values of correlation time $\tauct \in \{ 7.1, 14.1, 28.2 \}\ \mu\mathrm{s}$. In both panels (a) and (b), the measurement of $\kzt\ellzt$ is made by fitting (\ref{eff_time_delay_shape}) to the auto-correlation function (\ref{correlation_def}) of the numerically generated fluctuating density field. In (b), the fit is performed after correcting the auto-correlation function for its temporal decay $\tauct$, by multiplying it by $\exp(\dt^2/\tauct^2)$. \label{fig:kyly} }
\end{figure}
The temporal auto-correlation function in the laboratory frame (\ref{lab_temporal_corr_func}) contains both spatial and temporal correlation parameters. In this Appendix, we demonstrate that it is possible to extract the product of the poloidal wavenumber $\kzt$ and poloidal correlation length $\ellzt$ from the temporal auto-correlation function (\ref{lab_temporal_corr_func}), provided that the correlation time $\tauct$ has been measured (using, for example, the CCTD method --- see Section~\ref{sec:meas_corr_time}). 

The product $\kzt\ellzt$ plotted in Figure~\ref{fig:kyly}(a) is measured by fitting (\ref{eff_time_delay_shape}) to the temporal auto-correlation function calculated using (\ref{correlation_def}), with the intensity field $\delta I$ replaced by a density field $\delta n$ generated numerically using our model of fluctuating fields (see Section~\ref{sec:toymodel} and \ref{sec:toymodel_appendix}).  Figure~\ref{fig:kyly}(a) shows that a finite value of the correlation time causes the measured product $\kzt\ellzt$ to be an underestimate of the true $\kzt\ellzt$ (input into the model).

The degree to which $\kzt\ellzt$ is underestimated increases as the correlation time becomes shorter. This is because the contribution of the correlation time $\tauct$ to the auto-correlation time (\ref{tauauto}) becomes large compared to the ratio of the poloidal correlation length to the toroidal velocity, $\ellzt/\vz$, which is held constant for this analysis. The comparison between $\tauct$ and $\ellzt/\vz$ can be understood by inspecting (\ref{tauauto}), using (\ref{lab_pol_length}) to relate the plasma-frame correlation parameters to the laboratory-frame correlation parameters, setting the poloidal velocity $\vZ=0$, and assuming that the parallel correlation length $\ellzp\gg\ellzt$. 

In Figure~\ref{fig:kyly}(b), we show that the underestimate of $\kzt\ellzt$ can be corrected for, as discussed in Section~\ref{sec:ztd_measurement}, by multiplying the measured auto-correlation function by $\exp(\dt^2/\tauct^2)$, before fitting (\ref{eff_time_delay_shape}). The quality of the reconstructed value of $\kzt\ellzt$ clearly decreases as the correlation time becomes shorter. This is because any small errors in the measurement of either the temporal auto-correlation function or the correlation time will be increased by using the correction factor $\exp(\dt^2/\tauct^2)$, which grows faster in $\dt$ for smaller $\tauct$.

We have not considered PSF effects in this Appendix, because we could perform the exact same analysis to extract the product $\kza\ellza$ of the intensity field, provided we have a measurement of the correlation time of the intensity field $\tauca$. This is possible because the technique does not depend on how the correlation function decays in time (see discussion in Section~\ref{sec:fluc_rad_velocity}), but rather just that we are able to correct for that effect (i.e., all the information about the temporal decay is contained in $\tauca$).

Finally, we estimate whether this procedure is able to reconstruct accurately the product $\kza\ellza$ from experimental data obtained using the MAST BES. As discussed above, and from (\ref{tauauto}), we require that
\begin{equation}\label{elly_inequality}
\tauct \gg \frac{\ellzt}{\vz},
\end{equation}
in order for us to have confidence in the resulting value of $\kzt\ellzt$.  We can translate (\ref{elly_inequality}) into BES-relevant quantities by changing $\tauct\rightarrow\tauca$ and $\ellzt\rightarrow\ellza$. In MAST, $\tauca \simeq 10\ \mu\mathrm{s}$, $\ellza \lesssim 10\ \mathrm{cm}$, and $\vz \gtrsim 2\ \mathrm{cm}/\mu\mathrm{s}$. These values satisfy (\ref{elly_inequality}) and, therefore, we should be able to use this procedure to measure $\kza\ellza$ with reasonable accuracy for real MAST BES data.

% ------------------------------------------------------------------------------------------------------------------------------------------------------
% ------------------------------------------------------------------------------------------------------------------------------------------------------
% ------------------------------------------------------------------------------------------------------------------------------------------------------

% ------------------------------------------------------------------------------------------------------------------------------------------------------
% ------------------------- APPENDIX B: TOY MODEL OF TURBULENCE ------------------------------------------------------------
% ------------------------------------------------------------------------------------------------------------------------------------------------------

% ------------------------------------------------------------------------------------------------------------------------------------------------------
% ------------------------------------------------------------------------------------------------------------------------------------------------------
% ------------------------------------------------------------------------------------------------------------------------------------------------------

\section{Modelling a fluctuating field}\label{sec:toymodel_appendix}
In this Appendix, we provide the detailed description of the model of fluctuating fields introduced in Section~\ref{sec:toymodel}, extended in Section~\ref{sec:fluc_rad_velocity} and used by us to test our data reconstruction methodology. First, in \ref{sec:basic_equations}, we describe the equations used to generate the two-dimensional fluctuating field. In \ref{sec:plasma_frame_corr}, the plasma-frame correlation function of this fluctuating field is calculated, followed by the laboratory-frame correlation function in \ref{sec:lab_frame_corr}. In \ref{sec:toyflucamp}, the fluctuation amplitude of this model is discussed. Then, in \ref{sec:toymodel_algorithm}, details of the algorithm for generating the described fluctuating field are presented.

\subsection{Basic equations}\label{sec:basic_equations}
We generate a fluctuating field in the radial-poloidal plane as a function of time:
\begin{eqnarray}\label{density_field_output}
\frac{\delta n}{n}(r, Z, t) = \sum_{i=1}^N s_i(r,Z,\zeta=0,t, r_{i0}, Z_{i0}, \zeta_{i0}, t_{i0}),
\end{eqnarray}
where $N$ is the number of `individual perturbations' $s_i$, which are functions of the laboratory-frame coordinates: the radial $r$, poloidal $Z$, and toroidal $\zeta$ positions and the time $t$, as well as the initial position of the perturbation $(r_{i0}, Z_{i0}, \zeta_{i0}, t_{i0})$ in space and time. From here on, we suppress the dependence of $s_i$ on the initial position. The toroidal coordinate of the radial-poloidal plane at which the fluctuating field is generated is specified to be $\zeta=0$. 

The structure of each individual perturbation is specified in the plasma frame. The plasma frame is defined in terms of field-aligned coordinates ($x$ radial, $z$ parallel to the magnetic field, $y$ binormal). In this Appendix, we define the transformations that take us from the plasma frame into the laboratory frame, so that an individual perturbation
 \begin{eqnarray}\label{3dfun}
s_i(x, y, z, t) = A_i &\exp&\left\{-\frac{[\Delta x_i(t)]^2}{\lambda_x^2}-\frac{[\Delta y_i(t)]^2}{\lambda_y^2}-\frac{[\Delta z_i(t)]^2}{\lamz^2}-\frac{[\Delta t_i(t)]^2}{\tau_{c}^2}\right\} \nonumber \\
 &\quad& \times \cos[k_{x}(t) \Delta x_i(t) + k_{y} \Delta y_i(t) + \phi_i],
\end{eqnarray}
can be written in terms of the laboratory-frame coordinates, and, therefore, making it possible to evaluate (\ref{density_field_output}). In (\ref{3dfun}), $A_i$ is the amplitude of the perturbation taken from a Gaussian random distribution with zero mean and standard deviation $\sigma_\mathrm{amp}$, $\lambda_x$ is the characteristic radial length, $\lambda_y$ is the characteristic binormal length, $\lamz$ is the characteristic parallel length, $\tau_{c}$ is the characteristic lifetime, $k_{x}(t)$ is the radial wavenumber, $k_y$ is the binormal wavenumber, $\phi_i$ is a random phase uniformly distributed in the range $[0,2\pi]$. Note that the term `characteristic' is used  because, as we will find, these differ non-trivially from the correlation lengths and times that are measured in the laboratory frame. 

If the position of the centre (maximum) of the $i$th perturbation is ($x_i(t), y_i(t), z_i(t)$), then, in (\ref{3dfun}),
\begin{equation}\label{field_aligned_coords}
\Delta x_i(t) = x - x_i(t), \quad 
\Delta y_i(t) = y - y_i(t), \quad
\Delta z_i(t) = z - z_i(t).
\end{equation}
The relative plasma-frame coordinates (\ref{field_aligned_coords}) are related to the laboratory-frame coordinates $(r,Z,\zeta)$ via
\begin{eqnarray}\label{relation_plasma_lab_coords}
\Delta x_i(t) &=& \Delta r_i(t), \nonumber \\ 
\Delta y_i(t) &=& \Delta \zeta_i(t) \sin\alpha + \Delta Z_i(t) \cos\alpha, \nonumber \\
\Delta z_i(t) &=& \Delta \zeta_i(t) \cos\alpha - \Delta Z_i(t) \sin\alpha,
\end{eqnarray}
where $\alpha$ is the field-line pitch angle and
\begin{equation}
\Delta r_i(t) = r - r_i(t), \quad 
\Delta Z_i(t) = Z - Z_i(t), \quad
\Delta \zeta_i(t) = \zeta - \zeta_i(t).
\end{equation}
The centre of the perturbation ($r_i(t), Z_i(t), \zeta_i(t)$) is related to  ($x_i(t), y_i(t), z_i(t)$) by expressions analogous to (\ref{relation_plasma_lab_coords}). 

The time evolution of each individual perturbation is determined by five effects: 
\begin{enumerate}
\item the growth and decay of the perturbation controlled by the characteristic lifetime, $\tau_c$:
\begin{equation}\label{delta_ti}
\Delta t_i = t - \left(t_{i0} + \frac{1}{2} C \tau_c\right),
\end{equation}
where $t_{i0}$ is the time at which the perturbation is created and $C$ is a dimensionless constant that describes how long a perturbation exists: time $C \tau_c/2$ passes before it reaches its peak amplitude;
\item the advection of the perturbation in the toroidal direction by the toroidal flow $\vz$, so the toroidal position of the centre of the perturbation evolves as
\begin{equation}
\zeta_i(t) = \zeta_{i0} + \vz\Delta t_i,
\end{equation}
where $\zeta_{i0}$ is the initial toroidal position of the perturbation;
\item the advection of the perturbation in the poloidal direction by a (small) poloidal flow $\vZ$, so the poloidal position evolves as
\begin{equation}
Z_i(t) = Z_{i0} + \vZ\Delta t_i,
\end{equation}
where $Z_{i0}$ is the initial poloidal position of the perturbation;
\item the advection of the perturbation in the radial direction by a (fluctuating) radial flow $\vr$, so that the radial position evolves as
\begin{equation}
r_i(t) = r_{i0} + \vr\Delta t_i,
\end{equation}
where $r_{i0}$ is the initial radial position of the perturbation;
\item the increase in $k_x(t)$ caused by the radial gradient of the toroidal velocity $S$ (the flow shear):
\begin{equation}
k_x(t) = k_y S \Delta t_i + k_{x0},
\end{equation}
where $k_{x0}$ is the radial wavenumber of the perturbation (whose meaning is discussed below).
\end{enumerate}
Summarising the above expressions, we can write the time and spatial dependence in (\ref{3dfun}) explicitly as
\begin{eqnarray}
\Delta t_i &=& t-(t_{i0} + \frac{1}{2} C \tau_c), \\
\Delta x_i &=& r - r_{i0} - \vr\Delta t_i, \label{deltax} \\
\Delta y_i(t) &=& \left( \zeta - \zeta_{i0} - \vz \Delta t_i \right) \sin\alpha + (Z - Z_{i0} - \vZ\Delta t_i)\cos\alpha, \label{deltay}  \\
\Delta z_i(t) &=& \left( \zeta - \zeta_{i0} - \vz \Delta t_i \right) \cos\alpha - (Z - Z_{i0} - \vZ\Delta t_i)\sin\alpha, \label{deltaz} \\ 
k_{x}(t) &=& k_y S \Delta t_i + k_{x0}.
\end{eqnarray}

It is worth discussing the implications of the choice of the constant $C$. When $C = 0$, the perturbations are created at their peak amplitude, with their specific radial wavenumber, $k_{x0}$, and then evolve in the sheared velocity field. When $C > 0$, each perturbation grows and then decays and $k_{x0}$ is then the radial wavenumber when the perturbation reaches its maximum amplitude. The choice of $C$ does not affect the results of the analytic calculations of the correlation function in \ref{sec:plasma_frame_corr}, but it does affect the results calculated from the numerical implementation of this model described in \ref{sec:toymodel_algorithm}. In order to ensure that the time series generated by this model is smooth around the peak amplitude of single perturbations, which is necessary for the CCTD method described in Section~\ref{sec:meas_corr_time} to work, we use a finite value of $C$.

% ------------------------------------------------------------------------------------------------------------------------------------------
% ------------------------------------------------------------------------------------------------------------------------------------------
% ------------------------------------------------------------------------------------------------------------------------------------------

\subsection{Plasma-frame correlation function}\label{sec:plasma_frame_corr}
In the calculation of the correlation function, we need only consider the correlation of each individual perturbation (\ref{3dfun}) with itself, because, in our model, each individual perturbation is independent of all other such perturbations. The two-point covariance function of a single perturbation, $s_i$, evaluated at $(x_a, y_a, z_a, t_a)$ and $(x_b, y_b, z_b, t_b)$ in the plasma frame is
\begin{eqnarray}
\fl \langle s_{ai}s_{bi} \rangle_P &\equiv& \langle s_i(x_a, y_a, z_a, t_a) s_i(x_b, y_b, z_b, t_b) \rangle_P \nonumber \\
 &=& \int \frac{s_{ai}s_{bi} \ \mathrm{d}A_i\mathrm{d}\phi_i\mathrm{d}x_{i0}\mathrm{d}y_{i0}\mathrm{d}z_{i0}\mathrm{d}t_{i0}\mathrm{d}t \mathrm{d}\vr}{(2 \pi)^2 L_x L_y L_z T^2 \sigma_\mathrm{amp}\dvr}   \exp\left(-\frac{A_i^2}{2\sigma_\mathrm{amp}^2}-\frac{\vr^2}{2\dvr^2}\right) \label{plasma_frame_integral},
\end{eqnarray}
where $L_x, L_y, L_z, T$ are the compact supports of uniform probability distributions for the initial positions $x_{i0}, y_{i0}, z_{i0}$, and $ t_{i0}$. The integration ranges for each of the variables are: $A_i \in [-\infty, \infty]$, $v_r \in [-\infty, \infty]$, $\phi_i \in [0, 2\pi]$, $x_{i0} \in [-L_x/2, L_x/2]$, $y_{i0} \in [-L_y/2, L_y/2]$, $z_{i0} \in [-L_z/2, L_z/2]$, $t_i$ and $t \in [0, T]$. The radial velocity for each perturbation is taken from a Gaussian random distribution with zero mean and standard deviation $\dvr$. The toroidal and poloidal velocities are both assumed to be constant in space and time (in Section~\ref{sec:numeric} we set $\vZ=0$). We compute the integrals in (\ref{plasma_frame_integral}) under the assumption that the integration ranges are much larger than the typical correlation lengths/times in all directions (e.g., $L_x \gg \lamx$). Then it is possible to extend to infinity the limits of integration with respect to the random variables $x_{i0}, y_{i0}, z_{i0}$ and $t_{i0}$, so that the integrals then become standard Gaussian integrals. 

The two-point covariance function in the plasma frame is the sum of (\ref{plasma_frame_integral}) over all individual perturbations
\begin{equation}\label{plasma_cov_def}
C^\mathrm{cov}_P(\Delta x, \Delta y, \Delta z, \Delta t) = \sum_{i=0}^{N-1} \langle s_{ai}s_{bi} \rangle_P.
\end{equation}
It is a function only of the distance between the two points' locations and the time difference between them $(\Delta x = x_a - x_b, \Delta y = y_a-y_b, \Delta z = z_a-z_b, \Delta t=t_a-t_b)$. After completing the integration in (\ref{plasma_frame_integral}), (\ref{plasma_cov_def}) takes the form
\begin{eqnarray}\label{plasma_frame_cov_fun}
\fl C^\mathrm{cov}_P(&\Delta x&, \Delta y, \Delta z, \Delta t) \nonumber \\ 
&=& N \frac{\pi^2 \sigma^2_\mathrm{amp}\lamx\lamy\lamz\tau_c}{4 L_x L_y L_z T} \sqrt{G(\dt)}\nonumber \\
&\quad& \times \exp\left\{ - \frac{\Delta x^2}{\ellxp^2}G(\dt) - \frac{\Delta y^2}{\ellyp^2} - \frac{\Delta z^2}{\ellzp^2}  - \left[\frac{1}{\taucpl^2} + \frac{\kxp^2\dvr^2}{2}G(\dt)\right]\dt^2 \right\} \nonumber \\
&\quad& \times\cos \left[ \kxp G(\dt) \Delta x + \kyp \Delta y \right].
\end{eqnarray}
The plasma-frame correlation function is then
\begin{eqnarray}\label{plasma_frame_correlation_function_with_dvr}
\fl  C_P(\Delta x, \Delta y, \Delta z, \Delta t) &\equiv& \frac{C^\mathrm{cov}_P(\Delta x, \Delta y, \Delta z, \Delta t)}{C^\mathrm{cov}_P(0,0,0,0)} \nonumber \\
 &=& \sqrt{G(\dt)}\exp\left\{ - \frac{\Delta x^2}{\ellxp^2}G(\dt) - \frac{\Delta y^2}{\ellyp^2} - \frac{\Delta z^2}{\ellzp^2} \right. \nonumber \\
 &\quad& \qquad\qquad\qquad\qquad \left. - \left[\frac{1}{\taucpl^2} + \frac{\kxp^2\dvr^2}{2}G(\dt)\right]\dt^2 \right\} \nonumber \\
&\quad&\qquad\quad \times\cos \left[ \kxp G(\dt) \Delta x + \kyp \Delta y \right],
\end{eqnarray}
where
\begin{eqnarray}
\ellxp &=& \sqrt{2} \lamx' =  \frac{\sqrt{2} \lamx}{\sqrt{1 + (k_y\lamx S \tau_c/2)^2}}, \label{ellxp}\\
\ellyp &=& \sqrt{2} \lamy, \label{ellyp_from_toy} \\
\ellzp &=& \sqrt{2} \lamz, \label{ellzp_from_toy}\\
\taucpl &=& \sqrt{2} \tau'_c = \frac{\sqrt{2} \tau_c}{\sqrt{1 + (k_y\lamx S \tau_c/2)^2}}, \label{tau_life} \\
\kxp &=& k_{x0}, \\
\kyp &=& k_y, \label{kyp_from_toy} \\
G(\dt) &=& \frac{\ellxp^2}{\ellxp^2 + 2 \dvr^2\dt^2}\label{Gterm}.
\end{eqnarray}
The plasma-frame auto-correlation function has the form
\begin{equation}\label{plasma_autocorrelation}
C_P(0,0,0, \Delta t) = \sqrt{G(\dt)}\exp\left[  - \left(\frac{1}{\taucpl^2} + \frac{\kxp^2\dvr^2}{2}G(\dt)\right)\dt^2 \right].
\end{equation}
To define the plasma-frame correlation time $\taucp$, we have to be able to compare the expression (\ref{plasma_autocorrelation}) with the plasma-frame auto-correlation function defined in (\ref{plasma_frame_correlation_function}), i.e., $\exp(-\dt^2/\taucp^2)$. To do this, we expand the $G(\dt)$ term (\ref{Gterm}) for $\dvr\dt \ll \ellxp$, which is consistent with extending the asymptotic ordering (\ref{fullordering}-\ref{endfullordering}) to include $\dvr\ll\vz$ (this is formalised in \ref{sec:toy_CCTD}). Retaining terms to the lowest order $\mathcal{O}(\epsilon^2)$ in the exponent of (\ref{plasma_autocorrelation}), the plasma-frame correlation time of our model is then
\begin{equation}\label{taucp_from_toy}
\frac{1}{\taucp^2} = \frac{1}{\taucpl^2} + \left(\kxp^2\ellxp^2 + 2 \right)\frac{\dvr^2}{2 \ellxp^2}.
\end{equation}

The expressions for the plasma-frame correlation parameters defined by (\ref{plasma_frame_correlation_function}) are related to the characteristic lengths and times of the individual perturbations (\ref{3dfun}) via (\ref{ellxp}-\ref{taucp_from_toy}). When $\dvr=0$, (\ref{plasma_frame_correlation_function_with_dvr}) becomes exactly the same as the plasma-frame correlation function (\ref{plasma_frame_correlation_function}), and, as there are no $G(\dt)$ terms, there is no need to use the asymptotic ordering to find an expression for the plasma-frame correlation time (i.e., $\taucp = \taucpl$). When $\dvr\ne0$, as is discussed in Section~\ref{sec:fluc_rad_velocity}, the time dependence in   (\ref{Gterm}) introduces cross-terms between spatial and temporal coordinates, and therefore, (\ref{plasma_frame_correlation_function_with_dvr}) is not equal to (\ref{plasma_frame_correlation_function}).

% ------------------------------------------------------------------------------------------------------------------------------------------
% ------------------------------------------------------------------------------------------------------------------------------------------
% ------------------------------------------------------------------------------------------------------------------------------------------

\subsection{Laboratory-frame correlation function}\label{sec:lab_frame_corr}
To calculate the two-point covariance function in the laboratory frame, we follow the same procedure as in \ref{sec:plasma_frame_corr}, but write the individual perturbation (\ref{3dfun}) in laboratory frame coordinates using the transformations (\ref{deltax}-\ref{deltaz}) and then integrate over the range of initial positions in the laboratory-frame coordinates $(r_{i0}, Z_{i0}, \zeta_{i0})$, rather than in the plasma-frame coordinates $(x_{i0}, y_{i0}, z_{i0})$. Explicitly,
\begin{eqnarray}
\fl \langle s_{ai}s_{bi} \rangle_L &\equiv& \langle s_i(r_a, Z_a, \zeta_a=0, t_a) s_i(r_b, Z_b, \zeta_b=0, t_b) \rangle_L \nonumber \\
 &=& \int \frac{s_{ai}s_{bi}\ \mathrm{d}A_i\mathrm{d}\phi_i\mathrm{d}r_{i0}\mathrm{d}Z_{i0}\mathrm{d}\zeta_{i0}\mathrm{d}t_{i0}\mathrm{d}t \mathrm{d}\vr}{(2 \pi)^2 L_r L_Z L_\zeta T^2 \sigma_\mathrm{amp}\dvr}  \exp\left(-\frac{A_i^2}{2\sigma_\mathrm{amp}^2}-\frac{\vr^2}{2\dvr^2}\right) \label{Timeintegral},
\end{eqnarray}
where $L_r, L_Z, L_\zeta$ are the compact supports of uniform probability distributions for the initial positions of the perturbations $r_{i0}, Z_{i0}, \zeta_{i0}$. We complete this integration, in the same way as we did in \ref{sec:plasma_frame_corr}, by assuming that the integration ranges are large compared to the size of the perturbations, e.g. $L_r \gg \lambda_x$. Then the covariance function in the laboratory frame, defined analogously to (\ref{plasma_cov_def}), is
\begin{eqnarray}
\fl C^\mathrm{cov}_L(\dr, \dz, \dt) &=& F_0(\dt)\exp \left\{ -\left[ F_1(\dt)\dr^2 + F_2 \dz^2 + F_3 \dz \vz \dt + F_4(\dt) \vz^2 \dt^2  \right]\right\}\nonumber \\
&\quad& \times\cos\left[F_5(\dt) \dr+ F_6 \dz + F_7 \vz \dt  \right], \label{fulltimedelay}
\end{eqnarray}
where, in terms of the plasma-frame correlation parameters (\ref{ellxp}-\ref{kyp_from_toy}),  % See perp_para_and_radial_Gaussian_vr_average.nb
\begin{eqnarray}
F_0(\dt) &=& N\frac{\pi^2 \sigma^2_\mathrm{amp}\lamx\lamy\lamz\tau_c}{4 L_r L_Z L_\zeta T}  \sqrt{G(\dt)},\label{F0_cov_lab} \\
F_1(\dt) &=& \frac{G(\dt)}{\ellxp^2}, \label{F1_cov_lab} \\
F_2 &=& \frac{ \ellyp^2 \sin^2\alpha+ \ellzp^2 \cos^2\alpha }{ \ellyp^2 \ellzp^2}, \\
F_3 &=&  \frac{(\ellyp^2 - \ellzp^2)  \sin(2\alpha)}{ \ellyp^2 \ellzp^2} + \frac{2\vZ}{\vz} \frac{(\ellyp^2 \sin^2\alpha - \ellzp^2\cos^2\alpha) }{\ellyp^2 \ellzp^2}, \\
F_4(\dt) &=& \frac{1}{ \vz^2 \taucpl^2} +  \frac{\dvr^2}{2\vz^2}\kxp^2 G(\dt) \nonumber \\
&\quad& + \frac{(\vz\cos\alpha + \vZ\sin\alpha)^2}{\ellzp^2 \vz^2} + \frac{(\vz\sin\alpha + \vZ\cos\alpha)^2}{\ellyp^2 \vz^2},  \\
F_5(\dt) &=&  \kxp G(\dt),\label{F5_cov_lab} \\
F_6 &=& \kyp  \cos\alpha, \\
F_7 &=& -  \kyp \left(\sin\alpha + \frac{\vZ}{\vz}\cos\alpha\right),
\end{eqnarray}
and where $G(\dt)$ is defined in (\ref{Gterm}). The laboratory-frame correlation function is defined analogously to the plasma-frame correlation function (\ref{plasma_frame_correlation_function_with_dvr}):
\begin{eqnarray}
\fl C_L(\dr, \dz, \dt) &=& \sqrt{G(\dt)}\exp \left\{ -\left[ F_1(\dt)\dr^2 + F_2 \dz^2 + F_3 \dz \vz \dt + F_4(\dt) \vz^2 \dt^2  \right]\right\}\nonumber \\
&\quad& \times\cos\left[F_5(\dt) \dr+ F_6 \dz + F_7 \vz \dt  \right]. \label{lab_corr_function_full}
\end{eqnarray}
When $\dvr=0$ and $\dt=0$, the laboratory-frame correlation function (\ref{lab_corr_function_full}) is exactly the same as (\ref{lab_frame_correlation_function}), and the relationships (\ref{ellrt_ellxp}-\ref{kzt_kyp}) between the laboratory-frame and plasma-frame correlation parameters are determined by comparing these two expressions. 

The procedure for calculating the relationship between the CCTD correlation time in the laboratory frame $\tauct$ and the plasma-frame correlation time $\taucp$ is described in Section~\ref{sec:asymtotics}, for the case when $\dvr=0$. The detailed calculation of the laboratory-frame correlation time $\tauct$ for the general case when $\dvr\ne0$ is given in \ref{sec:toy_CCTD}, the results of which are discussed in Section~\ref{sec:fluc_rad_velocity}.

% ------------------------------------------------------------------------------------------------------------------------------------------
% ------------------------------------------------------------------------------------------------------------------------------------------
% ------------------------------------------------------------------------------------------------------------------------------------------

\subsection{Fluctuation amplitude}\label{sec:toyflucamp}
The plasma-frame mean square fluctuation amplitude is the amplitude of the plasma-frame covariance function (\ref{plasma_frame_cov_fun}) with $\Delta x=\Delta y =\Delta z=\dt=0$, i.e., 
\begin{equation}\label{fluc_amp_plasma}
\sigma_{\mathrm{amp}|P}^2 \equiv N \frac{\pi^2 \sigma^2_\mathrm{amp}}{4} \frac{\lamx\lamy\lamz\tau_c}{L_x L_y L_z T},
\end{equation}
which is proportional to the mean square fluctuation amplitude of our model, $\sigma_\mathrm{amp}^2$, introduced in (\ref{3dfun}). Similarly, the laboratory-frame mean square fluctuation amplitude is the amplitude of the laboratory-frame covariance function, $F_0(\dt=0)$, given by (\ref{F0_cov_lab})
\begin{equation}\label{fluc_amp_lab}
\sigma_{\mathrm{amp}|L}^2 \equiv N \frac{\pi^2 \sigma_\mathrm{amp}^2}{4} \frac{\lamx \lamy \lamz \tau_c}{L_r L_Z L_\zeta T}.
\end{equation}
As the volumes $L_x L_y L_z$ and $L_r L_Z L_\zeta$ are equal, the fluctuation amplitudes in both frames are the same. 

For the plasma-frame (or laboratory-frame) fluctuation amplitude to be independent of the size of the domain, the number of individual perturbations $s_i$ is chosen to be $N = \kappa L_x L_y L_z T/\lamx \lamy \lamz \tau_c$, where $\kappa$ is an arbitrary constant. The exact choice of $\kappa$ does not affect the linear relationship between the plasma-frame (or laboratory-frame) fluctuation amplitude, $\sigmap$ ($\sigma_{\mathrm{amp}|L}$), and the fluctuation amplitude $\sigma_\mathrm{amp}$ of the model field.

The finiteness of the spatial domain in which the model fluctuating signals are generated can cause the plasma- and laboratory-frame fluctuation amplitudes to be lower when calculated from numerically generated time series than those expected from (\ref{fluc_amp_lab}), because the integration limits in (\ref{Timeintegral}) cannot be extended to infinity. However, because the integration domain, spatial scales and velocity of the perturbations are all kept constant throughout the study of varying fluctuation amplitudes in Section~\ref{sec:modelflucamp}, these effects simply cause the resulting fluctuation amplitude to be reduced by a constant factor~\cite{Kim2016}. Therefore, because we already have an arbitrary free parameter $\kappa$, we can rescale the analytic expression for the fluctuation amplitude (\ref{fluc_amp_lab}) to match the value measured from the numerically generated fluctuating field. The effect of Gaussian-model PSFs on the laboratory-frame fluctuation amplitude can then be determined with (\ref{PSFflucamp}) and compared with the effect of real PSFs (see Section~\ref{sec:gauss_fluc_amp}).

% ------------------------------------------------------------------------------------------------------------------------------------------
% ------------------------------------------------------------------------------------------------------------------------------------------
% ------------------------------------------------------------------------------------------------------------------------------------------

\subsection{Method for generating a fluctuating field}\label{sec:toymodel_algorithm}
To generate a time series from the model of fluctuating fields described in \ref{sec:basic_equations}, one can simply use (\ref{density_field_output}) and sum over $N$ perturbations, each with its own random amplitude, phase, radial velocity, start time and position. However, $N$ is determined from (\ref{fluc_amp_lab}), and takes typical values of $N >1000$, therefore evolving this number of perturbations over the whole four-dimensional (3 space + 1 time) domain requires long computation times.

In order to reduce the computation time, we use the fact that the contribution of a perturbation to a time series is localised in time, because $\tau_c \ll T$. Then we only have to calculate the contribution to the time series in a reduced time range centred around the time at which the perturbation reaches its peak amplitude. This is achieved through the introduction of the parameter  $C$ in (\ref{delta_ti}), which determines the duration of time that a perturbation exists before reaching its peak amplitude. In \ref{sec:plasma_frame_corr} and \ref{sec:lab_frame_corr}, it was shown that the correlation function does not depend on $C$, provided that the integration range (in time) is taken to infinity. This is equivalent to ensuring that $C \gg 1$, so that each perturbation has sufficient time to grow and decay from and to near-zero amplitude. 

Thus, for each individual perturbation, the density field is calculated on a sub-domain of the full domain $(r, Z, t) = (L_r, L_Z, C \tau_c)$. Then this sub-domain is added to the full domain (\ref{density_field_output}) by positioning the sub-domain at the start time of the perturbation set by the random variable $t_{i0}$. The toroidal length of the domain is set by $L_\zeta = C \vz\tau_c$ so that only perturbations that will pass through $\zeta = 0$ during the reduced time period $C\tau_c$ are considered. 

For typical turbulence parameters, using this approach to construct a fluctuating time series allows for the calculation of a $2\ \mathrm{ms}$, $30\ \mathrm{cm}\times20\ \mathrm{cm}$ two-dimensional radial-poloidal time series on a $0.5\ \mu\mathrm{s}$ and $0.5\ \mathrm{cm} \times0.5\ \mathrm{cm}$ grid in less than 5 minutes using a desktop computer. 

\subsection{Input parameters}
\begin{table}[h]
	\begin{tabular}{ l | l | l }
		Input parameter & Value range & Description  \\
		\hline                       
		$\lamx$ / cm & 0.75 --- 5.0 & Characteristic length in the radial direction   \\
		$\lamy$ / cm & 1.5 --- (10) --- 15 & Characteristic length in the binormal direction  \\
		$\lamz$ / cm& 150.0 & Characteristic length in the parallel direction   \\
		$\tau_c$  / $\mu s$ & 1 --- (10) --- 25 & Characteristic life time \\
		$\Theta$ /deg. & -45  --- (25) ---  45 & Tilt angle of the correlation function \\
		$k_{x0}$ / cm$^{-1}$& $- k_y \cos\alpha \tan\Theta$  & Radial wavenumber  \\
		$k_y$ / cm$^{-1}$& (\ref{ky_fixed}) & Binormal wavenumber \\
		$\sigma_\mathrm{amp}$ & 0.1 --- (1.0) --- 2.0 & Std. dev. of the amplitude of perturbations \\
		$\vz$  / cm/$\mu$s & 0.5 --- (1.5)$^*$ --- 8.0 & Bulk toroidal velocity (10 km/s) \\
		$\vZ$  / cm/$\mu$s & 0.0 & Poloidal velocity \\
		$\dvr$ / cm/$\mu$s & $0.3$ & Standard deviation of radial velocity \\
		$S$  / $\mu \mathrm{s}^{-1}$& 0.02 & Shearing rate \\
		$\alpha$  / deg. & 30 & Pitch angle of the magnetic field \\
		$C$ & 8 & $C\tau_c$: existence time of a perturbation \\
		$L_r$ / cm & 30 & Radial length of the domain  \\
		$L_Z$ / cm & 20 &Poloidal length of the domain   \\
		$L_\zeta$ / cm & $C \vz \tau_c$ & Toroidal length of the domain  \\
		$T$/ $\mu$s & 2000$^{**}$   & Duration of the simulation in time  \\
	\end{tabular}
	\caption[Input parameters for generating time series from the model of fluctuating fields]{Input parameters for generating time series from the model of fluctuating fields described in Section~\ref{sec:toymodel}. Values in brackets are the fixed values used when varying other parameters. $^*$In Figure~\ref{fig:blobby_plots}(e), $\vz=3.0$ was used. $^{**}$Except for Figure~\ref{fig:zoomtau}, where $T=20\ \mathrm{ms}$.}
	\label{tab:blobby_inputs}
\end{table}
The input parameters required to generate a time series of fluctuations from our model are listed in Table~\ref{tab:blobby_inputs} with the range of values used in this work also specified, for reference. The tilt angle $\Theta = -\arctan(\krt/\kzt)$, has been made an input parameter for the model by using this to specify the radial wavenumber, $k_x = - k_y \cos\alpha \tan\Theta$, where (\ref{kzt_kyp}) has been used.  

For our model field, we have chosen to fix the laboratory-frame poloidal wavenumber to the laboratory-frame poloidal correlation length: $\kzt\ellzt=2\pi$. This then requires that the input parameter $k_y$ is fixed using
\begin{equation}\label{ky_fixed}
k_y = 2\pi\left( \frac{1}{2\lamy^2} + \frac{\tan^2\alpha}{2\lamz^2} \right)^{1/2},
\end{equation}
where (\ref{lab_pol_length}), (\ref{kzt_kyp}), (\ref{ellyp_from_toy}), and (\ref{ellzp_from_toy}) have been used. This choice was influenced by the requirement to fix the product $\kza\ellza$ in order to constrain a fit to the poloidal correlation function, which was done in previous studies by assigning the product a value of $2\pi$~\cite{Ghim2013,FieldPPCF2014}. We note that, by using the method described in Section~\ref{sec:ztd_measurement} and \ref{sec:poloidal_corr_lengths}, we can now experimentally measure the product $\kza\ellza$  from the auto-correlation function.

% ------------------------------------------------------------------------------------------------------------------------------------------
% ------------------------------------------------------------------------------------------------------------------------------------------
% ------------------------------------------------------------------------------------------------------------------------------------------
% Appendix C - CORRELATION PARAMETERS & CCTD METHOD

% ------------------------------------------------------------------------------------------------------------------------------------------
% ------------------------------------------------------------------------------------------------------------------------------------------
% ------------------------------------------------------------------------------------------------------------------------------------------
\section{CCTD method: apparent poloidal velocity and correlation time}\label{sec:appendix_CCTD}
In \ref{sec:toy_CCTD}, we derive the expression for the laboratory-frame correlation time and apparent poloidal velocity that would be measured using the CCTD method (see Section~\ref{sec:meas_corr_time}) assuming that the correlation function has the form (\ref{lab_corr_function_full}). The results include the effect of a Gaussian-distributed fluctuating radial velocity $\dvr$. They reduce to the results of Section~\ref{sec:asymtotics} when $\dvr=0$. In \ref{sec:PSFcorrtime}, we discuss the details of the calculation for the BES-measured correlation time and apparent poloidal velocity, by starting from the correlation function including PSF effects.

\subsection{CCTD method in the laboratory frame}\label{sec:toy_CCTD}
As the principle of the CCTD method was described in Section~\ref{sec:asymtotics}, here we focus on details, including how to perform the calculation accounting for a fluctuating radial velocity. The goal is to find expressions for the amplitude of the peak of the cross-correlation function $A_{\mathrm{peak}|L}(\dz)$ and to find the time delay $\dtpeakt(\dz)$ of this peak, both of which quantities are functions of the distance $\dz$ between poloidally separated channels. Furthermore, we require the functional form of $A_{\mathrm{peak}|L}(\dz)$ to be a Gaussian in $\dz$, so that it is possible to compare  $A_{\mathrm{peak}|L}(\dz)$ with the Gaussian fitting function used in (\ref{correlation_timedelay_fit}) to measure the laboratory-frame correlation time $\tauct$.

To proceed, we adopt the same asymptotic ordering (\ref{fullordering}-\ref{endfullordering}) as in the main text, but include also a small radial velocity, which is modelled as a Gaussian-distributed random variable with zero mean and standard deviation $\dvr$. This standard deviation is then ordered small compared to the toroidal velocity $\dvr/\vz \sim \epsilon$, and the same size as $\vZ$. Additionally, we express the time delay of the peak as $\dtpeakt = \dto + \ddt$, where $\ddt \sim \epsilon\dto$. The asymptotic ordering becomes
\begin{eqnarray}\label{fullordering_vr}
\vz \taucp, \vth \taucpl \sim \ellzp &\sim& \mathcal{O}(R),\\
\vz\dto \sim \dz, \dr, \ellxp, \ellyp \quad\quad &\sim& \mathcal{O}(\epsilon R),  \\
\vZ \dto, \dvr \dto, \vz \ddt &\sim&\mathcal{O}(\epsilon^2R), \\
\vZ \ddt, \dvr \ddt  &\sim& \mathcal{O}(\epsilon^3R).\label{endfullordering_vr}
\end{eqnarray}

\subsubsection{Zeroth order.}\label{sec:zerothordercctd}
We assume that the correlation function has the form (\ref{lab_corr_function_full}). We start by setting $\dr=0$ in (\ref{lab_corr_function_full}) and then neglect all terms that are smaller than $\mathcal{O}(\epsilon^0)$ in the arguments of the exponential and the cosine. The inclusion of a radial velocity has introduced a non-Gaussian dependence of the correlation function on the time delay $\dt$ through the $G(\dt)$ function (\ref{Gterm}). However, realising that $\dvr\dto/\ellxp \sim \epsilon \ll 1$, we find that $G(\dt)=1 + \mathcal{O}(\epsilon^2)$, and therefore, at the lowest order, we can neglect this effect. This gives the correlation function to zeroth order:
\begin{eqnarray}
\fl C_0(\dz, \dto) &=& \exp\left[ - \frac{(\dz \cos\alpha - \vz \dto \sin\alpha)^2}{\ellyp^2} \right] \nonumber \\
&\quad&\times  \cos\left[ \kyp (\dz \cos\alpha - \vz \dto \sin\alpha) \right].\label{zeroth_order_correlation_function}
\end{eqnarray}
This is clearly a wave travelling at the apparent poloidal velocity $\vz \tan\alpha$, with maxima occurring at 
\begin{equation}\label{cctd_peak}
\dto \equiv \dz \cot\alpha/\vz.
\end{equation}
Note that these maxima do not decay in time. Therefore, we need to consider the small terms that we have neglected to determine how the amplitude of the peak of the correlation function (\ref{lab_corr_function_full}) changes in time. 

\subsubsection{Second order corrections to the peak amplitude of the correlation function.}\label{sec:cctd_second_order}
To do this, we consider small deviations $\ddt \sim \epsilon\dto$ around the position of the maximum (\ref{cctd_peak}) of the zeroth-order correlation function (\ref{zeroth_order_correlation_function}). Then, by writing $\dt = \dto + \ddt$, and substituting this into (\ref{lab_corr_function_full}), we find that the lowest-order terms that remain are $\mathcal{O}(\epsilon^2)$ (the peaks of the correlation function (\ref{lab_corr_function_full}) do not decay at $\mathcal{O}(\epsilon^1)$). When carrying out this expansion, care must be taken, as the argument of the cosine is, in isolation, $\mathcal{O}(\epsilon^1)$, but, in order to compare this with the argument of the exponential in (\ref{lab_corr_function_full}), we have to expand the cosine for small argument, which results in an $\mathcal{O}(\epsilon^2)$ contribution from the argument of the cosine.  The correlation function (\ref{lab_corr_function_full}), at second order in $\epsilon$, is 
\begin{eqnarray}
\fl C_2(\dz, \ddt) &=& \left(1 + \frac{2\dvr^2\dz^2}{\ellxp^2\vz^2\tan^2\alpha}\right)^{-1/2}\exp\left[ 
- \left(\frac{1}{\taucpl^2} + \frac{\kxp^2\ellxp^2\dvr^2}{2\ellxp^2}\right)\frac{\dz^2}{\vz^2\tan^2\alpha}
 \right. \nonumber \\
&\quad& \qquad\qquad \left.  - \frac{ (\vz\ddt \tan\alpha + \vZ \dz \cot\alpha /\vz)^2\cos^2\alpha}{\ellyp^2} - \frac{\dz^2}{\ellzp^2\sin^2\alpha}\right] \nonumber \\
&\quad&\times\cos\left[ - \left(\kyp \vz \ddt \tan\alpha +  \kyp\vZ \dz \cot\alpha/\vz\right)\cos\alpha\right]. \label{secondordercorrfun}
\end{eqnarray}
At this order, the effects of the $G(\dt)$ term (\ref{Gterm}) first become evident. The coefficient of (\ref{secondordercorrfun}) has a dependence on $\dz$, which causes a non-Gaussian decay with $\dz$. As the second term in this coefficient is $\mathcal{O}(\epsilon^2)$ small, we make the following approximation
\begin{equation}\label{GtermApprox}
\fl \left(1 + \frac{2\dvr^2\dz^2}{\ellxp^2\vz^2\tan^2\alpha}\right)^{-1/2} \simeq 1 - \frac{\dvr^2\dz^2}{\ellxp^2\vz^2\tan^2\alpha}+ \mathcal{O}(\epsilon^4) \simeq \exp\left( - \frac{\dvr^2\dz^2}{\ellxp^2\vz^2\tan^2\alpha} \right),
\end{equation}
which means that this can be incorporated into the exponent of (\ref{secondordercorrfun}).

After using the approximation (\ref{GtermApprox}) in (\ref{secondordercorrfun}), we eliminate $\ddt$ from the resulting expression by calculating the value of $\ddt$ at the maximum of (\ref{secondordercorrfun}). The maximum of the correlation function occurs when
\begin{equation}\label{ddtsolve}
\ddt =- \frac{\vZ \dto}{\vz\cot\alpha} - \frac{\kyp \ellyp^2 }{2\vz \sin\alpha}\tan\left[  \kyp\vz\ddt\sin\alpha + \kyp\vZ \dto \cos\alpha  \right] ,
\end{equation}
which can be solved for $\ddt$ by realising that the argument of the tangent is $\mathcal{O}(\epsilon^1)$ and, therefore, we may expand in this argument and obtain
\begin{equation}\label{ddt}
\frac{\ddt}{\dto} =  - \frac{\vZ}{\vz}\cot\alpha.
\end{equation}
This implies that the peak of the cross-correlation function is shifted in time in comparison with the zeroth-order solution (\ref{cctd_peak}). Thus, the peak of the cross-correlation function occurs at
\begin{equation}\label{peak_dt}
\dtpeakt(\dz) = \left(1 - \frac{\vZ}{\vz}\cot\alpha \right) \frac{\dz \cot\alpha}{\vz} + \mathcal{O}(\epsilon^2\dto).
\end{equation}

\subsubsection{Apparent poloidal velocity.}
We see immediately that the apparent poloidal velocity in the laboratory frame is then
\begin{eqnarray}
v_{\mathrm{pol}|L} &\equiv& \frac{\dz}{\dt_{\mathrm{peak}|L}} = \vz \tan\alpha \left(1 -  \frac{\vZ}{\vz}\cot\alpha\right)^{-1} \nonumber \\
&=&  \vz \tan\alpha  + \vZ +\mathcal{O}(\epsilon^2\vz) \label{apparent_pol_vel},
\end{eqnarray}
which is simply the sum of the projection of the toroidal velocity on to the poloidal plane, $\vz \tan\alpha$, and the `true' poloidal velocity $\vZ$ (which is $\mathcal{O}(\epsilon^1)$ compared to $\vz$).

\subsubsection{Correlation time.}
Using the result (\ref{ddt}) in (\ref{secondordercorrfun}), we find the expression for the  amplitude of the peak of the cross-correlation function between two detector channels separated by a distance $\dz$ to be
\begin{equation}\label{cpeak}
\fl A_{\mathrm{peak}|L}(\dz) = \exp\left\{ - \left[\frac{1}{\ellzp^2 \cos^2\alpha} + \frac{ 1}{\vz^2\taucpl^2}  +  \frac{(2 + \kxp^2\ellxp^2)}{2\ellxp^2}\frac{\dvr^2}{\vz^2} \right] \dz^2\cot^2\alpha \right\}.
\end{equation}

To find the laboratory-frame correlation time, we compare (\ref{cpeak}) with the fitting function (\ref{correlation_timedelay_fit}). This fitting function is given as a function of $\dtpeakt$ by (\ref{peak_dt}), however, because the exponent in (\ref{cpeak}) is only accurate to $\mathcal{O}(\epsilon^2)$, it is only necessary (and correct) to use the lowest-order part of (\ref{peak_dt}). Therefore, the laboratory-frame correlation time is
\begin{eqnarray}
\frac{1}{\tau_{c|L}^2} &=& \frac{1}{\taucpl^2}  + \frac{\vz^2}{\ellzp^2 \cos^2\alpha} +  \left( \frac{2  + \kxp^2\ellxp^2}{2\ellxp^2}\right)\dvr^2 + \mathcal{O}\left(\frac{\epsilon^2}{\taucpl^2}\right), \nonumber \\
&=&\frac{1}{\taucp^2}  + \frac{\vz^2}{\ellzp^2 \cos^2\alpha}. \label{labcorrtime}
\end{eqnarray}
We see that the measurement of the correlation time is contaminated by the radial and toroidal velocities of the turbulent perturbations, but not by the poloidal velocity. Further discussion of this result is given in the main text; Sections~\ref{sec:lab_frame_corr_time} and \ref{sec:gs2temporal}. 

% ------------------------------------------------------------------------------------------------------------------------------------------------------
% ------------------------------------------------------------------------------------------------------------------------------------------------------
% ------------------------------------------------------------------------------------------------------------------------------------------------------

\subsection{PSF effects on the correlation time and apparent poloidal velocity}\label{sec:PSFcorrtime}
In this Appendix, we describe how to calculate the effect of PSFs on the correlation time and apparent poloidal velocity, including radial velocity effects. We start from the laboratory-frame covariance function (\ref{fulltimedelay}), apply the PSFs using the integral (\ref{Cbes3}) and define the BES correlation function analogously to (\ref{Cbes4}). This expression is too cumbersome to reproduce here: it is similar in form to (\ref{lab_corr_function_full}), but includes a cross-term between $\dr$ and $\dz$ similar to that in (\ref{Cbesfinal}), arising due to the PSF integral.

With this BES correlation function, the exact same procedure is used as was described in \ref{sec:toy_CCTD}, starting from (\ref{lab_corr_function_full}), though with the additional ordering of the PSF lengths to be the same order as the radial and binormal correlation lengths, $L_1, L_2 \sim \ellxp, \ellyp$, so (\ref{fullordering_vr}-\ref{endfullordering_vr}) becomes:
\begin{eqnarray}\label{fullordering_PSF}
\vz \taucp, \vth \taucpl \sim \ellzp &\sim& \mathcal{O}(R),\\
\vz\dto \sim \dz, \dr, \ellxp, \ellyp, L_1, L_2 \quad\quad &\sim& \mathcal{O}(\epsilon R),  \\
\vZ \dto, \dvr \dto, \vz \ddt &\sim&\mathcal{O}(\epsilon^2R), \\
\vZ \ddt, \dvr \ddt  &\sim& \mathcal{O}(\epsilon^3R).\label{endfullordering_PSF}
\end{eqnarray}
The calculation of the time-delay $\dtpeaka(\dz)$ and amplitude $\ApeakB(\dz)$ of the peak of the BES cross-correlation function becomes more extended than in \ref{sec:toy_CCTD}, because there are more $G(\dt)$ terms (\ref{Gterm}) that exist in the exponent of the BES correlation function (with $\dr=0$). They arise because the PSF integral (\ref{Cbes3}) brings (\ref{F1_cov_lab}) into the coefficients of the $\dz$ terms. Therefore, to get the correct expression for the second-order correlation function, analogous to (\ref{secondordercorrfun}), it is necessary to approximate $G(\dt)$ using (\ref{GtermApprox}). However, other than this detail, the process is no different from that described in \ref{sec:toy_CCTD}.

The resulting expression for the time at which the peak of the cross-correlation function occurs is the same as (\ref{peak_dt}), and is, therefore, unaffected by the PSFs. Thus, the apparent poloidal velocity that would be measured by the BES, $v_{\mathrm{pol}|B}$, is equal to the laboratory-frame apparent poloidal velocity (\ref{apparent_pol_vel}), $v_{\mathrm{pol}|L}$. In contrast, the BES correlation time is affected by the PSFs:
\begin{eqnarray}\label{corrtimewithPSFs}
\fl\frac{1}{\tauca^2} &=& \frac{1}{\tauct^2} 
  + (2+  \krt^2 \ellrt^2)\left(1-\frac{\ellra^2}{\ellrt^2}\right)\frac{\dvr^2}{2 \ellra^2} \nonumber \\
 &\quad&\qquad+ \left(\frac{\kzt^2 \ellzt^2}{2 \ellra^2} - \frac{\kzt^2 \ellzt^4}{2 D^4}  \right) \dvr^2 
  + \mathcal{O}\left(\frac{\epsilon^2}{\taucp^2}\right),
\end{eqnarray}
where we have written the expression in terms of both laboratory-frame (subscript $L$) and BES parameters (subscript $B$), and $D$ is given by (\ref{Dterm}). The mixed notation is used in order to be concise, and should not cause concern, as the zero-time-delay expressions for the relationship between the two parameter sets are determined by (\ref{an_ellrt_inv}-\ref{kzt_inv}), independent of any radial-velocity contributions. The result (\ref{corrtimewithPSFs}) is discussed in Section~\ref{sec:fluc_rad_velocity}.

% ------------------------------------------------------------------------------------------------------------------------------------------------------
% ------------------------------------------------------------------------------------------------------------------------------------------------------
% ------------------------------------------------------------------------------------------------------------------------------------------------------

% ------------------------------------------------------------------------------------------------------------------------------------------------------
% ------------------- APPENDIX C: PSF testing how close to Gaussian       ----------------------------------------
% ------------------------------------------------------------------------------------------------------------------------------------------------------

% ------------------------------------------------------------------------------------------------------------------------------------------
% ------------------------------------------------------------------------------------------------------------------------------------------
% ------------------------------------------------------------------------------------------------------------------------------------------

\section{Quantifying the difference between real and Gaussian-model PSFs}\label{sec:psf_testing}
\begin{figure}
	\centering
	\includegraphics[width=\textwidth]{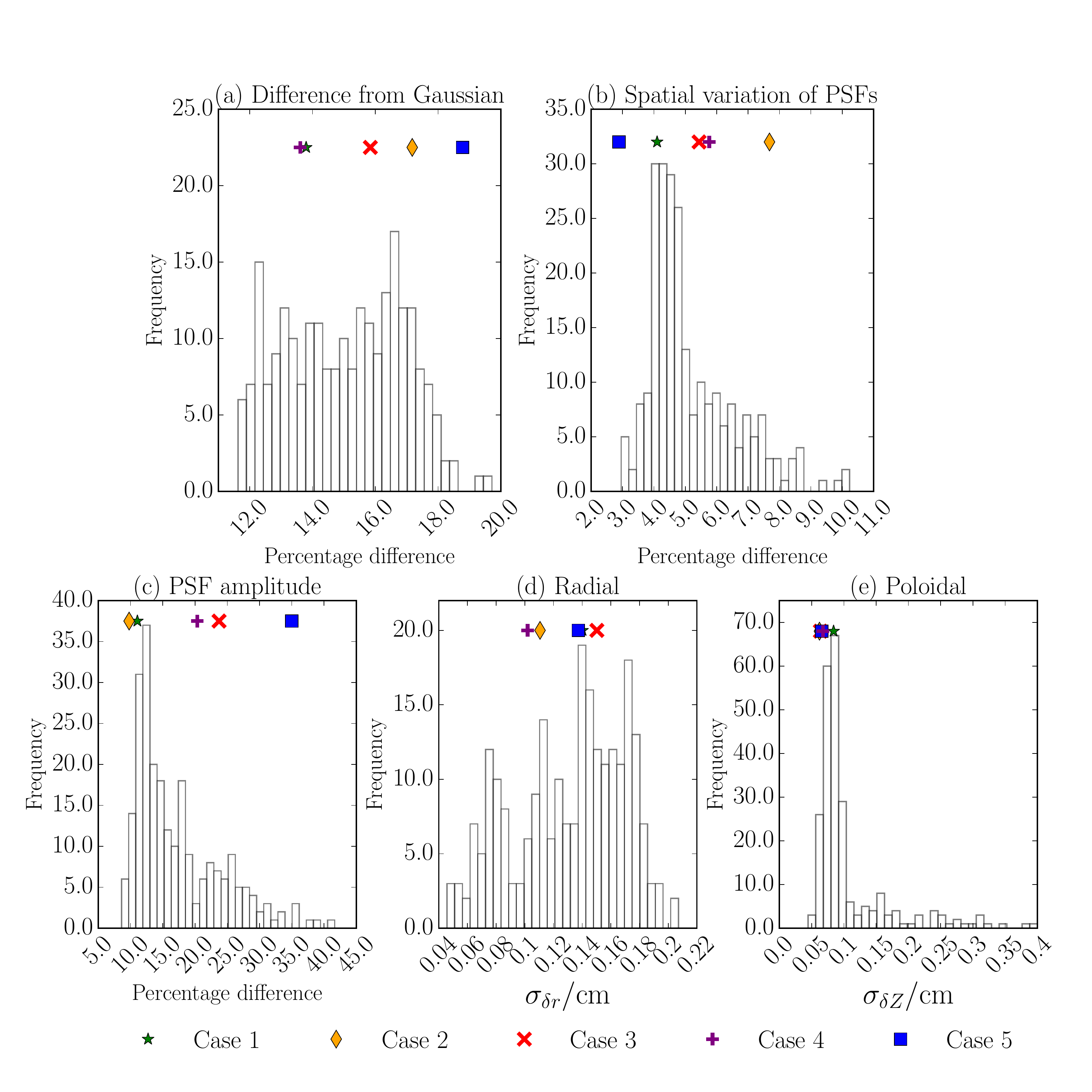}
	\caption[Quantifying the difference between real and Gaussian-model PSFs]{Variation of real PSFs from Gaussian-model PSFs for 245 sets of PSFs calculated at various times in 20 MAST shots in DND configuration. The different measures are designed to measure: (a) the difference of the real PSFs from a Gaussian (\ref{sec:app_diff_gaus_psf}); (b) the spatial variation of the shape of the PSFs (\ref{sec:app_spatial_var_psf}); (c) the spatial variation of the amplitudes of the PSFs (\ref{sec:app_amp_var_psf}); and (d) and (e) the variation across the channels of the weighted centre of the response compared to the focal centre (\ref{sec:app_centre_var_psf}). The values for each of the cases, considered in Figure~\ref{fig:psf_alpha}, are indicated with the coloured symbols. Case 5 is not included in the calculation of the histograms, yet is labelled for comparison. \label{fig:comp_full_gauss_psf} }
\end{figure}

In Section~\ref{sec:PSFapproxTest}, we compared sets of real PSFs with sets of Gaussian-model PSFs, where each set of Gaussian-model PSFs had the same parameters $L_1, L_2$ and $\apsf$. These parameters are, respectively, the means of the sets of parameters $\{L_{1i}\}, \{L_{2i}\}$ and $\{\apsfi\}$ measured from each of the real PSFs, at channel $i$, using principal component analysis as described in Section~\ref{sec:MeasuringrealPSFs}. The purpose of this Appendix is to assess which part of the assumptions encoded in the Gaussian-model PSFs cause the biggest difference between these Gaussian-model PSFs and the real PSFs.

\subsection{Difference of the shape of real PSFs from a tilted Gaussian function}\label{sec:app_diff_gaus_psf}
The first assumption made by us was that we could describe the real PSFs by a Gaussian function. We measure the quality of this assumption by comparing each real PSF with a single Gaussian that has the parameters $L_{1i}, L_{2i}$ and $\apsfi$ measured from that real PSF. We use the measure
\begin{equation}\label{deltaPdef}
\Delta P_i = \frac{\mathrm{RMS}\left[ P_i^\mathrm{real}(\vec{r}-\vec{r}_i) - P_i^\mathrm{Gauss}(\vec{r}-\vec{r}_i,L_{1i}, L_{2i}, \alpha_{\mathrm{PSF}i}, A_{\mathrm{PSF}i}, r_i, Z_i) \right]_\vec{r}}{A_{\mathrm{PSF}i}},
\end{equation}
which gives the root-mean-square (RMS) difference between the real PSF, $P_i^\mathrm{real}$, and the Gaussian model (\ref{GaussianPSF}) of the same PSF, relative to the peak amplitude of the real PSF. The coordinates of the centre of the real PSF, $\vec{r}_i$, are determined by considering the line defined by $\apsf$ and the point of the maximum of the PSF. The point on this line a distance $L_1$ from the $e^{-1}$ contour towards the maximum of the PSF is taken to be $\vec{r}_i$. 

Averaging $\Delta P_i$ given by (\ref{deltaPdef}) over the set of channels in a BES sub-array gives the measure plotted in the histogram of Figure~\ref{fig:comp_full_gauss_psf}(a), for 245 different sets of PSFs. This shows that there is a variation of 12\% to 20\% in this `difference from Gaussian' measure. Indeed, this Figure is very similar to Figure~\ref{fig:comp_full_gauss_psf_simple}, and, therefore, most of the difference between using the Gaussian-model (with all PSFs the same) and the real PSFs is due to the assumption of a Gaussian shape for the PSFs. 

\subsection{Spatial variation of PSFs}\label{sec:app_spatial_var_psf}
To calculate the PSF effects analytically (Section~\ref{sec:PSF_analytic}), we had to assume that the shape of the PSFs was the same for all the detector channels. We measure how the PSFs vary in space, in violation of this assumption, using 
\begin{equation}\label{overlineP}
\Delta \overline{P}_i = \mathrm{RMS}\left[P_i^\mathrm{Gauss}(\vec{r}-\vec{r}_i,L_{1i}, L_{2i}, \alpha_{\mathrm{PSF}i}) - P_i^\mathrm{Gauss}(\vec{r}-\vec{r}_i,L_1, L_2, \apsf)\right]_\vec{r},
\end{equation}
which is the difference between the Gaussian-model PSFs for each channel and the Gaussian-model PSFs using the mean PSF parameters ($L_1, L_2, \apsf$) over the set of channels being considered.  For this measure, we have assumed that all PSFs have the same amplitude and central locations. The average over a set of channels of $\Delta \overline{P}_i$ is plotted in the histogram of Figure~\ref{fig:comp_full_gauss_psf}(b). This shows that most sets of PSFs have a spatial variation of  4\% to 5\%, suggesting that the spatial variation of the PSFs across the sub-array is of secondary importance compared to the difference of the real PSF shape from a Gaussian.

\subsection{Variation of PSF amplitudes}\label{sec:app_amp_var_psf}

In Figure~\ref{fig:comp_full_gauss_psf}(c), we consider the variation in amplitude of the PSFs across the array. The Figure shows the standard deviation of the $A_{\mathrm{PSF}i}$, relative to the mean of this quantity, over a set of channels. There is a large variation compared to the previous two measures, with some cases varying up to 40\%, however this is the tail of the distribution, with the most frequent variation around 10\%. We note, however, the amplitudes of the PSFs cancel in the calculation of the covariance function (\ref{covfun}), as shown in Section~\ref{sec:PSF_an_2D}. Therefore, the variation in the amplitudes of the PSFs can have no effect on the correlation function (\ref{correlation_def}).

\subsection{Shifted centres of PSFs}\label{sec:app_centre_var_psf}

When using the Gaussian-model PSFs, we assumed that the peak response was located at the focal points $(r_{\mathrm{focal}i},Z_{\mathrm{focal}i})$ of the detector channels (black dots in Figure~\ref{fig:psf_alpha}). However, as can be seen in Figure~\ref{fig:psf_alpha}, the PSF contours are in fact shifted from their focal centres. In order to measure this discrepancy, we first calculate the location of the weighted mean response of a PSF $P_i^\mathrm{real}$ in the radial direction, relative to the focal point of that PSF:
\begin{equation}
\delta r_i = \frac{\int \mathrm{d} r  \mathrm{d} Z (r-r_{\mathrm{focal}i}) P_i^\mathrm{real}(r-r_{\mathrm{focal}i},Z-Z_{\mathrm{focal}i}) }{ \int \mathrm{d} r  \mathrm{d} Z P_i^\mathrm{real}(r_k-r_{\mathrm{focal}i},Z-Z_{\mathrm{focal}i}) }.
\end{equation}
We consider the standard deviation $\sigma_{\delta r}$ of $\delta r_i$ over a sub-array of BES channels as our measure of the difference between the focal point and the location of the mean response of the PSFs. This is because we are interested only in the irregularity of the PSF response locations, and not in mean shifts to the position of the PSF response, as the latter would not alter the shape of the correlation function.

A similar definition is adopted for the poloidal difference $\delta Z_i$. The two measures $\sigma_{\delta r}$ and $\sigma_{\delta Z}$ are plotted in Figures~\ref{fig:comp_full_gauss_psf}(d) and \ref{fig:comp_full_gauss_psf}(e), respectively. The radial and poloidal variation of the mean PSF response from the regular grid of focal points is at most a few millimetres, which is an order of magnitude smaller than the typical sizes of the PSFs and of turbulent structures. Therefore, these effects can be neglected.

% ------------------------------------------------------------------------------------------------------------------------------------------
% ------------------------------------------------------------------------------------------------------------------------------------------
% ------------------------------------------------------------------------------------------------------------------------------------------

\section*{References}
\bibliographystyle{phaip}
\bibliography{references_thesis.bib}
%\bibliography{PSF_Paper_Bib}

\end{document}